%% file: main.tex
\documentclass[12pt]{article}

\usepackage{arxiv}
\usepackage[T1]{fontenc}
\usepackage[utf8]{inputenc}
\usepackage{palatino}
\usepackage{microtype}

\usepackage{graphicx}
\usepackage{booktabs}
\usepackage{verbatim}
\usepackage{amsmath}
\usepackage{amssymb}
\usepackage{xcolor}
\usepackage{multirow}
\usepackage[numbers]{natbib}

\usepackage{longtable}
\usepackage{listings}
\usepackage{wrapfig}
\usepackage[skins,breakable]{tcolorbox}

\usepackage[font=small, labelfont=bf, labelsep=period]{caption}

\usepackage{titlesec}
\titleformat{\section}{\large\bfseries}{\thesection}{1em}{}
\titleformat{\subsection}{\normalsize\bfseries}{\thesubsection}{1em}{}
\titleformat{\subsubsection}{\normalsize\itshape}{\thesubsubsection}{1em}{}

\usepackage{xspace}
\makeatletter
\DeclareRobustCommand\onedot{\futurelet\@let@token\@onedot}
\def\@onedot{\ifx\@let@token.\else.\null\fi\xspace}
\def\eg{e.g\onedot}

\makeatother

\definecolor{colHyp}{HTML}{80A2D8}
\definecolor{colP1}{HTML}{97C1A0}
\definecolor{colP2A}{HTML}{BB978F}
\definecolor{colP2B}{HTML}{B3ACCB}
\definecolor{colP3}{HTML}{E9A7A9}
\definecolor{colEco}{HTML}{E1B787}
\definecolor{colAudit}{HTML}{93B4C1}
\definecolor{linkblue}{RGB}{15,70,140}

\newtcolorbox{phasebox}[2][]{%
  breakable,
  enhanced,
  width=\phaseboxwidth,
  center,
  colback=#2!8, colframe=#2!80!black,
  fonttitle=\bfseries, title={#1},
  boxrule=0.6pt, arc=2pt,
  left=4pt, right=4pt, top=2pt, bottom=2pt,
  before skip=6pt, after skip=6pt}

\newcommand{\framework}{\textsc{Veritas}}

\newcommand{\fullframeworkbold}{\textbf{V}erifiable \textbf{E}pistemic \textbf{R}easoning for \textbf{I}mage-Derived Hypothesis \textbf{T}esting via \textbf{A}gentic \textbf{S}ystems}
\newcommand{\yes}{\textsc{Yes}}
\newcommand{\no}{\textsc{No}}
\newcommand{\inconclusive}{\textsc{Inconclusive}}

\newcommand{\badgexs}{0.15\linewidth}
\newcommand{\badgesm}{0.22\linewidth}
\newcommand{\badgemd}{0.30\linewidth}
\newcommand{\badgelg}{0.45\linewidth}

\newcommand{\phaseboxwidth}{0.82\textwidth}

\usepackage[
  colorlinks=true,
  linkcolor=linkblue,
  citecolor=linkblue,
  urlcolor=linkblue,
  breaklinks=true,
  pdftitle={VERITAS: Verifiable Epistemic Reasoning for Image-Derived Hypothesis Testing via Agentic Systems},
  pdfauthor={Lucas Stoffl, Benedikt Wiestler, Johannes Paetzold},
  pdfkeywords={Multi-Agent Systems, LLMs, Hypothesis Testing, Medical Image Analysis}
]{hyperref}

\usepackage[capitalize]{cleveref}

\newtcolorbox{codebox}{%
  breakable,
  enhanced,
  colback=white, colframe=black!60,
  boxrule=0.4pt, arc=0pt,
  left=4pt, right=2pt, top=2pt, bottom=2pt,
  before skip=6pt, after skip=6pt}

\newtcolorbox{promptbox}[1][]{%
  breakable,
  enhanced,
  colback=black!4, colframe=black!35,
  fonttitle=\bfseries, title={#1},
  boxrule=0.5pt, arc=2pt, left=4pt, right=4pt, top=2pt, bottom=2pt,
  before skip=6pt, after skip=6pt}

\graphicspath{{./}{figures/}{figures_supp/}}

\renewcommand{\shorttitle}{\framework{}}

\begin{document}

\title{\framework{}: A Multi-Agent Co-Scientist for\\ Verifiable Image-Derived Hypothesis Testing}

\author{%
  \textbf{Lucas Stoffl}$^{1,2}$, \textbf{Benedikt Wiestler}$^{3,4,5}$, \textbf{Johannes C. Paetzold}$^{1,2}$ \\[2mm]
  $^{1}$Department of Radiology, Weill Cornell Medicine, New York, USA \\
  $^{2}$Cornell Tech, New York, USA \\
  $^{3}$TUM University Hospital, Munich, Germany \\
  $^{4}$Technical University of Munich, $^{5}$Munich Center for Machine Learning \\[1mm]
  \texttt{lms465@cornell.edu, jpaetzold@med.cornell.edu}
}

\date{}

\maketitle

\begin{abstract}
Scientific research based on multimodal clinical data (including medical imaging) requires coordinating clinical, radiological, programming, and biostatistical expertise, a fragmented process that bottlenecks discovery. We present \framework{} (\fullframeworkbold{}), a clinical co-scientist: a multi-agent system that \textbf{autonomously} tests natural-language hypotheses and produces a fully auditable evidence trail, tracing every conclusion through executable outputs from analysis plan to segmentation masks to statistical code to final verdict. Unlike prior AI-scientist systems, which mainly operate on tabular or text data, \framework{} grounds autonomous discovery directly in medical images. It decomposes the workflow into four phases handled by role-specialized agents, and introduces an \textit{epistemic evidence label framework} that mechanically classifies outcomes as \textsc{Supported}, \textsc{Refuted}, \textsc{Underpowered}, or \textsc{Invalid} by jointly evaluating significance, effect direction, and study power. This distinction is critical in medical imaging, where non-significant results often reflect insufficient sample size rather than absent effects. We construct a tiered benchmark of 64 hypotheses spanning six complexity levels across cardiac and brain glioma MRI datasets. \framework{} reaches 81.4\% verdict accuracy with frontier models and 71.2\% with locally-hosted open-weight models (8--30B), outperforming all single-model baselines in both classes. It also produces the highest rate of independently verifiable statistical outputs (86.6\%), so even its failures remain diagnosable through artifact inspection. Structured multi-agent decomposition thus substitutes for model scale while preserving the verifiability that scientific discovery demands. We release code, hypothesis bank, and evaluation pipeline at \url{https://github.com/LucZot/veritas}.
\end{abstract}


\input{sec/01_introduction}

\input{sec/02_related_work}

\input{sec/03_methods}

\input{sec/04_results}

\input{sec/05_discussion}

\section*{Acknowledgments}
We thank Adina Scheinfeld for help with figures and writing.

\bibliographystyle{unsrtnat}
\bibliography{main}

\newpage

\appendix
\input{sec/supp_body}

\end{document}

%% file: sec/01_introduction.tex
\section{Introduction}
\label{sec:intro}

Scientific knowledge in medicine advances when clinicians translate bedside observations into formal hypotheses and test them in observational studies. Doing so requires coordinating expertise across clinical knowledge, patient data curation and analysis, programming, and biostatistics. In the data domain, medical images are indispensable; they
encode rich clinical information, including volumetric measurements, morphological features, and their associations with patient outcomes. Yet turning these image data into statistically valid insights, i.e., the gold standard for scientific evidence, remains a particularly fragmented, expert-intensive process. For example, a neurooncologist who wants to study if the more aggressive tumor biology of glioblastoma is reflected in larger contrast-enhancing tumors compared to lower-grade gliomas must coordinate image segmentation, metric extraction, statistical test selection, and result interpretation across multiple software tools and disciplinary boundaries. This pipeline bottlenecks discovery and limits who can participate in image-based research.

Recent ``AI scientist'' and ``co-scientist'' efforts apply large language models (LLMs) and multi-agent systems to automate scientific discovery~\cite{lu2024ai,gottweis2025towards,ghareeb2025robin}. Extending them to medical imaging hypothesis testing poses three challenges:
(i)~the pipeline must bridge complex and distinct tasks, including \emph{cohort definition} (selecting the right patients from a given dataset), \emph{vision} (segmenting anatomical structures), \emph{computation} (deriving biomarkers), and \emph{statistics} (selecting, executing and interpreting tests), requiring models to reason jointly across all tasks; (ii)~deriving clinical conclusions demands full \emph{auditability}, where every step from segmentation mask to p-value calculation must be inspectable and reproducible; and (iii)~small clinical cohorts require \emph{epistemic awareness}, which is the ability to distinguish ``no effect detected'' from ``insufficient power to detect an effect.''

\paragraph{Our main contribution} is \framework{} (\fullframeworkbold{}), a multi-agent framework inspired by clinician-scientist workflows. With role-specialized agents (PI, Imaging Specialist, Statistician), \framework{} addresses the above challenges and reaches the following capabilities:

\begin{enumerate}

    \item \textbf{Fully autonomous clinical hypothesis testing.} \framework{} transforms natural language hypotheses into statistically validated conclusions with zero human intervention, managing the full study cycle to generate structured plans, executable code, statistics, and visualizations.

    \item \textbf{Multimodal clinical data understanding.} The framework integrates medical imaging (e.g., MRI) and clinical metadata using specialized tools, including promptable segmentation models, to derive quantitative biomarkers directly from raw image data.

    \item \textbf{Auditable and verifiable at each step.} \framework{} produces a complete, transparent execution trail. All results are derived from executable Python scripts and verifiable segmentation masks, allowing clinicians to reproduce and audit both the underlying logic and visual evidence.

    \item \textbf{High performance in validation.} Validated on 64 hypotheses across cardiac and neuro-oncology domains, \framework{} achieves 71.2\% majority-vote verdict accuracy with locally-deployed open-weight models and 81.4\% with frontier models, outperforming all baselines in each model class.

    \item \textbf{Structured decomposition over scale.} By utilizing role-specialized agents, \framework{} enables small, locally-deployed models (8--30B) to perform complex reasoning that typically requires frontier models, facilitating secure, on-site clinical deployment.

\end{enumerate}

\paragraph{Further contributions:} To systematize the conclusions our agentic system produces, we propose the \textbf{epistemic evidence label framework}. Here, we introduce a four-label classification \textsc{Supported}, \textsc{Refuted}, \textsc{Underpowered}, \textsc{Invalid} that mechanically resolves the epistemic status of a hypothesis from observed statistics and a smallest effect size of interest (SESOI), without relying on agent judgment. This distinction is essential for medical imaging benchmarks, where mistaking underpowered subgroup analyses as refutations inflates apparent accuracy (Sec.~\ref{sec:evidence}).

A central question is whether a multi-agent system like \framework{} can deliver correct conclusions and verifiable evidence across increasing complexity; to answer it, we contribute a tiered \textit{evaluation benchmark} spanning six complexity levels, from metadata-only queries (L1) to multivariate survival models (L5), and untestable hypotheses (L0). For all hypotheses, the ground truth was established through dataset-derived reference statistics validated by domain experts. The benchmark includes positive, negative, no-effect, underpowered, and untestable controls, thereby probing the full range of the evidence-label framework. We evaluate on two public datasets across distinct clinical domains: cardiology (ACDC~\cite{bernard2018deep}, cardiac MRI, 150 subjects) and neuro-oncology (UCSF-PDGM~\cite{calabrese2022university}, brain glioma MRI, 501 subjects).

\begin{figure}[t]
\centering
\includegraphics[width=\textwidth]{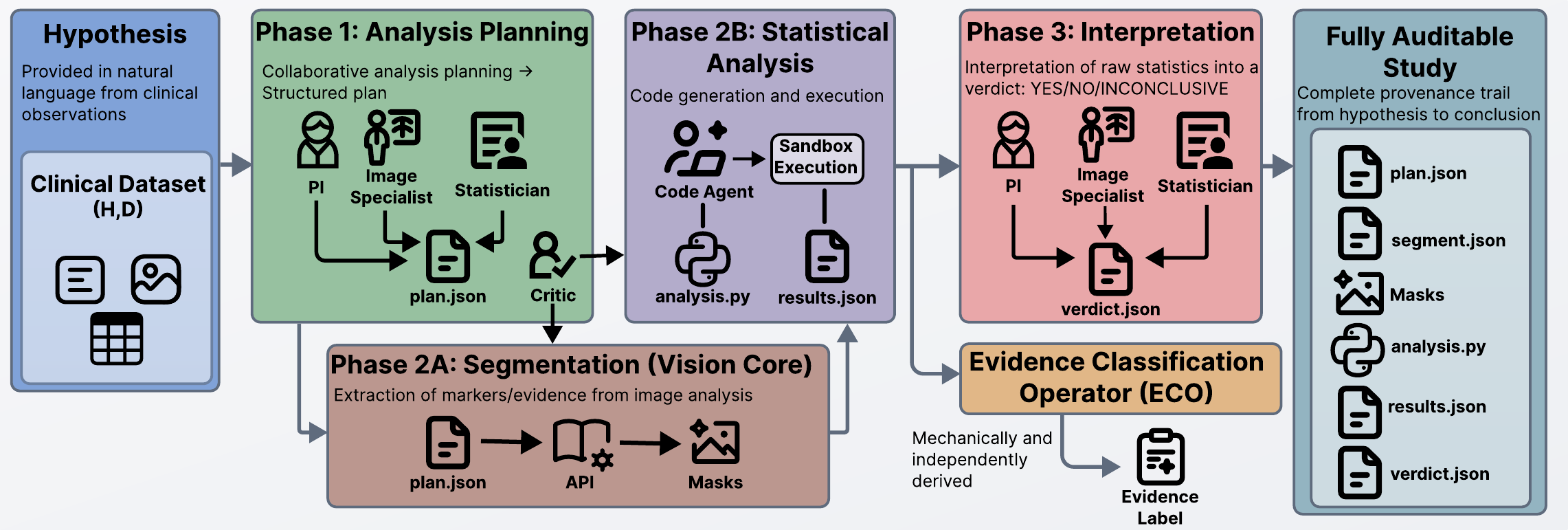}
\caption{\textbf{\framework{} architecture and evidence flow.} Given a natural language hypothesis and dataset, three role-specialized agents (PI, Imaging Specialist, Statistician) with a Critic collaborate through four phases: (1)~collaborative analysis planning producing a structured plan; (2A)~neural segmentation, grounding all evidence in image-derived masks; (2B)~sandboxed code generation and execution yielding statistical outputs; and (3)~interpretation of the Phase~2B statistical outputs in the context of the accumulated workflow history to produce a verdict. The Evidence Classification Operator (ECO) mechanically derives the evidence label from Phase~2B outputs. Every phase produces a versioned artifact (right), forming a complete provenance trail from hypothesis to conclusion.}
\label{fig:overview}
\end{figure}

%% file: sec/02_related_work.tex
\section{Related Work}
\label{sec:related}

Recent advances span medical image foundation models, agentic scientific systems, tool-augmented language models, and statistical best practices. Each addresses a component of the scientific workflow, but they have largely evolved independently; we review them below and position \framework{} as a unifying, auditable hypothesis-testing framework.

\paragraph{Medical image analysis and vision--language models.}

Foundation models have advanced promptable, generalizable medical segmentation across anatomies and modalities. The Segment Anything Model (SAM)~\cite{kirillov2023segment} introduced spatial-prompt segmentation, since extended to video and concept prompts~\cite{ravi2024sam,carion2025sam} and adapted to the medical domain by MedSAM and successors~\cite{ma2024segment,ma2025medsam2,zhu2024medical}. SAT~\cite{zhao2025large} and VoxTell~\cite{rokuss2025voxtell} further allow targets to be specified in natural language, unlike specialist architectures such as nnU-Net~\cite{isensee2021nnu} that require dedicated training per structure. This makes segmentation programmatically accessible to the code-writing agents our framework relies on.
 
In parallel, biomedical vision–language models reason jointly over images and clinical text~\cite{bannur2023learning,li2023llava,zhao2024biomedparse}. Closest to an agent-driven imaging pipeline, VoxelPrompt~\cite{hoopes2024voxelprompt} uses language-guided code generation to analyze 3D medical images. Yet these systems operate at the perception level (segmentation, retrieval, question answering) rather than cohort-level inference: none derives statistically validated conclusions from image-derived measurements across patients.

\paragraph{LLMs for scientific discovery and hypothesis testing.}

Large language models increasingly support scientific workflows such as literature synthesis, hypothesis generation, and experimental planning~\cite{openai2025gpt52,anthropic2024claude,comanici2025gemini,wei2025ai,gao2024empowering}, and several systems automate full research cycles spanning ML experimentation, manuscript writing, and wet-lab validation~\cite{lu2024ai,gottweis2025towards,ghareeb2025robin,mall2025disciple}. Popper~\cite{huang2025automated} is particularly related, validating hypotheses through sequential falsification with formal error control, but operates on tabular and text data without medical imaging or epistemic power analysis. More broadly, programmatic reasoning, where models generate and execute code rather than reasoning in text, improves reliability on quantitative tasks~\cite{chen2022program,gao2023pal,wang2024executable}, motivating \framework{}'s code-first statistical analysis.

In the medical domain, agent-based frameworks apply LLM collaboration to diagnostic reasoning and clinical decision support~\cite{chen2025radfabric,tang2024medagents,kim2024mdagents,jin2025agentmd,li2024agent,nori2025sequential}. Most closely related, MedAgent-Pro~\cite{wang2025medagent} pairs segmentation tools with coding agents to compute clinical indicators from images, but targets per-patient diagnosis rather than population-level hypothesis testing. None of these systems address vision-grounded hypothesis testing, which requires bridging image-derived measurements with cohort-level statistical analysis and epistemic classification.

\paragraph{Multi-agent systems for scientific reasoning.}

Multi-agent systems decompose complex scientific workflows across specialized roles, with teams that collaboratively plan, code, and execute experiments~\cite{swanson2025virtual,zhang2026virtual,schmidgall2025agent,su2025many,baek2024agent,ghafarollahi2025sciagents} built on general-purpose orchestration frameworks~\cite{wu2024autogen,hong2023metagpt}; MetaGPT's insight that structured intermediate artifacts curb cascading hallucinations directly informs our design. A recent independent evaluation, however, found that none of several such frameworks completed an end-to-end research cycle, with agents frequently hallucinating results during implementation~\cite{agrawal2026can} (detailed in the supplementary material). Closest to our setting, MESHAgents~\cite{zhang2025multi} orchestrates agents for phenotype-wide association studies in cardiovascular imaging, but reasons over pre-extracted tabular phenotypes rather than deriving measurements from images, and provides no epistemic classification of evidence strength.

\paragraph{Statistical evidence and epistemic classification.}
 
Equivalence testing via the TOST procedure and the smallest effect size of interest (SESOI) provides a principled way to distinguish ``no meaningful effect'' from ``insufficient power to detect''~\cite{lakens2017equivalence}; Lakens et al.~\cite{lakens2018equivalence} formalized this into a four-outcome decision framework that is the direct conceptual precursor of our evidence label taxonomy (\textsc{Supported}, \textsc{Refuted}, \textsc{Underpowered}, \textsc{Invalid}). \framework{} operationalizes this power-aware classification within an automated pipeline: the Evidence Classification Operator mechanically maps observed statistics to epistemic labels without relying on agent judgment, automating a framework that prior work proposed only as manual analytical guidance (see the supplementary material for an extended discussion).

\medskip
\noindent Across these four directions, no prior system unifies vision-grounded measurement, executable statistical validation, and power-aware evidence calibration within a single auditable pipeline: imaging models stop at perception, scientific and medical agents operate largely on tabular or textual data, and statistical-evidence frameworks remain manual analytical guidance. \framework{} closes this gap by linking segmentation outputs, statistical code, and final conclusions through verifiable artifacts, enabling autonomous, transparent, and reproducible hypothesis testing on medical imaging datasets.

%% file: sec/03_methods.tex
\section{Methods}
\label{sec:methods}

We present \framework{}, a multi-agent framework that transforms natural-language hypotheses about medical imaging datasets and their associated metadata into statistically validated conclusions through autonomous, auditable execution.
\subsection{Problem Formulation}
\label{sec:problem}

Given a dataset $\mathcal{D} = \{(x_i, m_i)\}_{i=1}^{N}$ where $x_i$ represents medical images for patient $i$ and $m_i$ denotes patient metadata (including cohort variables, clinical covariates, and outcomes such as survival endpoints), together with a natural-language hypothesis $\mathcal{H}$, we seek to autonomously determine whether $\mathcal{H}$ is \textsc{Supported}, \textsc{Refuted}, \textsc{Underpowered}, or \textsc{Invalid} given the available evidence.

Unlike traditional computational pipelines that require explicit metric definitions and test specifications, our framework interprets $\mathcal{H}$ semantically, checks feasibility against available data, designs an appropriate analysis plan, and executes it end-to-end with an auditable artifact trail.

\subsection{Framework Architecture}
\label{sec:architecture}

\framework{} operates through four sequential phases, each producing structured, inspectable outputs that feed into subsequent phases (Fig.~\ref{fig:overview}). Three role-specialized agents collaborate: a \textbf{Principal Investigator} (scientific validity, feasibility), a \textbf{Medical Imaging Specialist} (segmentation targets, imaging protocols), and a \textbf{Statistician} (test selection, effect sizes, power analysis). A separate \textbf{Critic} agent operates by default in execution phases (2A/2B).
We evaluate two model deployments with the same workflow and prompts: (i) a local open-weight team (GPT-OSS-20B~\cite{openai2025gptoss} for PI/imaging, Qwen3-8B~\cite{yang2025qwen3} for discussions and critic, Qwen3-Coder-30B~\cite{cao2026qwen3} for coding), and (ii) a frontier team (GPT-5.2~\cite{openai2025gpt52} for PI/imaging/coding and GPT-5 Mini~\cite{openai2025gpt5mini} for discussions and critic).

\paragraph{Dataset API abstraction.}
To keep the framework dataset-agnostic and auditable, agents never access raw files directly during analysis. Instead, they interact through a constrained API layer that exposes cohort discovery, per-patient metadata, observation identifiers, segmentation masks, and geometry-aware measurement utilities. This API boundary standardizes provenance and enables automated checks for off-contract data access. Detailed API functions are described in the supplementary material.

\paragraph{Phase 1: Analysis Planning.}
Given $\mathcal{H}$ and dataset metadata (available groups, observations, patient fields), the agents collaboratively produce a structured analysis plan specifying the target cohorts, population restriction, required anatomical structures and observations, derived measurements, and the planned statistical analysis. The plan includes a feasibility decision (\texttt{TESTABLE}/\texttt{UNTESTABLE}) with subtype and missing requirements; a programmatic validator checks schema completeness and consistency against the available dataset metadata before execution. If marked untestable, the run terminates as \textsc{Invalid}. The Statistician also computes \textit{a priori} power from the queried cohort sizes under a fixed planning SESOI.

\paragraph{Phase 2A: Segmentation.}
When the hypothesis requires image-derived measurements, the Imaging Specialist generates a segmentation request specifying the relevant patients, target structures, and observations. A segmentation backend (SAT~\cite{zhao2025large}) processes this request and stores the resulting binary masks in a shared database. This phase is the vision core of the framework: subsequent quantitative analyses are grounded in measurements extracted from these masks. For hypotheses that require no imaging structures, this phase is bypassed.

\paragraph{Phase 2B: Statistical Analysis.}
The Statistician writes executable code that retrieves the required masks and metadata through the API, computes the planned measurements, performs the statistical test, and reports effect size $\delta$, 95\% confidence interval $\text{CI}_{95}$, p-value $p$, and sample sizes. The code is executed in a sandboxed environment with timeout constraints. Failed or invalid attempts trigger iterative revision using execution traces and validator feedback, with optional Critic input, up to 8 attempts per round.

\paragraph{Phase 3: Interpretation.}
Agents synthesize the prior workflow outputs through structured discussion, producing a verdict $v \in \{\texttt{YES}, \texttt{NO}, \texttt{INCONCLUSIVE}\}$ with supporting rationale. Phase~3 has access to the relevant planning and statistical outputs from earlier phases, while the evaluator-computed evidence label (see Sec.~\ref{sec:evidence}) is explicitly withheld.

\begin{figure}[t]
\centering
\includegraphics[width=\textwidth]{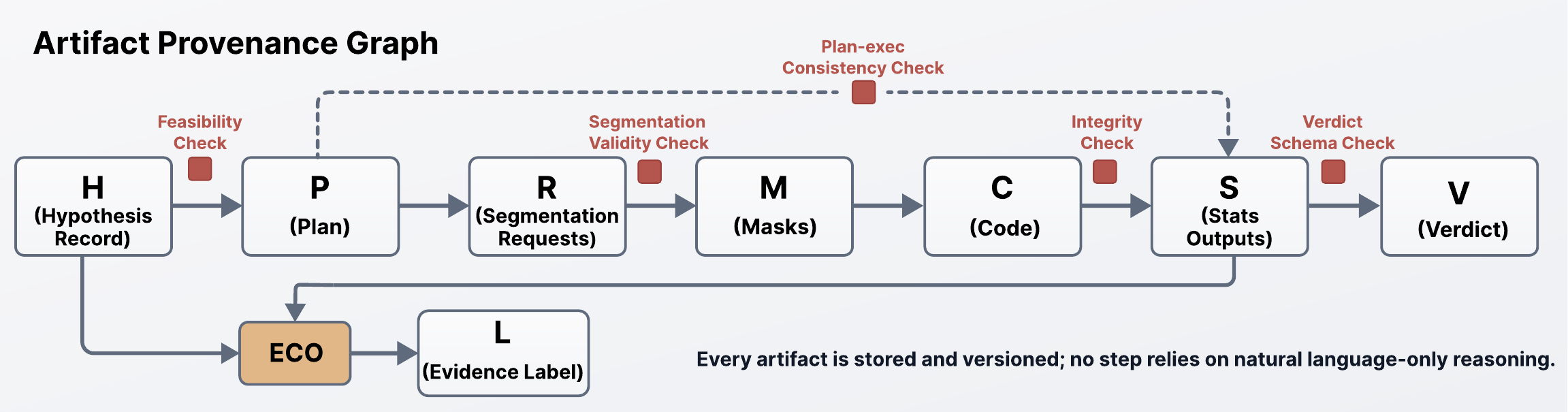}
\caption{\textbf{Auditability through artifact provenance.} Every phase produces versioned, inspectable artifacts. Solid checks are in-workflow validity gates; the dashed check is a post-hoc plan--execution consistency test applied by the evaluator. The evidence label is derived mechanically from Phase~2B statistics via the Evidence Classification Operator (ECO; Sec.~\ref{sec:evidence}), independent of agent judgment.}
\label{fig:auditability}
\end{figure}

\paragraph{Auditability.}
A key design goal is that every analysis is fully auditable. Each phase produces structured artifacts (e.g., JSON plans, segmentation requests, Python code, statistical outputs, verdicts) rather than natural language summaries (Fig.~\ref{fig:auditability}). Validity checks enforce consistency at each boundary, and the final statistical output is retained in a form that enables independent re-verification (see the \emph{evidence grounding} metric in Sec.~\ref{sec:evaluation}). The supplementary material provides the full list of per-phase validity checks.

\subsection{Evidence Label Framework}
\label{sec:evidence}

A central contribution is our four-label evidence classification that jointly evaluates statistical significance, effect direction, and study power (Table~\ref{tab:evidence}).

\begin{table}[t]
\centering
\caption{Evidence label definitions. Labels are mechanically derived from Phase~2B statistics with no agent judgment involved.}
\label{tab:evidence}
\resizebox{0.85\textwidth}{!}{
\begin{tabular}{lll}
\toprule
\textbf{Label} & \textbf{Condition} & \textbf{Interpretation} \\
\midrule
\textsc{Supported} & $p < \alpha \land $ direction matches & Evidence confirms hypothesis \\
\textsc{Refuted} & $(p \geq \alpha \land \pi \geq 0.80)$ & Adequate power, no effect \\
                 & $\lor\, (p < \alpha \land$ opposite dir.) & Significant opposite effect \\
\textsc{Underpowered} & $p \geq \alpha \land \pi < 0.80$ & Insufficient power to conclude \\
\textsc{Invalid} & Untestable or execution failure & Results unreliable \\
\bottomrule
\end{tabular}
}
\end{table}

\noindent Here, $\pi$ denotes statistical power computed at a smallest effect size of interest (SESOI) and $\alpha = 0.05$. This functions as a \emph{decision-theoretic epistemic operator}: given observed statistics $(p, \delta, n)$ and a pre-specified SESOI $\delta_0$, the evidence label is a deterministic function, fully removing subjective judgment from the classification. The epistemic distinction between \textsc{Refuted} and \textsc{Underpowered} is critical for medical imaging benchmarks: a non-significant result with power $\pi = 0.38$ (e.g., $n=13$ patients) provides essentially no evidence against the hypothesis, yet standard binary evaluation would score it as a correct rejection. Details on power computation are in the supplementary material.

\subsection{Benchmark \& Evaluation}
\label{sec:evaluation}

For reproducible benchmarking, we construct a bank of 64 hypotheses (32 per dataset), spanning six complexity tiers (Table~\ref{tab:tiers}). Each tier contains a mix of positive controls (37), negative/no-effect controls (16), and underpowered or untestable controls (11). Ground-truth evidence labels are established through canonical reference statistics computed from ground-truth segmentation masks and patient metadata, validated by a domain expert. Full per-hypothesis specifications are in the supplementary material.

\begin{table}[t]
\centering
\caption{Tiered hypothesis bank. Six complexity tiers probe progressively harder capabilities. \textbf{Neg.} includes negative, no-effect, and nonsense controls. \textbf{Undp./Unt.\ }= underpowered ($\pi < 0.80$) or untestable hypotheses.}
\label{tab:tiers}
\resizebox{0.85\textwidth}{!}{
\begin{tabular}
{@{}cl@{\hspace{6pt}}r@{\hspace{8pt}}l@{\hspace{8pt}}c@{\hspace{6pt}}c@{\hspace{6pt}}c@{}}
\toprule
\textbf{Tier} & \textbf{Name} & $n$ & \textbf{Capability probed} & \textbf{Pos.} & \textbf{Neg.} & \textbf{Undp./Unt.} \\
\midrule
L0 & Untestable       &  5 & Feasibility detection           & --  & --  & 5 \\
L1 & Metadata-only    &  9 & Statistical reasoning           & 3   & 5   & 1 \\
L2 & Single imaging   & 23 & Vision-to-statistics pipeline    & 15  & 8   & -- \\
L3 & Engineered feat. &  6 & Feature engineering              & 6   & --  & -- \\
L4 & Mixed            & 16 & Multi-modal integration          & 8   & 3   & 5 \\
L5 & Multivariate     &  5 & Advanced statistical reasoning   & 5   & --  & -- \\
\midrule
   & \textbf{Total}   & \textbf{64} &                          & \textbf{37} & \textbf{16} & \textbf{11} \\
\bottomrule
\end{tabular}
}
\end{table}

Unless stated otherwise, all accuracy metrics are computed on \emph{testable} hypotheses (L1--L5); L0 feasibility accuracy is reported separately. We evaluate with:

\begin{itemize}
    \item \textbf{Evidence-label accuracy}: agreement between evaluator-computed evidence labels (\textsc{Supported}, \textsc{Refuted}, \textsc{Underpowered}) and ground truth, requiring executable statistical output.
    \item \textbf{Verdict accuracy}: agreement between final verdicts (\yes, \no, \inconclusive) and ground truth.
    \item \textbf{Majority-vote accuracy}: hypothesis-level aggregation over repeated runs; the majority verdict/label across runs is compared to ground truth.
    \item \textbf{L0 feasibility accuracy}: fraction of untestable hypotheses correctly classified as \textsc{Invalid}, reported separately from testable hypothesis performance.
\end{itemize}

\noindent For repeated runs, we report both run-level (micro-averaged) and majority-vote (hypothesis-level) metrics. Evidence-label and verdict accuracy are each computed over runs producing valid output for that metric; these pools may differ when a run yields a verdict but incomplete statistical output. We report completion rate separately for \framework{} to quantify execution reliability. We additionally report \textbf{diagnostic rates} across three categories: \emph{conclusion quality} (overclaim, false refutation), \emph{analysis integrity} (hallucinated significance, synthetic data, literal $p$-value assignment), and \emph{auditability} (evidence grounding rate, measuring whether runs report the four core statistical outputs required for independent verification). Complete metric definitions are in the supplementary material.

%% file: sec/04_results.tex
\section{Experiments}
\label{sec:experiments}
\subsection{Experimental Setup}
\label{sec:setup}

\subsubsection{Datasets}
\label{sec:datasets}

We evaluate on two public medical imaging datasets from cardiology and neuro-oncology (per-dataset tier coverage in the supplementary material).

\textbf{ACDC}~\cite{bernard2018deep} contains 150 subjects (100 training + 50 testing) with short-axis cardiac cine MRI across five diagnostic groups (30 per group): dilated cardiomyopathy (DCM), hypertrophic cardiomyopathy (HCM), myocardial infarction (MINF), right ventricular abnormality (RV), and normal controls (NOR). Each subject includes end-diastolic (ED) and end-systolic (ES) phases with expert-annotated ground-truth segmentations of left ventricle (LV), right ventricle (RV cavity), and myocardium. Patient metadata includes height, weight, and cardiac timing parameters. Hypotheses span L0--L4.

\textbf{UCSF-PDGM}~\cite{calabrese2022university} is a preoperative diffuse glioma MRI dataset with 501 subjects across WHO grades 2-4 (2: 56, 3: 43, 4: 396) and molecular markers (IDH mutation, MGMT methylation, 1p/19q codeletion). Ground-truth tumor segmentation masks, following the BraTS conventions~\cite{menze2014multimodal, bakas2018identifying}, are available for 495 subjects. Patient metadata includes age, sex, overall survival (395 with survival data; 228 events), and extent of resection. The SAT model~\cite{zhao2025large} segments four brain tumor subregions: whole tumor (WT), necrotic core (NCR), peritumoral edema (ED), and enhancing tumor (ET). Hypotheses span L0--L5, including survival analyses and multivariate regression.
\subsubsection{Baselines}
\label{sec:baselines}
We compare \framework{} against a structured ablation of five baselines.

\paragraph{Single-model baselines (SMa--SMe).}
To isolate the contributions of code execution, data access strategy, iterative self-correction, and structured phase guidance, we define our baselines forming an ablation progression:
\begin{itemize}
    \item \textbf{SMa: Direct reasoning.} The model receives per-group summary statistics pre-computed from ground-truth segmentation and produces a verdict without any code.
    \item \textbf{SMb: Code on pre-computed features.} The model receives a per-patient CSV of pre-computed imaging features (volumes, masses, etc.) and writes one Python script executed automatically in the sandbox.
    \item \textbf{SMc: Code via API.} The model receives only the hypothesis and API documentation. It writes a single end-to-end script covering patient discovery, mask loading, metric computation, and statistical testing.
    \item \textbf{SMd: Agentic (iterative).} Same API access as SMc, but the model operates in an iterative loop (up to 3 rounds): write code $\rightarrow$ execute $\rightarrow$ observe output $\rightarrow$ refine or conclude.
    \item \textbf{SMe: Pipeline (structured).} Same API access as SMd, but guided by \framework{}'s exact phase agendas (Phase~1 planning, Phase~2 coding, Phase~3 interpretation) as a single-model chain-of-thought.
\end{itemize}

\noindent These baselines isolate four capability dimensions: (i)~code execution vs.\ pure reasoning (SMa$\to$SMb); (ii)~pre-extracted features vs.\ raw API access (SMb$\to$SMc); (iii)~iterative self-correction (SMc$\to$SMd); and (iv)~structured phase guidance (SMd$\to$SMe). Comparing SMe to \framework{} then directly isolates the contribution of multi-agent discussion over a single model following identical phase instructions.

\paragraph{Model and fairness.}
We evaluate all five baselines with two model families: \textbf{frontier} (GPT-5.2~\cite{openai2025gpt52}) and \textbf{local/open-weight} (GPT-OSS-20B~\cite{openai2025gptoss}). For fairness, all baselines that use the API (SMc--SMe) receive \emph{identical} API documentation to \framework{} agents, extracted directly from the same prompt templates. The full dataset-specific structure catalogue (\eg, \texttt{"LV"}, \texttt{"RV"}, \texttt{"Myo"} for ACDC) is also provided, equivalent to what \framework{}'s Phase~2A agents discover through tool calls; the model must still determine which structures are relevant to the hypothesis. Code execution uses the same sandboxed environment as \framework{}. For SMd and SMe, a single interpretation step (equivalent to \framework{}'s Phase~3) is always invoked after successful code execution; the model's conclusion from this step is used as the final verdict.

\subsection{Main Results}
\label{sec:main_results}

\textbf{Overall performance.} Each method was evaluated with 10 repeated runs per hypothesis (64 hypotheses, 640 runs per method), and we report run-level and majority-vote accuracy across both model families (Table~\ref{tab:main_results}).

\begin{table}[t]
\centering
\caption{Main results on the 64-hypothesis tiered benchmark. Each cell shows accuracy (\%). Evidence accuracy requires executable statistical output (not applicable for SMa). Metrics are computed on testable hypotheses (L1--L5); L0 feasibility is reported separately. \textbf{Compl.}: fraction of runs completing all workflow phases (Sec.~\ref{sec:evaluation}); baselines always produce output by construction.}
\label{tab:main_results}
\resizebox{0.8\textwidth}{!}{
\begin{tabular}{l|cc|cc|c}
\toprule
 & \multicolumn{2}{c|}{\textbf{Run-level}} & \multicolumn{2}{c|}{\textbf{Majority vote}} & \\
\textbf{Method} & \textbf{Evidence} & \textbf{Verdict} & \textbf{Evidence} & \textbf{Verdict} & \textbf{Compl.} \\
\midrule
\multicolumn{6}{l}{\textit{Local/Open-weight models (GPT-OSS-20B)}} \\
SMa: Direct reasoning       & --   & 56.1 & --   & 55.9 & -- \\
SMb: Code on features       & 58.1 & 63.8 & 61.0 & 62.7 & -- \\
SMc: Code via API       & 39.0 & 63.2 & 40.7 & 59.3 & -- \\
SMd: Agentic    & 46.3 & 66.0 & 52.5 & 66.1 & -- \\
SMe: Pipeline  & 55.1 & 58.0 & 57.6 & 61.0 & -- \\
\textbf{\framework{} (local)} & \textbf{63.4} & \textbf{71.4} & \textbf{67.8} & \textbf{71.2} & 78.1 \\
\midrule
\multicolumn{6}{l}{\textit{Frontier models (GPT-5.2)}} \\
SMa: Direct reasoning       & --   & 55.9 & --   & 55.9 & -- \\
SMb: Code on features       & 63.7 & 70.7 & 69.5 & 69.5 & -- \\
SMc: Code via API       & 52.9 & 72.8 & 61.0 & 76.3 & -- \\
SMd: Agentic    & 64.7 & 76.6 & 64.4 & 74.6 & -- \\
SMe: Pipeline   & 60.8 & 65.5 & 59.3 & 61.0 & -- \\
\textbf{\framework{} (frontier)} & \textbf{70.0} & \textbf{81.0} & \textbf{76.3} & \textbf{81.4} & 87.5 \\
\bottomrule
\end{tabular}
}\\[0.2cm]
\footnotesize{\framework{} (local) uses locally-deployed 8--30B models. L0 feasibility: \framework{} 74.0\% (local), 100\% (frontier); all baselines 100\% except SMb-frontier (98\%).}
\end{table}

\paragraph{Key findings.}
\framework{} achieves the highest accuracy in both model families: 71.2\% majority-vote verdict accuracy with locally-deployed open-weight models (8--30B) and 81.4\% with frontier models, outperforming all single-model baselines in each class. The ablation progression reveals consistent patterns:

\textbf{Code execution is essential.} SMa (direct reasoning) achieves only ${\sim}$56\% verdict accuracy regardless of model scale, confirming that pure LLM reasoning over summary statistics is insufficient. Adding code execution (SMb) lifts verdict accuracy by 8--15 percentage points.

\textbf{Iterative self-correction helps selectively.} The SMc$\to$SMd gain is modest for verdict accuracy (+2.8 local/open-weight, +3.8 frontier) but more pronounced for evidence accuracy (+7.3, +11.8), suggesting that self-correction primarily improves statistical analysis quality rather than final interpretation.

\textbf{Multi-agent decomposition adds value.} Comparing SMe to \framework{} isolates the multi-agent contribution: +10.2pp verdict majority vote for local/open-weight (71.2\% vs.\ 61.0\%) and +20.4pp for frontier (81.4\% vs.\ 61.0\%). The larger frontier gap suggests that multi-agent discussion becomes \emph{more} valuable as base model capability increases.

\textbf{Per-dataset gap.} Evidence-label accuracy drops sharply from ACDC to UCSF-PDGM across all methods (Table~\ref{tab:stratified}), reflecting genuine complexity differences (survival analyses, multivariate regression, molecular-marker stratification). On ACDC, frontier \framework{} achieves 90.0\% evidence accuracy and 78.9\% verdict at 88.8\% completion. Pipeline failures on a subset of runs reduce verdict performance on easier hypotheses where baselines rarely fail. Local \framework{}, despite much smaller model capacities, remains competitive and attains the highest L5-tier scores. \framework{}'s advantage emerges on harder UCSF-PDGM hypotheses, where it leads in both evidence (+16pp over SMd) and verdict accuracy (+29pp).

\textbf{Stratified results.} Table~\ref{tab:stratified} provides complementary breakdowns by dataset and by complexity tier, comparing both \framework{} configurations against the three strongest frontier baselines.

\begin{table}[t]
\centering
\caption{Stratified run-level accuracy (\%) by dataset and complexity tier. Per-dataset: evidence / verdict. Per-tier: evidence-label accuracy. Best non-\framework{} result per tier is \underline{underlined}. L0 = feasibility detection. Full per-method table in supplementary.}
\label{tab:stratified}
\resizebox{0.85\textwidth}{!}{
\begin{tabular}{l|cc|cc|c|ccccc|c}
\toprule
 & \multicolumn{2}{c|}{\textbf{ACDC}} & \multicolumn{2}{c|}{\textbf{UCSF-PDGM}} & & \multicolumn{5}{c|}{\textbf{Evidence by tier}} & \\
\textbf{Method} & \textbf{Ev.} & \textbf{Vd.} & \textbf{Ev.} & \textbf{Vd.} & \textbf{L0} & \textbf{L1} & \textbf{L2} & \textbf{L3} & \textbf{L4} & \textbf{L5} & \textbf{All} \\
\midrule
\framework{} (local) & 91.0 & 87.9 & 36.7 & 49.0 & 74 & 54.4 & 87.0 & 51.7 & 49.4 & \textbf{30.0} & 63.4 \\
\midrule
SMc: Code (GPT-5.2)        & 79.0 & 85.1 & 27.7 & 60.9 & 100 & 75.6 & 70.9 & 43.3 & 34.4 &  0.0 & 52.9 \\
SMd: Agentic (GPT-5.2)     & \textbf{\underline{95.9}} & \textbf{\underline{96.6}} & \underline{34.7} & 53.9 & 100 & \textbf{\underline{94.4}} & \underline{85.2} & \underline{55.0} & \textbf{\underline{42.5}} &  0.0 & \underline{64.7} \\
SMe: Pipeline (GPT-5.2)    & 90.3 & 96.9 & 32.3 & 35.1 & 100 & 56.7 & 89.1 & 56.7 & 43.1 &  0.0 & 60.8 \\
\textbf{\framework{} (frontier)} & 90.0 & 78.9 & \textbf{50.7} & \textbf{83.2} & 100 & 90.0 & 90.4 & \textbf{73.3} & 45.0 & 16.0 & \textbf{70.0} \\

\bottomrule
\end{tabular}
}
\end{table}

\textbf{Tier behavior.} L2 (single imaging metric) is the strongest tier across all methods (71--90\%), as these hypotheses require straightforward segmentation feature comparisons. L4 and L5 are hardest, driven by execution fragility in survival analyses and multi-covariate regression. Critically, \framework{} is the \emph{only} method to achieve non-zero L5 evidence accuracy (16--30\% across configurations), while all single-model baselines score 0\% on L5 despite achieving up to 92\% verdict accuracy on these hypotheses. This indicates they produce correct YES/NO guesses without valid statistical backing. The small L5 sample (5~hypotheses) limits precision, but the qualitative gap (non-zero vs.\ zero) is robust.

\begin{table}[t]
\centering

\caption{Diagnostic rates (\%) for code-producing methods on the combined benchmark. \textbf{Conclusion quality:} \textit{Overclaim}: YES verdicts where evidence is not \textsc{Supported}; \textit{False ref.}: predicted \textsc{Refuted} labels disagreeing with ground truth. \textbf{Analysis integrity:} \textit{Hall.\ sig.}: YES verdict with $p \geq 0.05$ or missing $p$-value (output-level check); \textit{Synth.}: code generating mock/random data instead of using real measurements (code-level); \textit{Lit.\ $p$}: significance threshold hard-coded in analysis code (code-level). \textbf{Auditability:} \textit{Verif.}~($\uparrow$): fraction of all runs reporting the four core statistical outputs (test type, sample sizes, effect size, $p$-value) required for independent verification.}
\label{tab:diagnostics}
\resizebox{0.78\textwidth}{!}{
\begin{tabular}{l cc ccc c}
\toprule
& \multicolumn{2}{c}{\textbf{Conclusion quality}} & \multicolumn{3}{c}{\textbf{Analysis integrity}} & \textbf{Audit.} \\
 \cmidrule(lr){2-3} \cmidrule(lr){4-6} \cmidrule(lr){7-7}
\textbf{Method} & \textbf{Overcl.} & \textbf{False ref.} & \textbf{Hall.\ sig.} & \textbf{Synth.} & \textbf{Lit.\ $p$} & \textbf{Verif.}~$\uparrow$ \\
\midrule
\multicolumn{7}{l}{\textit{Local/Open-weight (GPT-OSS-20B)}} \\
SMb: Features    & 44.7 &  7.4 & 10.9 & 0.2 &  0.7 & 81.9 \\
SMc: Code        & 65.1 &  6.7 &  8.7 & 0.2 &  1.4 & 82.3 \\
SMd: Agentic     & 25.6 & 10.2 &  1.5 & 0.2 & 12.2 & 53.6 \\
SMe: Pipeline    &  3.8 &  8.5 &  0.0 & 0.9 & 17.2 & 57.7 \\
\framework{}     &  7.2 & 24.2 &  2.7 & 10.9 &  2.4 & 77.3 \\
\midrule
\multicolumn{7}{l}{\textit{Frontier (GPT-5.2)}} \\
SMb: Features    & 34.7 & 15.3 & 11.7 & 2.2 &  4.1 & 82.8 \\
SMc: Code        & 45.4 &  8.6 & 11.0 & 1.6 &  2.7 & 81.2 \\
SMd: Agentic     & 14.8 & 11.7 &  0.0 & 0.6 &  3.3 & 70.5 \\
SMe: Pipeline    & 13.0 &  7.9 &  0.0 & 0.5 & 10.9 & 68.1 \\
\framework{}     & 20.0 & 12.6 &  0.4 & 3.6 & 15.0 & 86.6 \\
\bottomrule
\end{tabular}
}
\end{table}

\textbf{L1 scaling behavior.} Evidence accuracy on L1 (metadata-only) reveals a pronounced model-capacity dependence: local \framework{} drops to 54.4\% on L1 versus 87.0\% on L2, while frontier \framework{} maintains near-parity (90.0\% vs.\ 90.4\%). Although L1 hypotheses require no imaging computation, they demand nontrivial statistical design capabilities (survival model specification, covariate adjustment, and appropriate test selection) that benefit disproportionately from larger model capacity. By contrast, L2 hypotheses (single imaging metric comparisons) rely on more mechanical code generation that local models already handle effectively.

\subsection{Diagnostic Analysis}
\label{sec:diagnostics}

Table~\ref{tab:diagnostics} reports diagnostic rates that characterize \emph{how} methods fail, beyond aggregate accuracy; these are conditional rates over method-specific pools, not a flat quality ranking. SMa is excluded because it produces no executable statistical output, making code-level diagnostics inapplicable.

One-shot baselines (SMb, SMc) show high overclaim rates (35--65\%) and hallucinated significance (9--12\%), producing YES verdicts unsupported by their own statistical outputs. Iterative and structured baselines (SMd, SMe) largely eliminate hallucinated significance ($\leq$1.5\%) through self-correction, but trade this for higher literal $p$-value assignment (3--17\%), where models hard-code $p$-values rather than computing them from data.

Local \framework{} shows the lowest overclaim (7.2\%) but the highest false refutation (24.2\%) and synthetic-data rate (10.9\%). Since evidence labels are assigned mechanically by the ECO, false refutation is not a conservative interpretation style but weaker code yielding null or invalid statistics on adequately powered, truly \textsc{Supported} hypotheses. Frontier \framework{} accordingly cuts false refutation to 12.6\% as code improves, while its higher overclaim (20.0\%) and literal-$p$ rate (15.0\%) follow from completing more, and more complex, analyses. The code-integrity scans also charge \framework{} for engaging the hardest tiers (it alone reaches non-zero L5), where baselines short-circuit. Both configurations keep hallucinated significance low ($\leq$2.7\%), the most egregious failure mode the phased pipeline prevents.

The verifiability column (Verif.) quantifies auditability over all runs. Among methods operating on raw data (SMd, SMe, \framework{}), \framework{} achieves the highest verifiability (77.3\% local, 86.6\% frontier), substantially exceeding SMd (53.6\%, 70.5\%) and SMe (57.7\%, 68.1\%). SMb and SMc achieve higher absolute rates (${\sim}$82\%) but operate on pre-computed features, yielding simpler code with higher completion rates, at the cost of zero L5 evidence accuracy. Thus \framework{}'s phased decomposition improves not only accuracy but inspectability, keeping even its failures \emph{recoverable} through artifact inspection.

\subsection{Ablation Studies}
\label{sec:ablations}
Table~\ref{tab:main_ablation} reports ablation results using the local \framework{} configuration (10 runs per hypothesis). We evaluate context window size, critic removal, and temperature zero decoding. Vision source ablations (GT masks, SAT-Pro) are reported in the supplementary material. 

\begin{table}[t]
\centering
\footnotesize
\caption{Ablation results for \framework{} (local, 10 runs per hypothesis). \textbf{Default}: 16k context, critic enabled in execution phases, $T{=}0.2$. Per-dataset split shows the ACDC vs.\ UCSF-PDGM gap across configurations.}
\label{tab:main_ablation}
\resizebox{0.65\textwidth}{!}{
\begin{tabular}{l|cc|cc|cc}
\toprule
 & \multicolumn{2}{c|}{\textbf{ACDC}} & \multicolumn{2}{c|}{\textbf{UCSF-PDGM}} & \multicolumn{2}{c}{\textbf{Combined}} \\
\textbf{Configuration} & \textbf{Evid.} & \textbf{Verd.} & \textbf{Evid.} & \textbf{Verd.} & \textbf{Evid.} & \textbf{Verd.} \\
\midrule
Default (16k context) & 91.0 & 87.9 & 36.7 & 49.0 & 63.4 & 71.4 \\
Context 8k            & 91.4 & 91.7 & 29.0 & 45.7 & 59.7 & 72.0 \\
Context 32k           & 90.0 & 87.1 & 31.7 & 51.8 & 60.3 & 72.7 \\
No Critic             & 90.7 & 90.1 & 32.0 & 50.5 & 60.8 & 73.2 \\
Temperature zero      & 90.3 & 88.7 & 27.7 & 47.5 & 58.5 & 72.6 \\
\bottomrule
\end{tabular}
}
\end{table}

ACDC evidence remains stable across all ablations (${\sim}$90--91\%), indicating that cardiac hypotheses are effectively solved by the base architecture; sensitivity lies almost entirely in UCSF-PDGM evidence, making it the discriminating benchmark. The default 16k context achieves the best UCSF evidence (36.7\%): reducing to 8k degrades it by $-$7.7pp (complex survival/regression analyses no longer fit), while 32k gives no uplift ($-$5.0pp). Deterministic decoding ($T{=}0$) similarly hurts it ($-$9.0pp), consistent with complex hypotheses benefiting from stochastic code exploration. Critic removal has minimal impact ($+$1.8pp combined verdict), indicating the iterative code-revision loop already provides sufficient error correction.

\subsection{Analysis}
\label{sec:analysis}

\textbf{Execution vs.\ reasoning.} Across all methods, verdict accuracy consistently exceeds evidence-label accuracy (\eg, \framework{} local: 71.4\% vs.\ 63.4\%; frontier: 81.0\% vs.\ 70.0\%), indicating that the primary bottleneck is producing correct statistical outputs (code generation), not reasoning about them (interpretation). The gap is most striking for frontier \framework{} on UCSF-PDGM: verdict accuracy (83.2\%) \emph{exceeds} ACDC verdict (78.9\%) despite much lower evidence accuracy (50.7\% vs.\ 90.0\%), identifying code generation for complex analyses (survival models, multivariate regression) as the primary improvement target. Notably, this verdict--evidence gap is much narrower for \framework{} (8--11\,pp overall) than for single-model baselines on comparable tiers (\eg, up to 92\% verdict vs.\ 0\% evidence accuracy on L5), confirming that \framework{}'s verdicts remain substantially grounded in computed statistics rather than prior knowledge alone.

\textbf{Epistemic label analysis.} The four-label evidence framework is essential for fair evaluation. Without the \textsc{Underpowered} label, several UCSF-PDGM hypotheses with small subgroups would be conflated with true negatives: glioma\_31 (1p/19q codeletion, $n{=}13$ vs.\ $n{=}86$, power${=}$0.38) is correctly classified as \textsc{Underpowered} rather than \textsc{Refuted}; glioma\_14 (IDH-mutant Grade~IV, $n{=}28$ vs.\ $n{=}367$, power${=}$0.72) is similarly borderline. In a binary evaluation, a system that conservatively labels every non-significant result as ``refuted'' would achieve spuriously high accuracy. The underpowered distinction forces systems to demonstrate genuine epistemic awareness.

%% file: sec/05_discussion.tex
\section{Discussion and Conclusion}
\label{sec:discussion}

We presented \framework{}, a multi-agent co-scientist for autonomous, auditable hypothesis testing inspired by clinician-scientist workflows. Its design targets three bottlenecks highlighted in the introduction: the fragmented coordination of clinical, imaging, programming, and statistical expertise; the lack of reproducible analysis trails in image-derived studies; and the misinterpretation of underpowered findings as true negatives. By decomposing the workflow into verifiable phases and code, \framework{} makes every step auditable and renders its failures \emph{recoverable} in a way that baseline failures are not.
It achieves strong performance even against baselines that receive human assistance (e.g., pre-computed per-group statistics or per-patient feature tables), and enables small, locally-deployable models to perform complex reasoning that typically requires frontier models.

Recall that \framework{} produces two complementary outcomes: the agents' \emph{verdict} (\yes/\no/\inconclusive), reflecting their synthesized judgment, and the \emph{evidence label} (\textsc{Supported}/\textsc{Refuted}/\textsc{Underpowered}/\textsc{Invalid}), derived mechanically by the Evidence Classification Operator and withheld from the agents. This separation is intentional: comparing the two reveals whether a conclusion is actually warranted by the statistics the agents produced, and distinguishes a true null from a merely underpowered one.
Beyond accuracy, the evidence label framework functions as a mechanistic decision rule: given observed statistics $(p, \delta, \pi)$, the label is a deterministic function with no free parameters beyond the significance level $\alpha$ and the smallest effect size of interest (SESOI, $\delta_0$). This eliminates evaluation ambiguities common in agent benchmarks. We view this as a contribution beyond medical imaging: any benchmark involving statistical testing could adopt this framework to avoid conflating negative results with underpowered designs.

Most agent evaluations report only end-to-end accuracy. For scientific applications, \emph{how} a conclusion is reached matters as much as \emph{whether} it is correct. Our evidence grounding metric (Verif., Table~\ref{tab:diagnostics}) quantifies this: among methods operating on raw data, \framework{} produces the highest rate of fully verifiable outputs (86.6\% frontier), meaning even its incorrect conclusions can be traced, diagnosed, and corrected through artifact inspection.

By reducing the coordination overhead and execution burden of patient selection, image analysis and biomarker extraction, and statistical testing, \framework{} can accelerate clinical research as an auditable collaborator in the human-led scientific discovery loop. Ultimately, a clinical observational study should always be finalized by a human expert, a process that is substantially facilitated through the availability and auditability of the reasoning steps, artifacts, and code.

\medskip
\noindent \textbf{Limitations}

\noindent \textit{Segmentation dependency.} All imaging-derived conclusions depend on segmentation quality. We quantify the gap between ground-truth and SAT segmentations in the supplement, but the framework treats segmentation as a point estimate, leaving uncertainty propagation to future work.

\noindent \textit{Benchmark scope.} Our evaluation covers two MRI domains analyzed by a specific segmentation model; extending to other modalities (CT, histopathology) requires new but widely available segmentation models. The hypothesis bank also does not yet cover longitudinal or causal analyses.

\noindent \textit{Ground truth.} Our ground truth verdicts are established from the given datasets using expert-validated reference computations. This was strictly necessary to ensure mechanical verifiability in our evaluation, but may reflect dataset-specific rather than population-level effects.

%% file: sec/supp_body.tex
\section{Example Workflow Walkthrough}
\label{sec:example-workflow}

This section illustrates the end-to-end \framework{} pipeline on two representative hypotheses: one imaging-based group comparison (ACDC) and one multivariate survival analysis (UCSF-PDGM). For each, we show the artifacts produced at each phase and demonstrate the full audit trail. Both examples use the frontier model configuration (GPT-5.2). The colored phase boxes below follow the phase decomposition in Figure~\ref{fig:overview}.

\subsection{Example 1: Cardiac LVEF Group Difference (L2)}
\label{sec:example-cardiac}

\vspace{2pt}

\begin{phasebox}[Hypothesis]{colHyp}
\begin{wrapfigure}{r}{\badgexs}
  \vspace{-2em}
  \hspace{0pt}\includegraphics[width=0.92\linewidth]{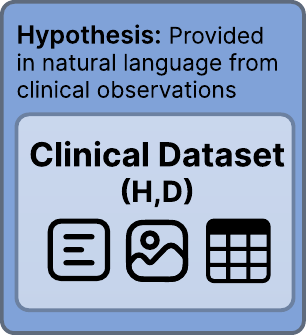}
  \vspace{-0.3em}
\end{wrapfigure}
\vspace{0.6em}
{\large\itshape
“DCM patients show significantly lower LVEF than normal controls.”
\par}
\vspace{0.5em}
{\normalsize
This is an L2 (imaging-only) hypothesis requiring LV cavity segmentation at two cardiac phases and a derived metric (ejection fraction).
\par}
\end{phasebox}

\vspace{2pt}

\begin{phasebox}[Phase 1: Planning]{colP1}
\begin{wrapfigure}{r}{\badgemd}
    \vspace{-0.5em}
    \hspace{0pt}\includegraphics[width=0.92\linewidth]{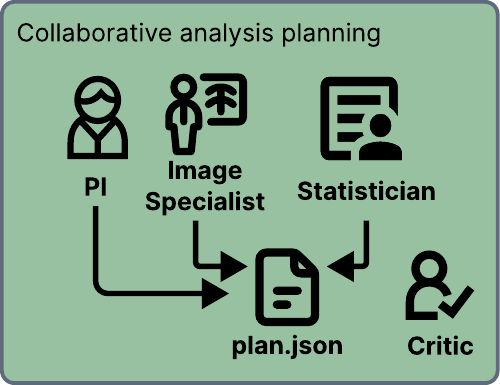}
    \vspace{-1em}
\end{wrapfigure}
The three-agent team aligns on the analysis plan in a structured discussion. The PI confirms the target quantity (\texttt{LVEF = ((LVEDV $-$ LVESV) / LVEDV) $\times$ 100}), rejects proxy substitution, and recommends an unadjusted primary analysis. The Imaging Specialist queries the Imaging Analysis API, confirms structure \texttt{"left heart ventricle"} is available, and retrieves cohort sizes (DCM: $n{=}30$, NOR: $n{=}30$). The Statistician proposes a Welch two-sample $t$-test (two-sided, $\alpha{=}0.05$) and computes planning power ($\pi{=}0.48$ at the fixed Phase~1 reference SESOI $d{=}0.5$), flagging the study as underpowered for a \emph{medium-effect planning target}. This does not stop execution: Phase~1 power is advisory and only affects interpretation if the subsequent result is non-significant. In the benchmark, \texttt{cardiac\_01} carries the evaluator SESOI profile \texttt{loose} ($d_0{=}0.8$), for which the ground-truth power is 0.86.

\smallskip\noindent The Phase~1 output is an executable JSON plan contract:
\begin{codebox}
\begin{lstlisting}[language={},basicstyle=\ttfamily\scriptsize,numbers=none]
{
  "feasibility": {"status": "TESTABLE"},
  "groups": ["DCM", "NOR"],
  "structures": ["left heart ventricle"],
  "observations": ["ED", "ES"],
  "metrics": ["LVEF"],
  "statistical_test": "Welch two-sample t-test",
  "a_priori_power": {"d": 0.5, "n1": 30, "n2": 30, "power": 0.478}
}
\end{lstlisting}
\end{codebox}
\end{phasebox}

\vspace{2pt}

\begin{phasebox}[Phase 2A: Segmentation]{colP2A}
\begin{wrapfigure}{r}{\badgelg}
  \vspace{-0.7em}
  \hspace{1em}\includegraphics[width=0.92\linewidth]{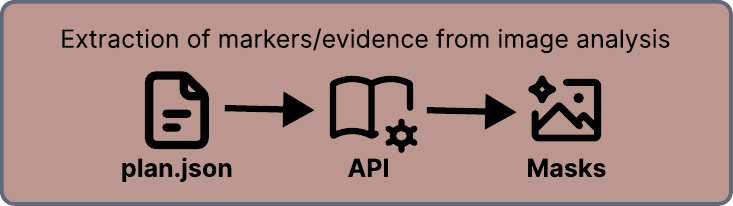}
  \vspace{-0.5em}
\end{wrapfigure}
The agents construct a segmentation request specifying the LV cavity structure at both ED and ES frames. The SAT backend segments all 60 patients (30 DCM + 30 NOR), producing per-patient binary masks. Figure~\ref{fig:cardiac-mri} shows exemplary segmentation overlays on short-axis cine MRI slices.

\smallskip
\begin{center}
\includegraphics[width=0.95\linewidth]{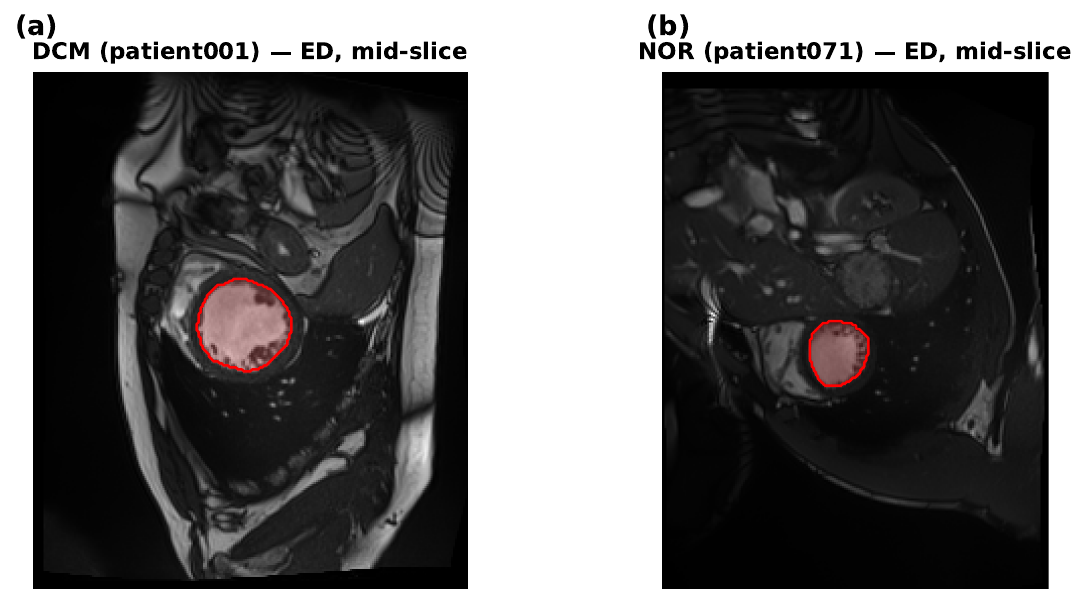}
\captionof{figure}{Our system's LV cavity segmentation (red overlay) on short-axis cardiac cine MRI at end-diastole for (a)~a DCM patient (patient001) showing marked LV dilation and (b)~a normal control (patient071).}
\label{fig:cardiac-mri}
\end{center}
\end{phasebox}

\vspace{2pt}

\begin{phasebox}[Phase 2B: Statistical Analysis]{colP2B}
\begin{wrapfigure}{r}{0.4\linewidth}
  \vspace{-0.3em}
  \hspace{0pt}\includegraphics[width=0.92\linewidth]{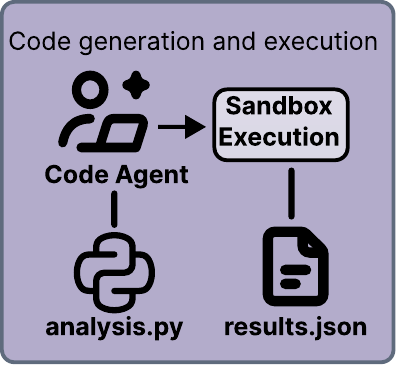}
  \vspace{-0.6em}
\end{wrapfigure}
The coding agent generates a Python script that:
\begin{enumerate}
    \item Iterates over all patients, loading LV masks at ED and ES via the Imaging Analysis API
    \item Computes voxel-level volumes using \texttt{sat.calculate\_volume(mask, spacing)}
    \item Derives per-patient LVEF from the exact formula
    \item Applies five prespecified QC checks (finite volumes, EDV $> 0$, ESV $\geq 0$, ESV $\leq$ EDV)
    \item Performs a Welch $t$-test and saves complete statistics to \texttt{statistical\_results.json}
\end{enumerate}

\noindent Key excerpt from the generated code:
\begin{codebox}
\begin{lstlisting}[language=Python,basicstyle=\ttfamily\scriptsize]
for pid in patients:
    md = sat.get_patient_metadata(pid)
    if md.get("group") not in ["DCM", "NOR"]:
        continue
    obs_map = sat.get_observation_identifiers(pid)
    for obs in ["ED", "ES"]:
        mask = sat.load_structure_mask(
            results_db_path, pid, "left heart ventricle",
            source_image_contains=obs_map[obs])
        vols[obs] = sat.calculate_volume(mask[0]["mask"], mask[0]["spacing"])
    lvef = ((vols["ED"] - vols["ES"]) / vols["ED"]) * 100
\end{lstlisting}
\end{codebox}

\noindent The analysis produces a boxplot visualization (Figure~\ref{fig:cardiac-boxplot}), and the following statistical output:

\begin{center}
\scriptsize
\begin{tabular}{ll}
\toprule
\textbf{Statistic} & \textbf{Value} \\
\midrule
Test & Welch two-sample $t$-test \\
$n$ (DCM / NOR) & 30 / 30 \\
Mean LVEF (DCM) & 18.6\% $\pm$ 8.2\% \\
Mean LVEF (NOR) & 61.2\% $\pm$ 5.3\% \\
Mean difference & $-$42.6\% (95\% CI: $-$46.1 to $-$39.1) \\
$t$-statistic & $-$23.83 (df $=$ 49.6) \\
$p$-value & $7.33 \times 10^{-29}$ \\
Cohen's $d$ & $-$6.15 \\
\bottomrule
\end{tabular}
\end{center}

\smallskip
\begin{center}
\includegraphics[width=0.55\linewidth]{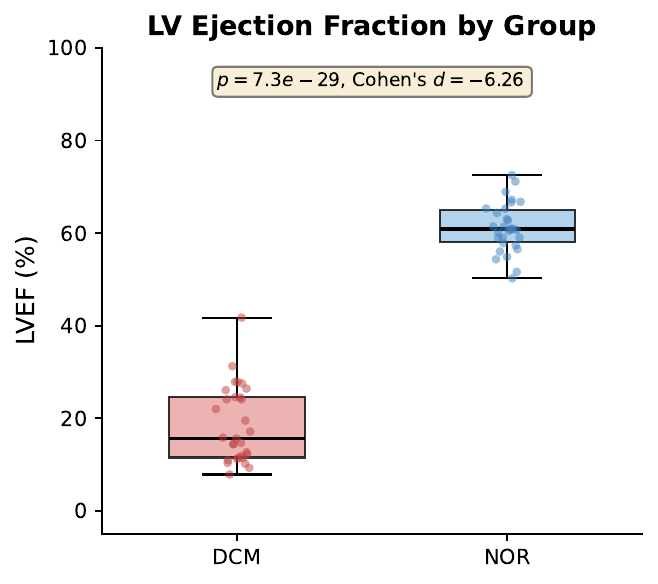}
\captionof{figure}{LVEF distribution by group, generated by the Phase~2B analysis code, showing clear separation between DCM (mean $18.6\%$) and NOR (mean $61.2\%$), $p = 7.33 \times 10^{-29}$, Cohen's $d = -6.15$.}
\label{fig:cardiac-boxplot}
\end{center}
\end{phasebox}

\vspace{2pt}

\begin{phasebox}[Phase 3: Interpretation]{colP3}
\begin{wrapfigure}{r}{\badgesm}
  \vspace{-1.4em}
  \hspace{1em}\includegraphics[width=0.92\linewidth]{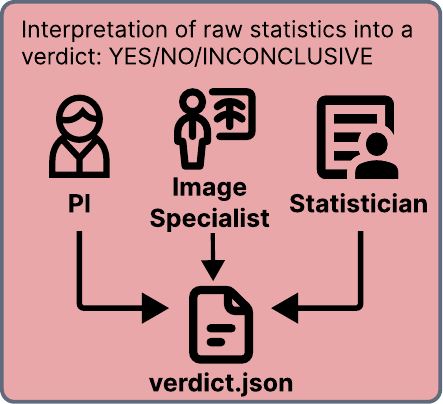}
  \vspace{-0.5em}
\end{wrapfigure}
The agent team reviews the Phase~2B output. Given the highly significant result ($p \ll 0.001$), large effect size ($|d| = 6.15$), and effect direction consistent with the hypothesis (DCM $<$ NOR), the team reaches a unanimous verdict: \textbf{YES} with evidence label \textsc{Supported}. This matches the ground-truth label, demonstrating correct end-to-end execution for a straightforward L2 hypothesis.
\end{phasebox}

\vspace{5pt}

\subsection{Example 2: MGMT Methylation Survival Analysis (L5)}
\label{sec:example-glioma}

\vspace{2pt}

\begin{phasebox}[Hypothesis]{colHyp}
\begin{wrapfigure}{r}{0.2\linewidth}
  \vspace{-1.7em}
  \hspace{0pt}\includegraphics[width=0.92\linewidth]{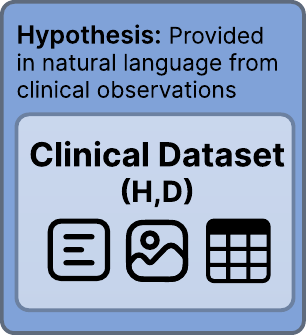}
  \vspace{-0.3em}
\end{wrapfigure}
\vspace{0.6em}
{\large\itshape
“In Grade~IV glioblastoma patients, MGMT promoter methylation is associated with longer overall survival after adjusting for age and extent of resection.”
\par}
\vspace{0.5em}
{\normalsize
This is an L5 (multivariate) hypothesis requiring Cox proportional hazards modeling with covariate adjustment, the most complex analysis tier in the benchmark.
\par}
\end{phasebox}

\newpage

\begin{phasebox}[Phase 1: Planning]{colP1}
\begin{wrapfigure}{r}{0.3\linewidth}
  \vspace{-1em}
  \hspace{0pt}\includegraphics[width=0.92\linewidth]{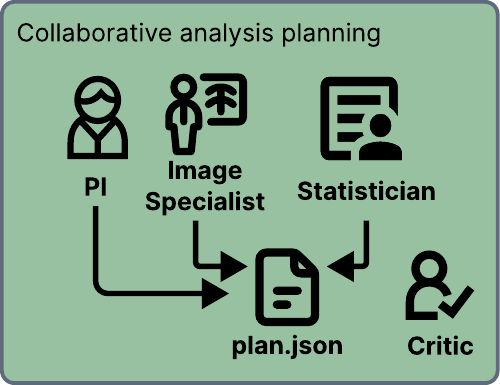}
  \vspace{-0.3em}
\end{wrapfigure}
The PI restricts the analysis to Grade~IV patients only and confirms the outcome variables (\texttt{survival\_days}, \texttt{survival\_status}). The Statistician specifies a Cox PH model with \texttt{mgmt\_status} as primary predictor and \texttt{age} + \texttt{extent\_of\_resection} as covariates, and recommends complete-case analysis with Schoenfeld residual checks for the proportional hazards assumption. No imaging structures are required (\texttt{structures: []}).
\end{phasebox}

\vspace{2pt}

\begin{phasebox}[Phase 2B: Statistical Analysis]{colP2B}
\begin{wrapfigure}{r}{0.4\linewidth}
  \vspace{-0.3em}
  \hspace{0pt}\includegraphics[width=0.92\linewidth]{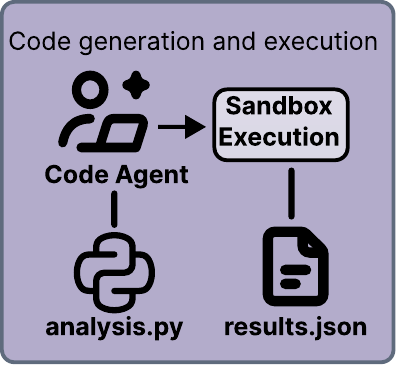}
  \vspace{-0.6em}
\end{wrapfigure}
The coding agent generates a survival analysis script using \texttt{lifelines}:
\begin{codebox}
\begin{lstlisting}[language=Python,basicstyle=\ttfamily\scriptsize]
from lifelines import CoxPHFitter, KaplanMeierFitter
# Restrict to GradeIV, encode MGMT, one-hot EOR (ref=GTR)
cph = CoxPHFitter()
cph.fit(df[["survival_days","survival_status",
            "mgmt_methylated","age","eor_Biopsy","eor_STR"]],
        duration_col="survival_days", event_col="survival_status")
\end{lstlisting}
\end{codebox}

\noindent The script also generates Kaplan-Meier survival curves stratified by MGMT status (Figure~\ref{fig:workflow-glioma}) and performs Schoenfeld residual tests for PH assumption validation. Key results:

\begin{center}
\scriptsize
\begin{tabular}{ll}
\toprule
\textbf{Statistic} & \textbf{Value} \\
\midrule
Test & Cox proportional hazards \\
Cohort & Grade~IV ($n = 378$, 217 events) \\
Primary predictor & MGMT methylation (methylated $= 1$) \\
Covariates & Age, extent of resection (Biopsy/STR vs.\ GTR) \\
Hazard ratio (MGMT) & 0.712 (95\% CI: 0.532--0.954) \\
$p$-value (MGMT) & 0.023 \\
Events per variable & 54.25 \\
Schoenfeld (MGMT) & $p = 0.277$ (PH satisfied) \\
\bottomrule
\end{tabular}
\end{center}

\smallskip
\begin{center}
\par\noindent\hspace*{0.4\linewidth}%
\includegraphics[width=0.53\textwidth]{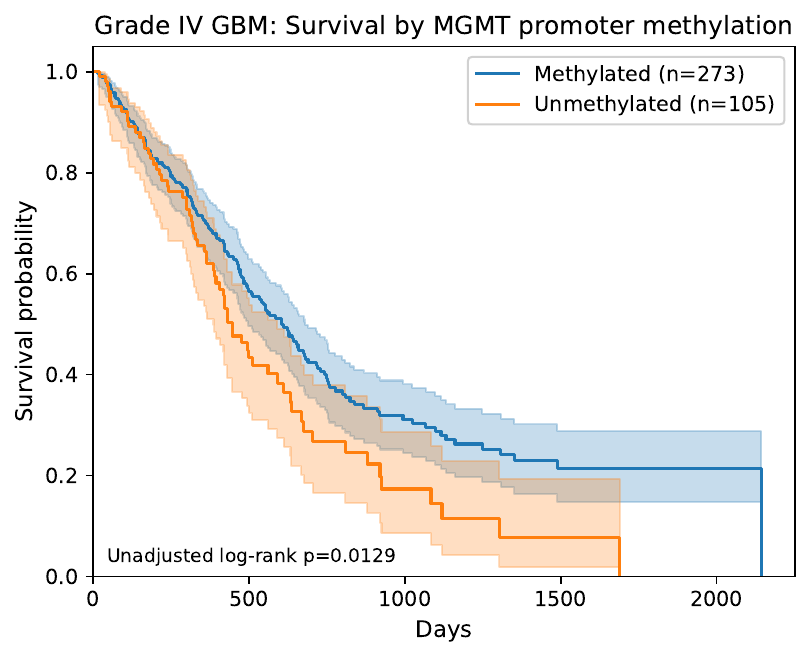}
\captionof{figure}{Kaplan-Meier survival curves by MGMT methylation status in Grade~IV glioblastoma patients (methylated $n = 273$, unmethylated $n = 105$), generated by the Phase~2B analysis code. Unadjusted log-rank $p = 0.013$; the Cox PH model with age and extent-of-resection covariates yields HR $= 0.712$ ($p = 0.023$).}
\label{fig:workflow-glioma}
\end{center}
\end{phasebox}

\vspace{2pt}

\begin{phasebox}[Phase 3: Interpretation]{colP3}
\begin{wrapfigure}{r}{\badgesm}
  \vspace{-0.7em}
  \hspace{0pt}\includegraphics[width=0.92\linewidth]{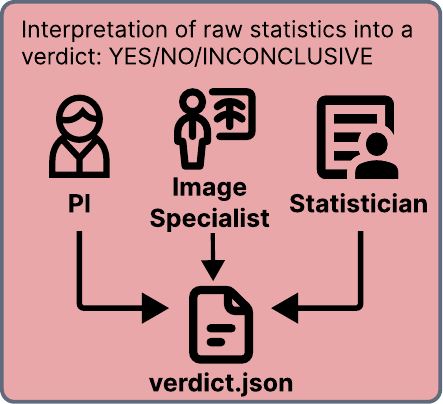}
  \vspace{-0.5em}
\end{wrapfigure}
The agents interpret the hazard ratio (HR $= 0.712 < 1$) as indicating that MGMT-methylated patients have ${\sim}29\%$ lower hazard of death after adjustment for age and extent of resection. The PH assumption holds for the primary predictor ($p = 0.277$). Verdict: \textbf{YES} with evidence label \textsc{Supported}. This example demonstrates \framework{}'s ability to handle the most complex analysis tier (L5), including correct covariate specification, assumption checking, and multivariate interpretation.
\end{phasebox}

\subsection{Artifact Trail and Auditability}
\label{sec:artifact-trail}

\begin{wrapfigure}{r}{\badgexs}
  \vspace{-3.5em}
  \hspace{0pt}\includegraphics[width=0.92\linewidth]{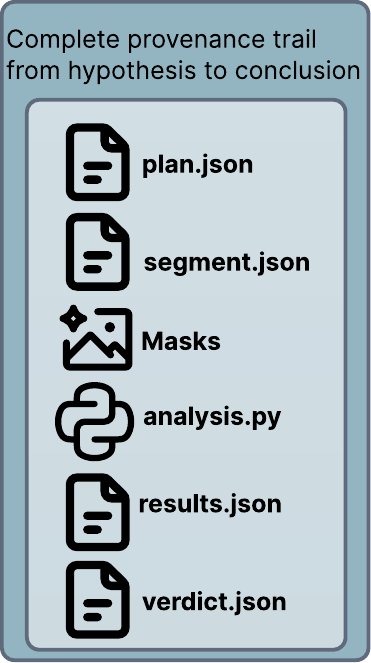}
  \vspace{-0.2em}
\end{wrapfigure}

Table~\ref{tab:artifact-trail} summarizes the complete set of artifacts produced by a single \framework{} run. Every phase produces human-readable, inspectable outputs that enable post-hoc verification of the entire analysis pipeline.

\begin{table}[t]
\centering
\caption{Artifacts produced per phase in a \framework{} run. All artifacts are saved to disk and available for post-hoc inspection.}
\label{tab:artifact-trail}
\scriptsize
\begin{tabular}{llll}
\toprule
\textbf{Phase} & \textbf{Artifact} & \textbf{Format} & \textbf{Enables} \\
\midrule
Phase 1 & Team discussion transcript & Markdown & Review of scientific reasoning \\
Phase 1 & Analysis plan contract & JSON & Verify test selection, feasibility \\
Phase 1 & Power analysis & JSON & Assess statistical adequacy \\
\midrule
Phase 2A & Segmentation request & JSON & Verify correct structures/observations \\
Phase 2A & SAT execution log & JSON & Confirm data provenance \\
\midrule
Phase 2B & Generated analysis code & Python & Full reproducibility \\
Phase 2B & Statistical results & JSON & Verify $p$, effect size, CIs \\
Phase 2B & Visualization plots & PNG & Visual sanity checks \\
Phase 2B & Execution logs & Text & Debug failures \\
\midrule
Phase 3 & Interpretation transcript & Markdown & Review verdict reasoning \\
Phase 3 & Final verdict + evidence label & JSON & End-to-end conclusion \\
\bottomrule
\end{tabular}
\end{table}

These artifacts collectively satisfy the auditability criteria defined in \cref{sec:auditability-supp}: a reviewer can trace from the natural-language hypothesis through the analysis plan, verify which imaging structures were queried, inspect the generated code line by line, check the statistical output against the code, and evaluate whether the interpretation correctly reflects the numerical evidence.

\section{Extended Related Work}
\label{sec:extended-related}
 
\subsection{Statistical Evidence and Epistemic Classification}
\label{sec:related-stats-supp}
 
Interpreting statistical evidence is a central challenge in empirical research. A prominent American Statistical Association editorial cautioned against using $p$-value thresholds as the sole basis for scientific conclusions~\cite{wasserstein2019moving}, while others have argued for redefining significance thresholds~\cite{benjamin2018redefine} or shifting focus to effect sizes~\cite{cohen1994earth}. Low statistical power further compounds these issues, inflating both false-positive rates and effect-size estimates~\cite{ioannidis2005most,button2013power}. Equivalence testing via the TOST procedure and the smallest effect size of interest (SESOI) provide a principled way to distinguish ``no meaningful effect'' from ``insufficient power to detect''~\cite{lakens2017equivalence}. Lakens et al.~\cite{lakens2018equivalence} describe the four inferential outcomes obtained by combining null-hypothesis significance testing with equivalence testing; \framework{} adapts this idea into an evidence label taxonomy (\textsc{Supported}, \textsc{Refuted}, \textsc{Underpowered}, \textsc{Invalid}), with \textsc{Invalid} additionally covering feasibility failures outside the equivalence-testing setting. In medical imaging, reproducibility is additionally influenced by variability in segmentation and feature extraction pipelines, motivating standardization efforts such as IBSI~\cite{zwanenburg2020image} and PyRadiomics~\cite{van2017computational}. Recent benchmarks further confirm that LLM-driven statistical analysis remains error-prone: StatLLM~\cite{song2026statllm} evaluates LLM-generated statistical code against human-verified analyses and expert scores, while StatEval~\cite{lu2025stateval} shows that strong closed-source models still struggle on research-level statistical reasoning tasks, underscoring the need for the sandboxed execution and iterative validation that \framework{} employs.

\subsection{Tool-Augmented and Programmatic Language Models}
\label{sec:related-tools-supp}

A complementary line of work augments language models with external tools and executable programs. Toolformer~\cite{schick2023toolformer}, ViperGPT~\cite{suris2023vipergpt}, ToolLLM~\cite{qin2023toolllm}, CodeAct~\cite{wang2024executable}, and Gorilla~\cite{patil2024gorilla} enable LLMs to call APIs and write executable code, while the ReAct paradigm~\cite{yao2022react} interleaves reasoning with tool use. More broadly, programmatic reasoning, where models generate and execute code rather than reasoning purely in text, has been shown to substantially improve reliability for quantitative tasks~\cite{chen2022program,gao2023pal}, motivating \framework{}'s code-first approach to statistical analysis. Building on these primitives, agentic systems increasingly target code- and data-centric workflows: DataInterpreter~\cite{hong2025data} for end-to-end data science, TaskWeaver~\cite{qiao2023taskweaver} for code-first analytics, and SWE-Agent~\cite{yang2024swe} for software engineering, while self-debugging techniques let models iteratively repair their own code from execution feedback~\cite{chen2024teaching}, mirroring \framework{}'s iterative Phase~2B revision loop. In the natural sciences, tool-augmented agents such as ChemCrow~\cite{m2024augmenting} and Coscientist~\cite{boiko2023autonomous} plan and execute experiments through external tools.

\subsection{Reliability of End-to-End Research Agents}
\label{sec:related-agenteval-supp}

A recent independent evaluation of AI research frameworks, including AgentLaboratory~\cite{schmidgall2025agent}, AutoGen~\cite{wu2024autogen}, MOOSE-Chem2~\cite{yang2025moose}, SciAgents~\cite{ghafarollahi2025sciagents}, SciMON~\cite{wang2024scimon}, and Virtual Lab~\cite{swanson2025virtual}, found that none successfully completed an end-to-end research cycle on two real-world tasks, with agents frequently hallucinating results during implementation~\cite{agrawal2026can}. This highlights the importance of executable, code-grounded reasoning~\cite{wang2024executable,chen2022program} for bridging the gap between planning and reliable analysis, a central design principle of \framework{}.

\subsection{Faithfulness and Auditability of Agent Reasoning}
\label{sec:related-faithfulness-supp}

A growing body of work shows that an LLM's stated reasoning need not reflect the computation that produced its answer: chain-of-thought explanations can be unfaithful or post-hoc rationalizations~\cite{turpin2023language}, complicating efforts to monitor agent reasoning for reliability~\cite{korbak2025chain}. This motivates \framework{}'s design choice to ground every conclusion in inspectable, re-executable artifacts (plans, code, statistical outputs) rather than in agents' narrated rationales, so that correctness can be verified independently of what the agents claim.

\section{Additional Experimental Results}
\label{sec:additional-results}

\subsection{Full Per-Method Stratified Results}
\label{sec:full-stratified}

Table~\ref{tab:full_stratified} extends Table~\ref{tab:stratified} to all methods and both model families, reporting non-L0 evidence-label and verdict accuracy per dataset.

\begin{table}[t]
\centering
\caption{Per-dataset non-L0 evidence-label (Ev) and verdict (Vd) accuracy (\%) for all methods and model families. Completion rate shown for \framework{} (baselines always produce output). Best per column in \textbf{bold}.}
\label{tab:full_stratified}
\scriptsize
\begin{tabular}{ll|cc|cc|c}
\toprule
& & \multicolumn{2}{c|}{\textbf{ACDC}} & \multicolumn{2}{c|}{\textbf{UCSF-PDGM}} & \\
\textbf{Method} & \textbf{Family} & Ev & Vd & Ev & Vd & Compl. \\
\midrule
\multicolumn{7}{l}{\textit{Local/Open-weight models}} \\
\midrule
SMa: Direct      & local & 0.0  & 83.0 & 0.0  & 30.1 & -- \\
SMb: Features     & local & 76.9 & 72.7 & 40.0 & 55.2 & -- \\
SMc: Code         & local & 76.2 & 67.9 & 3.0  & 58.5 & -- \\
SMd: Agentic      & local & 78.3 & 87.5 & 15.3 & 38.5 & -- \\
SMe: Pipeline     & local & 92.4 & 90.7 & 19.0 & 26.2 & -- \\
\framework{}      & local & 91.0 & 87.9 & 36.7 & 49.0 & 78.1 \\
\midrule
\multicolumn{7}{l}{\textit{Frontier models (GPT-5.2)}} \\
\midrule
SMa: Direct       & frontier & 0.0  & 85.2 & 0.0  & 27.7 & -- \\
SMb: Features     & frontier & 77.9 & 84.1 & 50.0 & 57.7 & -- \\
SMc: Code         & frontier & 79.0 & 85.1 & 27.7 & 60.9 & -- \\
SMd: Agentic      & frontier & \textbf{95.9} & \textbf{96.6} & 34.7 & 53.9 & -- \\
SMe: Pipeline     & frontier & 90.3 & 96.9 & 32.3 & 35.1 & -- \\
\framework{}      & frontier & 90.0 & 78.9 & \textbf{50.7} & \textbf{83.2} & 87.5 \\
\bottomrule
\end{tabular}
\end{table}

\noindent Key observations:
\begin{itemize}
    \item \textbf{Performance on ACDC dataset is near-saturated for frontier baselines}: SMd and SMe exceed 96\% verdict accuracy on the relatively straightforward cardiac dataset, leaving little room for improvement. \framework{}'s lower ACDC verdict accuracy (78.9\%) reflects pipeline completion failures rather than reasoning errors: completed runs achieve high accuracy (see Table~\ref{tab:e2e_gap}).
    \item \textbf{UCSF-PDGM differentiates methods}: On the more challenging neuro-oncology dataset (which includes survival analyses, multivariate models, and metadata-based group filtering), \framework{} frontier achieves the highest evidence-label (50.7\%) and verdict (83.2\%) accuracy, outperforming the best baseline by $>$16pp (evidence) and $>$29pp (verdict).
    \item \textbf{Open-weight gap}: Switching from frontier to open-weight models reduces UCSF-PDGM evidence accuracy by 10--20pp across methods, reflecting the difficulty of survival analysis and multivariate modeling for smaller models.
    \item \textbf{SMa confirms code necessity}: Pure reasoning from summary statistics (SMa) achieves 0\% evidence-label accuracy across all conditions, confirming that producing verifiable statistical outputs requires executable code generation.
    \item \textbf{Full autonomy}: \framework{} is the only method that executes the complete pipeline (from hypothesis intake through image segmentation, code generation, and statistical interpretation) without any human intervention or pre-computed inputs. All baselines rely on at least one shortcut: pre-extracted features (SMa, SMb), a single-shot prompt (SMc), or a simplified agentic loop without vision grounding (SMd, SMe).
\end{itemize}

\subsection{Additional Ablation Experiments}
\label{sec:supp_ablations}

The main paper reports the highest-impact ablations (context 8k/32k, no critic, temperature zero). We place the remaining ablations here to keep the main narrative focused.

\begin{table}[t]
\centering
\caption{Complete ablation results for \framework{} (local, 10 runs per hypothesis). \textbf{Default}: SAT-Nano masks, 16k context, critic in execution phases only, $T{=}0.2$, standard model assignment (20B PI/imaging, 8B statistician/critic, 30B coder). All metrics are non-L0 evidence-label (Ev) and verdict (Vd) accuracy (\%).}
\label{tab:supp_ablations}
\begin{tabular}{ll|cc|cc|cc}
\toprule
& & \multicolumn{2}{c|}{\textbf{ACDC}} & \multicolumn{2}{c|}{\textbf{UCSF-PDGM}} & \multicolumn{2}{c}{\textbf{Combined}} \\
\textbf{Scope} & \textbf{Configuration} & Ev & Vd & Ev & Vd & Ev & Vd \\
\midrule
\multicolumn{8}{l}{\textit{Vision source}} \\
\midrule
& Default (SAT-Nano) & 91.0 & 87.9 & 36.7 & 49.0 & 63.4 & 71.4 \\
& GT expert masks & 82.1 & 85.7 & 13.0 & 47.3 & 46.9 & 77.2 \\
& SAT Pro masks & 92.1 & 90.9 & 22.3 & 38.2 & 56.6 & 70.6 \\
\midrule
\multicolumn{8}{l}{\textit{Context window}} \\
\midrule
& Default (16k) & 91.0 & 87.9 & 36.7 & 49.0 & 63.4 & 71.4 \\
& 8k context & 91.4 & 91.7 & 29.0 & 45.7 & 59.7 & 72.0 \\
& 32k context & 90.0 & 87.1 & 31.7 & 51.8 & 60.3 & 72.7 \\
\midrule
\multicolumn{8}{l}{\textit{Critic}} \\
\midrule
& Default (execution phases) & 91.0 & 87.9 & 36.7 & 49.0 & 63.4 & 71.4 \\
& No critic & 90.7 & 90.1 & 32.0 & 50.5 & 60.8 & 73.2 \\
& Critic in all phases & 81.4 & 85.6 & 26.0 & 39.2 & 53.2 & 66.5 \\
\midrule
\multicolumn{8}{l}{\textit{Decoding}} \\
\midrule
& Default ($T{=}0.2$, uniform) & 91.0 & 87.9 & 36.7 & 49.0 & 63.4 & 71.4 \\
& Temperature zero ($T{=}0$) & 90.3 & 88.7 & 27.7 & 47.5 & 58.5 & 72.6 \\
& Role-specific $T$/top-$p$$^\dagger$ & 91.4 & 89.2 & 32.3 & 44.3 & 61.4 & 70.4 \\
\midrule
\multicolumn{8}{l}{\textit{Model assignment}} \\
\midrule
& Default (20B/8B/30B) & 91.0 & 87.9 & 36.7 & 49.0 & 63.4 & 71.4 \\
& Small model (all 8B) & 33.8 & 66.1 & 4.7 & 64.7 & 19.0 & 65.9 \\
& General coding (no coder model) & 90.0 & 88.5 & 29.5 & 62.8 & 59.5 & 80.0 \\
\bottomrule
\end{tabular}
\smallskip\\
\footnotesize{$\dagger$ Role-specific $T$/top-$p$: PI/imaging $T{=}1.0$/top-$p{=}1.0$; statistician/critic $T{=}0.6$/top-$p{=}0.95$; coder $T{=}0.7$/top-$p{=}0.8$.}
\end{table}

The structured multi-phase pipeline design makes \framework{} \emph{robust to component-level changes}: swapping vision backends, adjusting critic scope, or substituting model variants produces relatively moderate accuracy shifts, whereas the overall architecture consistently drives performance.
\begin{itemize}
    \item \textbf{Vision source}: GT expert masks show a notable evidence drop ($-$16.5pp combined) driven by UCSF-PDGM ($-$23.7pp), while SAT-Pro is closer to default ($-$6.8pp combined). On ACDC, GT masks slightly reduce evidence accuracy ($-$8.9pp) compared to SAT-Nano: most ACDC hypotheses are insensitive to segmentation quality, and the different data format may introduce minor integration friction. On UCSF-PDGM, where most hypotheses are metadata-driven, all vision sources perform similarly at low levels.
    \item \textbf{Critic scope}: Extending the critic beyond code execution into planning and interpretation phases degrades combined evidence accuracy ($-$10.2pp) and verdict ($-$4.9pp). Qualitative inspection reveals three failure modes: (i)~the critic raises spurious feasibility concerns during planning, causing the PI to mark testable hypotheses as \textsc{untestable} or to substitute proxy metrics; (ii)~it advocates covariate-adjusted tests (e.g.\ ANCOVA) over simple group comparisons, changing the test family and triggering evaluation mismatches; and (iii)~the additional discussion rounds inflate the context passed to downstream code-generation phases, degrading statistical code quality. These patterns suggest that the structured multi-phase pipeline already provides sufficient guardrails, and an additional adversarial critic during deliberative phases introduces noise rather than actionable corrections.
    \item \textbf{Model capacity (all 8B)}: Replacing 20B/30B models with 8B variants causes a large evidence-accuracy drop ($-$44.4pp combined) while verdict accuracy degrades only modestly ($-$5.5pp). Smaller models can still infer correct binary verdicts from partial outputs but lack the capacity to generate complete, well-formed statistical code.
    \item \textbf{Coding specialization}: Replacing the coding-specialized 30B model with a general-purpose 30B model yields comparable evidence accuracy ($-$3.9pp) but notably higher verdict accuracy (+8.6pp), suggesting that the general model produces simpler, more robust code that completes more often.
    \item \textbf{Role-specific decoding}: Assigning model-recommended temperature and nucleus sampling parameters per role (PI/imaging: $T{=}1.0$; statistician/critic: $T{=}0.6$/top-$p{=}0.95$; coder: $T{=}0.7$/top-$p{=}0.8$) yields a marginal ACDC uplift (+0.4pp evidence, +1.3pp verdict) compared to the uniform $T{=}0.2$ baseline, but degrades on UCSF-PDGM ($-$4.4pp evidence, $-$4.7pp verdict), consistent with the pattern observed for temperature zero. Complex multi-step reasoning tasks appear to benefit from conservative, low-variance sampling rather than role-tailored diversity.
\end{itemize}

\subsection{Run-to-Run Reproducibility}
\label{sec:reproducibility}

To characterize the stochastic variability of \framework{}, we repeat two hypotheses 100 times each under default settings (open-weight, SAT-Nano, 16k context, $T{=}0.2$):
\begin{enumerate}
    \item \texttt{cardiac\_01\_dcm\_lvef\_lower} (L2, positive control): a straightforward LV ejection-fraction group comparison.
    \item \texttt{glioma\_25\_mgmt\_survival\_longer\_adjust\_age\_eor} (L5, positive control): a multivariate Cox regression with confound adjustment.
\end{enumerate}

\noindent These span the easiest (L2, single imaging metric) and hardest (L5, multivariate survival) testable tiers, one per dataset.

\begin{table}[t]
\centering
\caption{Run-to-run reproducibility over 100 independent runs per hypothesis (open-weight, default settings). \texttt{cardiac\_01} is a straightforward L2 group comparison (DCM vs.\ NOR LVEF); \texttt{glioma\_25} is a challenging L5 multivariate Cox regression with covariate adjustment, the hardest testable tier. Majority \%: fraction of runs returning the most common verdict. $p$-value and effect size statistics are computed over successful runs only.}
\label{tab:reproducibility}
\scriptsize
\begin{tabular}{l|cc}
\toprule
& \textbf{cardiac\_01} (L2) & \textbf{glioma\_25} (L5) \\
\midrule
Ground truth label         & \textsc{Supported}     & \textsc{Supported}     \\
Runs (success / total)     & 95/100                 & 68/100                 \\
Verdict accuracy           & 100.0\%                & 37.7\%                 \\
Evidence-label accuracy    & 95.0\%                 & 20.0\%                 \\
Majority verdict (\%)      & YES (95\%)             & YES (26\%)             \\
\midrule
$p$-value (median [IQR])   & $7.3{\times}10^{-29}$ [$7.3{\times}10^{-29}$, $3.0{\times}10^{-11}$] & 0.050 [0.034, 0.70] \\
Effect size (median [IQR]) & $-6.15$ [$-6.26$, $-1.00$] & $0.99$ [$0.71$, $1.24$] \\
Runtime (median [IQR], s)  & 215 [192, 244]         & 418 [306, 571]         \\
\bottomrule
\end{tabular}
\end{table}

The L2 cardiac hypothesis exhibits near-perfect reproducibility: 100\% verdict accuracy, 95\% evidence-label accuracy (5 runs assigned \textsc{Underpowered} instead of \textsc{Supported} due to minor power estimation differences), and extremely tight $p$-value concentration ($<10^{-28}$). The L5 glioma hypothesis is substantially harder: only 68 of 100 runs complete successfully, and verdict accuracy drops to 37.7\%. The difficulty stems from the multivariate Cox regression, which requires correct covariate specification, Grade~IV cohort filtering, and proper survival data encoding, each introducing independent failure modes absent from simpler analyses. Despite this, the majority verdict remains correct (YES), and median runtime roughly doubles from 3.6 to 7.0 minutes. These results quantify the reproducibility--complexity trade-off inherent in agentic scientific workflows: simple analyses are highly reliable, while complex multivariate designs remain a frontier challenge.

\subsection{Phase-Level Reliability}
\label{sec:phase_reliability}

Table~\ref{tab:phase_reliability} reports phase-level pass rates for \framework{} under both model families, combining ACDC and UCSF-PDGM results. A phase ``passes'' when its output satisfies the programmatic validation checks (schema correctness, no synthetic data, valid sample sizes).

\begin{table}[t]
\centering
\caption{Phase-level pass rates and dominant failure modes for \framework{} (combined ACDC + UCSF-PDGM). Pool: number of runs entering that phase; runs that fail in an earlier phase do not enter subsequent phases. Phase~2B (code generation and execution) is the primary reliability bottleneck.}
\label{tab:phase_reliability}
\scriptsize
\begin{tabular}{l|ccc|ccc}
\toprule
& \multicolumn{3}{c|}{\textbf{Open-weight}} & \multicolumn{3}{c}{\textbf{Frontier (GPT-5.2)}} \\
\textbf{Phase} & Pool & Pass\% & Top failure & Pool & Pass\% & Top failure \\
\midrule
Phase 1 (Planning)    & 640 & 97.3 & Schema (2.5\%)     & 640 & 98.0 & Schema (2.0\%) \\
Phase 2A (Segmentation) & 604 & 97.8 & Obs.\ mismatch (1.0\%) & 587 & 100.0 & -- \\
Phase 2B (Statistics) & 591 & 85.1 & Synthetic (8.8\%), Schema (3.9\%) & 587 & 95.7 & Schema (3.7\%) \\
\bottomrule
\end{tabular}
\end{table}

Phase 2B is the primary bottleneck: open-weight models fail 14.9\% of the time, predominantly due to synthetic data generation (8.8\%, fabricating values instead of using the Imaging Analysis API) and results schema violations (3.9\%, incomplete JSON output). Frontier models reduce Phase 2B failures to 4.3\%, primarily schema issues, and achieve perfect Phase 2A pass rates. Phase 1 is highly reliable across both model families ($>$97\%), confirming that the planning phase's structured output format is well-handled.

\subsection{Coding Trial Efficiency}
\label{sec:coding_trials}

Table~\ref{tab:coding_trials} reports the number of code execution attempts per phase. Each coding round permits up to 8 executions; phases may span multiple rounds, so the effective per-phase limit is higher (observed max: 19).

\begin{table}[t]
\centering
\caption{Quantification of code execution attempts per phase (mean / max across all runs). Phase 2A writes the segmentation request; Phase 2B performs statistical analysis.}
\label{tab:coding_trials}
\scriptsize
\begin{tabular}{l|cc|cc|cc}
\toprule
& \multicolumn{2}{c|}{\textbf{Phase 2A}} & \multicolumn{2}{c|}{\textbf{Phase 2B}} & \multicolumn{2}{c}{\textbf{Total}} \\
\textbf{Configuration} & Mean & Max & Mean & Max & Mean & Max \\
\midrule
Local (ACDC)     & 1.4 & 13 & 1.7 & 8  & 2.9 & 13 \\
Local (UCSF)     & 1.4 & 10 & 3.7 & 8  & 4.7 & 13 \\
Frontier (ACDC)        & 1.1 & 4  & 4.9 & 16 & 5.8 & 17 \\
Frontier (UCSF)        & 1.4 & 6  & 6.4 & 17 & 7.3 & 19 \\
\bottomrule
\end{tabular}
\end{table}

Phase~2A typically requires 1--2 attempts regardless of model family, reflecting the constrained nature of the segmentation request task. Phase~2B shows greater variation: open-weight ACDC averages 1.7 attempts (simple cardiac metrics), while frontier UCSF-PDGM averages 6.4 (complex survival analyses with covariate adjustment). Notably, frontier models use \emph{more} iterations than open-weight on average, not due to higher failure rates but because the iterative code execution environment enables them to refine analyses across rounds (\eg, adding diagnostic plots, checking distributional assumptions, refining covariate encoding), whereas smaller models tend to terminate after fewer, less sophisticated attempts, partly due to lower model capacity and partly because their shorter context windows limit how many iterative rounds can be sustained before truncation.

\subsection{Accuracy by Control Type}
\label{sec:control_type}

Table~\ref{tab:control_type} stratifies accuracy by hypothesis control type, combining both datasets. Control types reflect the intended role of each hypothesis in the benchmark: \emph{positive controls} should yield YES/\textsc{Supported}, \emph{negative controls} should yield NO/\textsc{Refuted}, \emph{underpowered controls} are designed to have insufficient sample size, \emph{untestable controls} require data not present in the dataset, and \emph{nonsense/no-effect controls} test obviously false or null hypotheses.

\begin{table}[t]
\centering
\caption{Evidence-label (Ev) and verdict (Vd) accuracy (\%) by hypothesis control type for \framework{} (combined ACDC + UCSF-PDGM). $n$: number of valid evaluation instances.}
\label{tab:control_type}
\scriptsize
\begin{tabular}{l|rr|rr|rr|rr}
\toprule
& \multicolumn{4}{c|}{\textbf{Open-weight}} & \multicolumn{4}{c}{\textbf{Frontier}} \\
\textbf{Control type} & Ev & $n$ & Vd & $n$ & Ev & $n$ & Vd & $n$ \\
\midrule
Positive (37 hyps)      & 58.4 & 370 & 72.0 & 300 & 64.6 & 370 & 84.0 & 349 \\
Negative (12 hyps)      & 76.7 & 120 & 80.2 & 106 & 77.5 & 120 & 58.9 & 112 \\
Underpowered (6 hyps)   & 65.0 &  60 & 54.0 &  50 & 81.7 &  60 & 98.1 &  54 \\
Untestable (5 hyps)     & 74.0 &  50 & 50.0 &  16 & 100.0 &  50 & 55.6 &   9 \\
No-effect (2 hyps)      & 100.0 & 20 & 75.0 &  20 & 70.0 &  20 & 89.5 &  19 \\
Nonsense (2 hyps)       & 35.0 &  20 & 40.0 &  10 & 90.0 &  20 & 100.0 &  18 \\
\bottomrule
\end{tabular}
\end{table}

Key observations: (1)~Positive controls are the hardest for evidence accuracy (58--65\%), because generating correct statistical outputs with proper effect sizes and confidence intervals requires successful end-to-end code execution. (2)~Negative controls exhibit a counterintuitive frontier--open-weight reversal for verdict accuracy (59\% vs 80\%): frontier models more frequently produce hedging \textsc{Inconclusive} verdicts for reversed-direction hypotheses, whereas open-weight models issue decisive \textsc{No} verdicts more often. (3)~Untestable controls achieve 74--100\% evidence detection, with frontier models achieving perfect L0 identification (100\%) and open-weight models correctly identifying most cases (74\%), validating the feasibility-checking mechanism. (4)~Underpowered controls are reliably identified by the frontier model (82\% evidence, 98\% verdict), confirming that the \textsc{Underpowered} label captures genuine statistical ambiguity that a binary scheme would miss.

\subsection{End-to-End vs.\ Completed-Only Accuracy}
\label{sec:e2e_gap}

Table~\ref{tab:e2e_gap} contrasts two accuracy measures: \emph{completed-only} accuracy (excluding runs with execution failures) and \emph{end-to-end} (E2E) accuracy (treating failures as incorrect). The gap quantifies the ``execution tax'': how much accuracy drops when pipeline reliability is factored in.

\begin{table}[t]
\centering
\caption{Non-L0 verdict accuracy (\%): completed runs only vs.\ end-to-end (E2E, counting execution failures as incorrect). $\Delta$: accuracy penalty from incomplete runs, quantifying the ``execution tax'' imposed by pipeline reliability on effective accuracy.}
\label{tab:e2e_gap}
\scriptsize
\begin{tabular}{l|cc|cc|cc|cc}
\toprule
& \multicolumn{4}{c|}{\textbf{Open-weight}} & \multicolumn{4}{c}{\textbf{Frontier}} \\
\textbf{Dataset} & Vd & Vd$_\text{e2e}$ & $\Delta$ & & Vd & Vd$_\text{e2e}$ & $\Delta$ & \\
\midrule
ACDC       & 87.9 & 84.8 & $-$3.1 & & 78.9 & 77.2 & $-$1.7 & \\
UCSF-PDGM & 49.0 & 33.7 & $-$15.3 & & 83.2 & 74.0 & $-$9.2 & \\
Combined   & 71.4 & 58.8 & $-$12.6 & & 81.0 & 75.6 & $-$5.4 & \\
\bottomrule
\end{tabular}
\end{table}

The execution tax is most severe for open-weight models on UCSF-PDGM ($-$15.3pp), where complex survival analyses frequently fail to produce valid outputs. Frontier models substantially reduce the gap (5.4pp combined vs 12.6pp), indicating that improved code generation directly translates to higher effective accuracy.

\subsection{Underpowered Label Sensitivity}
\label{sec:underpowered_sensitivity}

To assess the importance of the \textsc{Underpowered} evidence label, we re-score all runs under a binary scheme that collapses \textsc{Underpowered} ground-truth labels to \textsc{Refuted} (i.e., treating low-powered null results identically to adequately-powered null results). Under this binary scheme, combined evidence accuracy increases by +3.6pp for open-weight and +1.7pp for frontier. The modest improvement confirms that (1)~the \textsc{Underpowered} category captures genuine ambiguity that would otherwise inflate the \textsc{Refuted} bucket, and (2)~the accuracy cost of maintaining this distinction is small, while the scientific benefit (distinguishing ``no evidence of effect'' from ``insufficient power to detect effect'') is substantial.

\subsection{SESOI Sensitivity}
\label{sec:sesoi_supp}

\begin{table}[t]
\centering
\caption{Effect of SESOI profile on ground-truth evidence-label distribution across the 64-hypothesis benchmark. Stricter profiles demand more power, shifting borderline hypotheses from \textsc{Refuted} to \textsc{Underpowered}. \textsc{Supported} and \textsc{Invalid} counts are invariant. Flips: number of label changes relative to the \texttt{standard} profile used as default.}
\label{tab:sesoi_supp}
\begin{tabular}{l|cccc|c}
\toprule
\textbf{SESOI Profile} & \textbf{Supported} & \textbf{Refuted} & \textbf{Underpowered} & \textbf{Invalid} & \textbf{Flips} \\
\midrule
Strict ($d=0.2$, $r=0.2$)  & 32 & 14 & 13 & 5 & 5 \\
Standard ($d=0.5$, $r=0.3$) & 32 & 19 &  8 & 5 & -- \\
Loose ($d=0.8$, $r=0.4$)    & 32 & 23 &  4 & 5 & 4 \\
\bottomrule
\end{tabular}
\end{table}

The number of \textsc{Supported} labels is invariant across profiles (32 of 64), because support depends only on $p < \alpha$ and correct direction, both independent of SESOI. Similarly, the 5 \textsc{Invalid} labels (L0 untestable) are fixed. All 9 label flips occur exclusively on the \textsc{Refuted}/\textsc{Underpowered} boundary: a stricter SESOI demands more power to declare a null result ``adequately powered,'' pushing borderline hypotheses from \textsc{Refuted} to \textsc{Underpowered}. No hypothesis changes to or from \textsc{Supported} under any profile, confirming that the evaluation's sensitivity to SESOI choice is limited and well-understood.

\subsection{Per-Hypothesis Detailed Results}
\label{sec:per_hypothesis_results}

Table~\ref{tab:per_hypothesis} reports per-hypothesis results for both the frontier (GPT-5.2) and open-weight \framework{} configurations. For each hypothesis, we show the majority-vote verdict and evidence label across 10 runs. Correct results (matching ground truth) are shown in \textbf{bold}. Hypothesis IDs are abbreviated for space; full descriptions are in Section~\ref{sec:hypotheses}.

{\scriptsize
\begin{longtable}{@{}l@{\hspace{4pt}}c@{\hspace{4pt}}c@{\hspace{4pt}}c@{\hspace{6pt}}cc@{\hspace{6pt}}cc@{}}
\caption{Per-hypothesis results for \framework{} under frontier (GPT-5.2) and open-weight configurations. Majority-vote verdict (Vd) and evidence label (Ev) across 10 runs. T = hypothesis tier (L0--L5), C = control type (P = positive, N = negative, U = null, L = low-$n$, X = untestable), GT = ground-truth label. Y = \yes{}, N = \no{}, I = \inconclusive{}, — = not completed. Correct results in \textbf{bold}.}
\label{tab:per_hypothesis}
\\
\toprule
& & & & \multicolumn{2}{c}{\textbf{Frontier}} & \multicolumn{2}{c}{\textbf{Open-weight}} \\
\textbf{Hypothesis} & \textbf{T} & \textbf{C} & \textbf{GT} & Vd & Ev & Vd & Ev \\
\midrule
\endfirsthead
\multicolumn{8}{c}{\scriptsize\textit{(continued)}} \\
\toprule
& & & & \multicolumn{2}{c}{\textbf{Frontier}} & \multicolumn{2}{c}{\textbf{Open-weight}} \\
\textbf{Hypothesis} & \textbf{T} & \textbf{C} & \textbf{GT} & Vd & Ev & Vd & Ev \\
\midrule
\endhead
\bottomrule
\endfoot
\multicolumn{8}{l}{\textit{ACDC: Cardiac MRI}} \\
\midrule
C27: LA volume HCM              & L0 & X & Inv & \textbf{—} & \textbf{Inv} & N & \textbf{Inv} \\
C28: GLS lower DCM              & L0 & X & Inv & \textbf{—} & \textbf{Inv} & N & \textbf{Inv} \\
C32: SBP higher DCM             & L0 & X & Inv & \textbf{—} & \textbf{Inv} & N & \textbf{Inv} \\
C24: Height--weight corr.       & L1 & P & Sup & \textbf{Y} & \textbf{Sup} & \textbf{Y} & \textbf{Sup} \\
C25: ED frame--weight corr.     & L1 & U & Ref & \textbf{N} & \textbf{Ref} & \textbf{N} & \textbf{Ref} \\
C26: Num frames--weight corr.   & L1 & U & Ref & \textbf{N} & \textbf{Ref} & \textbf{N} & \textbf{Ref} \\
C01: DCM LVEF lower             & L2 & P & Sup & \textbf{Y} & \textbf{Sup} & \textbf{Y} & \textbf{Sup} \\
C02: LV mass DCM                & L2 & P & Sup & \textbf{Y} & \textbf{Sup} & \textbf{Y} & \textbf{Sup} \\
C03: DCM LVEF higher            & L2 & N & Ref & I & \textbf{Ref} & \textbf{N} & \textbf{Ref} \\
C04: RV volume similar          & L2 & N & Ref & \textbf{N} & \textbf{Ref} & \textbf{N} & \textbf{Ref} \\
C07: DCM LVEDV higher           & L2 & P & Sup & \textbf{Y} & \textbf{Sup} & \textbf{Y} & \textbf{Sup} \\
C08: DCM LVESV higher           & L2 & P & Sup & \textbf{Y} & \textbf{Sup} & \textbf{Y} & \textbf{Sup} \\
C09: HCM LV mass higher         & L2 & P & Sup & \textbf{Y} & \textbf{Sup} & \textbf{Y} & \textbf{Sup} \\
C10: HCM LVEF $>$ DCM           & L2 & P & Sup & \textbf{Y} & \textbf{Sup} & \textbf{Y} & \textbf{Sup} \\
C11: MINF LVEF lower            & L2 & P & Sup & \textbf{Y} & \textbf{Sup} & \textbf{Y} & \textbf{Sup} \\
C12: RV RVEF lower              & L2 & P & Sup & \textbf{Y} & \textbf{Sup} & \textbf{Y} & \textbf{Sup} \\
C13: HCM LV mass lower          & L2 & N & Ref & \textbf{N} & \textbf{Ref} & \textbf{N} & \textbf{Ref} \\
C14: DCM LVEDV lower            & L2 & N & Ref & \textbf{N} & \textbf{Ref} & \textbf{N} & \textbf{Ref} \\
C15: DCM LVESV lower            & L2 & N & Ref & \textbf{N} & \textbf{Ref} & \textbf{N} & \textbf{Ref} \\
C16: RV RVEF higher             & L2 & N & Ref & I & \textbf{Ref} & \textbf{N} & \textbf{Ref} \\
C19: HCM LVEDV $<$ NOR          & L2 & P & Ref & I & \textbf{Ref} & I & \textbf{Ref} \\
C20: HCM LVESV $<$ NOR          & L2 & P & Sup & I & Ref & \textbf{Y} & \textbf{Sup} \\
C30: RV RVEDV $>$ NOR           & L2 & P & Sup & \textbf{Y} & \textbf{Sup} & \textbf{Y} & \textbf{Sup} \\
C31: RV RVEDV $<$ NOR           & L2 & N & Ref & I & \textbf{Ref} & \textbf{N} & \textbf{Ref} \\
C21: ESV/EDV ratio DCM          & L3 & P & Sup & \textbf{Y} & \textbf{Sup} & \textbf{Y} & \textbf{Sup} \\
C22: HCM mass/vol $>$ DCM       & L3 & P & Sup & \textbf{Y} & \textbf{Sup} & \textbf{Y} & \textbf{Sup} \\
C23: DCM mass/vol $<$ NOR       & L3 & P & Sup & \textbf{Y} & \textbf{Sup} & \textbf{Y} & \textbf{Sup} \\
C05: Weight--LVEF corr.         & L4 & L & Undp & \textbf{I} & \textbf{Undp} & N & \textbf{Undp} \\
C06: Height--LV volume          & L4 & P & Sup & \textbf{Y} & \textbf{Sup} & \textbf{Y} & \textbf{Sup} \\
C17: NOR weight--LVEF           & L4 & L & Undp & \textbf{I} & \textbf{Undp} & \textbf{I} & \textbf{Undp} \\
C18: DCM height--LVEF           & L4 & L & Undp & \textbf{I} & \textbf{Undp} & \textbf{I} & \textbf{Undp} \\
C29: LVEDVi (BSA-indexed) DCM   & L4 & P & Sup & \textbf{Y} & \textbf{Sup} & \textbf{Y} & \textbf{Sup} \\
\midrule
\multicolumn{8}{l}{\textit{UCSF-PDGM: Glioma MRI}} \\
\midrule
G30: ADC--survival assoc.       & L0 & X & Inv & \textbf{—} & \textbf{Inv} & I & \textbf{Inv} \\
G32: Frontal location--IDH      & L0 & X & Inv & I & \textbf{Inv} & I & \textbf{Inv} \\
G11: MGMT surv.\ longer         & L1 & P & Sup & \textbf{Y} & \textbf{Sup} & I & Ref \\
G13: GTR surv.\ longer          & L1 & P & Sup & \textbf{Y} & \textbf{Sup} & N & Ref \\
G14: IDH surv.\ longer (G4)     & L1 & L & Sup & \textbf{Y} & \textbf{Sup} & \textbf{Y} & Inv \\
G16: GTR surv.\ shorter         & L1 & N & Ref & \textbf{N} & \textbf{Ref} & \textbf{N} & \textbf{Ref} \\
G28: Men surv.\ longer (G4)     & L1 & U & Ref & \textbf{N} & \textbf{Ref} & I & \textbf{Ref} \\
G29: Men older (G4)             & L1 & U & Ref & \textbf{N} & \textbf{Ref} & I & Inv \\
G01: Volume by grade            & L2 & P & Sup & \textbf{Y} & \textbf{Sup} & \textbf{Y} & \textbf{Sup} \\
G05: G2 more enhancement        & L2 & N & Ref & \textbf{N} & \textbf{Ref} & \textbf{N} & \textbf{Ref} \\
G07: G4 more enhancement        & L2 & P & Sup & \textbf{Y} & \textbf{Sup} & \textbf{Y} & Inv \\
G08: G4 more necrosis           & L2 & P & Sup & \textbf{Y} & \textbf{Sup} & \textbf{Y} & \textbf{Sup} \\
G09: G4 more edema              & L2 & P & Sup & \textbf{Y} & \textbf{Sup} & \textbf{Y} & \textbf{Sup} \\
G19: Enh.\ fraction G4 higher   & L3 & P & Sup & \textbf{Y} & \textbf{Sup} & I & Inv \\
G20: Nec.\ fraction G4 higher   & L3 & P & Sup & N & Ref & N & Ref \\
G21: Edema fraction G4 higher   & L3 & P & Ref & \textbf{N} & \textbf{Ref} & \textbf{N} & Inv \\
G02: IDH--enhancement           & L4 & P & Sup & N & Inv & I & Inv \\
G03: Enhancement--survival      & L4 & P & Sup & \textbf{Y} & Inv & N & Inv \\
G04: IDH-mut more necrosis      & L4 & N & Ref & I & Inv & I & Undp \\
G06: Age--volume corr.          & L4 & N & Ref & \textbf{N} & \textbf{Ref} & \textbf{N} & \textbf{Ref} \\
G10: IDH-wt more necrosis       & L4 & P & Sup & N & Inv & N & Ref \\
G12: CET vol.\ surv.\ shorter   & L4 & P & Ref & \textbf{N} & Undp & \textbf{N} & Inv \\
G15: Enhancement surv.\ pos.    & L4 & N & Ref & \textbf{N} & Inv & \textbf{N} & Inv \\
G17: G2 IDH-wt enhancement      & L4 & L & Undp & \textbf{I} & \textbf{Undp} & \textbf{I} & \textbf{Undp} \\
G22: Enh.\ frac.\ IDH-wt        & L4 & P & Sup & I & Inv & \textbf{Y} & \textbf{Sup} \\
G24: Nec.\ frac.\ surv.\ neg.\ G4 & L4 & P & Ref & \textbf{N} & Inv & I & Inv \\
G31: 1p19q enh.\ fraction        & L4 & L & Undp & \textbf{I} & \textbf{Undp} & \textbf{I} & \textbf{Undp} \\
G18: CET surv.\ adj.\ EOR       & L5 & P & Sup & \textbf{Y} & Inv & I & Inv \\
G23: Surv.\ IDH adj.\ multi     & L5 & P & Sup & \textbf{Y} & Inv & I & Inv \\
G25: MGMT surv.\ adj.\ age      & L5 & P & Sup & \textbf{Y} & Inv & \textbf{Y} & Inv \\
G26: Age surv.\ adj.\ MGMT      & L5 & P & Sup & \textbf{Y} & Inv & \textbf{Y} & \textbf{Sup} \\
G27: IDH enh.\ adj.\ grade      & L5 & P & Sup & \textbf{Y} & \textbf{Sup} & I & Inv \\
\end{longtable}
}

\subsection{Baseline Summary}

Per-method aggregate metrics (evidence-label accuracy, verdict accuracy, completion rate, and L0 feasibility) are reported in Table~\ref{tab:full_stratified} above. All four coding baselines (SMb--SMe) use the same API documentation, and all five single-model baselines (SMa--SMe) use the same hypothesis bank as \framework{}, ensuring a fair comparison surface. SMa (direct reasoning) requires no code execution; SMb (pre-computed features), SMc (one-shot code) and SMd (agentic loop with up to 3 iterations) generate and execute analysis code; SMe (full pipeline) receives the same four-phase agenda structure as \framework{} but executes all phases within a single model context. Per-hypothesis baseline breakdowns are omitted for space; the evaluation scripts and all raw run directories are included in the supplementary code release for full reproducibility.

\subsection{Comparison to agentic systems for scientific discovery}
\label{sec:virtual_lab_supp}

In this section we are qualitatively comparing our work to agentic systems developed for scientific discovery. These works operate in a similar research space but are distinctly different in their objectives, implementation and autonomy. Most notably, the Virtual Lab~\cite{swanson2025virtual} is a multi-agent framework in which a PI and domain expert agents conduct structured discussion meetings and produce analysis code. However, a human operator must manually extract, execute, and debug all code between phases, relaying execution results back to the agents.
Table~\ref{tab:architecture_comparison} summarizes the key architectural differences between \framework{} and the Virtual Lab.

\begin{table}[t]
\centering
\caption{Architectural comparison: \framework{} vs.\ Virtual Lab.}
\label{tab:architecture_comparison}
\resizebox{\linewidth}{!}{
\begin{tabular}{lll}
\toprule
\textbf{Capability} & \textbf{\framework{} (Ours)} & \textbf{Virtual Lab~\cite{swanson2025virtual}} \\
\midrule
Code execution    & Sandboxed, closed-loop          & Human extracts \& runs \\
Data access       & Direct API queries (SAT)        & Textual descriptions only \\
Critic            & Phase-aware (schema, logic, direction) & Generic Scientific Critic \\
Tool integration  & MCP servers (segmentation, datasets) & PubMed search only \\
Models            & 8--30B local (Ollama)           & GPT-4o (API) \\
Output validation & Programmatic (JSON schema, CI checks) & Free-text summaries \\
Human intervention & None                           & Per-phase code execution \\
\bottomrule
\end{tabular}
}
\end{table}

\section{Failure Case Analysis}
\label{sec:failures}

We analyze two distinct failure modes using a single hypothesis, \emph{``IDH-wildtype tumors have significantly larger enhancing tumor volumes than IDH-mutant tumors''} (glioma\_02, L4), which exhibits one of the strongest ground-truth effects in the benchmark (Mann-Whitney U $p = 9.67 \times 10^{-29}$, rank-biserial $r = 0.71$). Despite this large effect, the frontier \framework{} configuration produced incorrect results in all five runs, through two qualitatively different mechanisms: one execution-level and one reasoning-level.

\subsection{Execution Failure: Schema Validation Mismatch}

In run~000, the coding agent generated statistically valid analysis code but populated the \texttt{variables\_tested.predictors} field of \texttt{statistical\_results.json} with \texttt{["idh\_status", "group"]}. The Phase~1 plan specified only \texttt{idh\_status} as the predictor (with \texttt{group} as a covariate). The schema validator caught this discrepancy (\texttt{P2B\_RESULTS\_SCHEMA\_INVALID}) and the pipeline terminated after 6 coding trials without reaching Phase~3.

\paragraph{Auditability value.} The full code and all 6 correction attempts are preserved in the run directory, enabling a researcher to identify the exact JSON key mismatch and either fix the schema or re-run with a relaxed validator.

\subsection{Reasoning Failure: Mediator Adjustment (Table~2 Fallacy)}

In runs~001--004, the pipeline completed all phases but reached the wrong verdict (\textbf{NO}/Refuted). During Phase~1, the agents correctly identified tumor grade as a potential confounder and chose OLS regression on log-transformed enhancing volume with grade adjustment:
$$\log(1 + \text{ET\_ml}) \sim \text{idh\_status} + \text{grade}$$
After adjustment, the IDH coefficient was $\beta = -0.137$ ($p = 0.593$, 95\% CI: $[-0.643, 0.368]$), leading Phase~3 to conclude ``no significant association.''

\paragraph{Root cause.} IDH status and tumor grade lie on the same causal pathway: IDH-wildtype tumors are predominantly Grade~IV, and Grade~IV tumors exhibit the most enhancement. Adjusting for grade blocks the biological mechanism through which IDH status produces enhancement differences, a classic mediator-adjustment error (``Table~2 fallacy''~\cite{westreich2013table}). The unadjusted analysis shows one of the strongest effects in the dataset ($r = 0.71$), but the grade-adjusted model sees no residual signal.

\paragraph{Auditability value.} Because \framework{} saves the complete Phase~1 plan, generated code, and statistical output, a domain expert reviewing the artifacts can immediately identify the covariate adjustment choice as the root cause, and re-run the analysis without grade adjustment to recover the correct result. A single-model baseline producing the same wrong answer would offer no such diagnostic trail.

\paragraph{Implication.} This case exposes a fundamental limitation of current LLM-based statistical agents: they apply generic epidemiological heuristics (``adjust for potential confounders'') without causal reasoning about whether a covariate is a confounder, mediator, or collider on the causal pathway between exposure and outcome. Integrating causal graph reasoning~\cite{Pearl2009causality} into the planning phase (for instance, by prompting agents to construct and evaluate directed acyclic graphs before selecting covariates) is a promising direction for future work.

\section{Framework Details}
\label{sec:framework-details}

This section provides detailed specifications of the framework components summarized in the main paper, including agent roles, phase-level design decisions, the imaging analysis API contract, and evaluation criteria. All hyperparameters and resource limits are listed in Table~\ref{tab:hyperparameters}.

\begin{table}[h]
\centering
\caption{Framework hyperparameters and limits.}
\label{tab:hyperparameters}
\begin{tabular}{lll}
\toprule
\textbf{Parameter} & \textbf{Value} & \textbf{Description} \\
\midrule
Max discussion rounds & 3 & Phase 1 and Phase 3 \\
Max code executions per round & 8 & Phases 2A and 2B \\
Phase timeout & 20 min & Wall-clock per phase \\
Temperature (discussion) & 0.2 & Phases 1 and 3 \\
Temperature (coding) & 0.2 & Phases 2A and 2B \\
Significance level ($\alpha$) & 0.05 & All statistical tests \\
Power threshold ($\pi_0$) & 0.80 & Underpowered classification \\
Standard SESOI ($d_0$) & 0.5 & Group differences \\
Standard SESOI ($r_0$) & 0.3 & Correlations \\
Loose SESOI ($d_0$) & 0.8 & Large-effect hypotheses \\
Strict SESOI ($d_0$) & 0.2 & Small-effect hypotheses \\
\bottomrule
\end{tabular}
\end{table}

\subsection{Agent Roles}
\label{sec:agents-supp}

We instantiate three specialized agents with distinct expertise:

\begin{itemize}
    \item \textbf{Principal Investigator (PI)}: Oversees scientific validity, adjudicates feasibility and confounds, and validates conclusions. Serves as team lead in discussion phases.
    \item \textbf{Medical Imaging Specialist}: Specifies segmentation targets, validates imaging protocols, queries cohort sizes, and interprets structural measurements.
    \item \textbf{Statistician}: Selects appropriate statistical tests based on data characteristics, computes effect sizes, performs power analysis, and assesses statistical validity.
\end{itemize}

Each agent is implemented as an LLM with role-specific system prompts encoding domain expertise. A separate \textbf{Critic} agent is enabled by default in execution phases (2A/2B), with optional ablation settings to also enable Critic participation in planning (Phase~1) and interpretation (Phase~3).

\subsection{Phase Details}

\paragraph{Phase 1: Feasibility Checks.}
During planning, the agent team must emit a structured feasibility block (\texttt{status}, \texttt{invalid\_subtype}, \texttt{missing\_requirements}). A deterministic validator then checks plan contract consistency against dataset metadata (groups, observations, metadata fields, and required plan keys). If the team marks a hypothesis as \texttt{UNTESTABLE}, the workflow terminates early with \textsc{Invalid}. This keeps feasibility decisions explicit and auditable while preserving agent flexibility for testable analyses.

\paragraph{Phase 1: A Priori Power Analysis.}
The Imaging Specialist queries actual cohort sizes using \texttt{list\_dataset\_patients()}. The Statistician computes \textit{a priori} power using a fixed planning SESOI ($d_0 = 0.5$ for group differences, $r_0 = 0.3$ for correlations). This planning power informs test selection and sample-adequacy assessment, but it is not the benchmark's final evidence-label power whenever a hypothesis carries a non-standard SESOI profile.

\paragraph{Phase 2B: Test Selection.}
The framework supports parametric (t-test, Pearson correlation, OLS regression), nonparametric (Mann-Whitney U, Spearman correlation), and survival analysis (log-rank test, Cox proportional hazards) tests. Agents select tests based on distributional assumptions and outcome type. For nonparametric group differences, rank-biserial correlation serves as the effect size; for survival analyses, hazard ratios with confidence intervals.

\paragraph{Phase 2B: Code Execution.}
Code execution occurs in an isolated sandboxed environment with timeout constraints. Each coding round permits up to 8 code execution attempts with a 20-minute wall-clock timeout. On each failed attempt, the Critic feeds error output back to the coding agent for revision. The number of attempts is logged as a measure of agent autonomy efficiency.

\paragraph{Phase 3: Direction Verification.}
For group difference tests, direction is verified from observed group means, not from the effect size sign, which is convention-dependent for nonparametric tests. For correlations and survival analyses, the effect size sign is used directly.

\subsection{Imaging Analysis API}
\label{sec:api-contract-supp}

All analysis phases operate through a constrained Imaging Analysis API rather than direct filesystem access, ensuring data provenance and enabling automatic detection of off-contract data loading. The API provides cohort discovery, metadata lookup, segmentation retrieval, and quantitative measurement utilities. Agents receive tool documentation, dataset-registry metadata (groups, observations, patient fields), and structure catalogs. The \texttt{sat.} namespace used in code examples is an implementation detail. Core API calls are:
\begin{itemize}
    \item \texttt{sat.list\_patients(results\_db\_path)} for cohort enumeration from segmentation outputs
    \item \texttt{sat.get\_patient\_metadata(patient\_id)} for group labels and clinical covariates
    \item \texttt{sat.get\_observation\_identifiers(patient\_id)} for timepoint/modality mapping
    \item \texttt{sat.load\_structure\_mask(...)} for structure-specific masks
    \item \texttt{sat.calculate\_volume(mask, spacing)} for geometry-aware volume extraction
\end{itemize}
This contract gives agents sufficient autonomy to engineer derived metrics (\eg, ejection fraction from end-diastolic and end-systolic volumes) while preserving provenance and enabling automatic detection of synthetic-data substitution.

\subsection{Auditability Criteria}
\label{sec:auditability-supp}

We define six auditability criteria that each ground-truth record and pipeline output must satisfy:

\begin{enumerate}
    \item \textbf{Test-family correctness}: the statistical test matches the hypothesis type (group difference $\rightarrow$ Mann-Whitney U; correlation $\rightarrow$ Spearman; survival $\rightarrow$ log-rank/Cox PH)
    \item \textbf{Cohort restriction correctness}: the correct subpopulation is selected before group splitting (\eg, Grade~IV only for GBM-specific hypotheses)
    \item \textbf{Censoring correctness}: survival analyses correctly encode time and event variables
    \item \textbf{Variable correctness}: the metric, group definitions, and covariates match the hypothesis specification
    \item \textbf{Reporting completeness}: full statistics are reported (sample sizes per group, effect size with CI, p-value, power at SESOI)
    \item \textbf{Plan--execution consistency}: the executed analysis matches the Phase~1 plan
\end{enumerate}

\subsection{Validity Checks}
\label{sec:validity-supp}

The evaluation pipeline performs automated validity checks:

\begin{enumerate}
    \item \textbf{Feasibility validation}: required fields/structures/observations exist
    \item \textbf{Numeric validation}: $p \in [0,1]$, effect sizes finite
    \item \textbf{CI consistency}: bounds ordered, effect size contained
    \item \textbf{Sign consistency}: effect size sign matches mean difference direction
    \item \textbf{Logic consistency}: verdict aligns with statistical evidence
    \item \textbf{Confound handling}: mixed-group correlations require adjustment or stratification
\end{enumerate}

Violations trigger warnings or \textsc{Invalid} labels, ensuring only well-formed results contribute to evaluation metrics.

\subsection{Evaluation Metric Definitions}
\label{sec:metrics-supp}

Table~\ref{tab:diagnostics} reports six diagnostic metrics organized into three categories. Below we define each metric and the additional diagnostics reported in this supplement.

\paragraph{Conclusion quality.}
\begin{itemize}
    \item \textbf{Overclaim rate}: fraction of YES verdicts where the evidence label is not \textsc{Supported}: the model claims support that its own statistics do not provide.
    \item \textbf{False-refutation rate}: fraction of \textsc{Refuted} evidence labels disagreeing with ground truth.
\end{itemize}

\paragraph{Analysis integrity.}
These metrics detect ``dishonest'' analyses at increasing severity: claiming significance without valid evidence (output-level), fabricating data (code-level), or hard-coding the decision threshold (code-level).
\begin{itemize}
    \item \textbf{Hallucinated significance}: YES verdict with $p \geq 0.05$ or missing $p$-value (output-level).
    \item \textbf{Synthetic-data violation}: code generating mock, random, or simulated data instead of using real measurements via the Imaging Analysis API (code-level). Benign stochastic utilities (\eg, plotting jitter, bootstrap resampling) are tracked separately as non-blocking warnings.
    \item \textbf{Literal $p$-value assignment}: significance threshold ($p = 0.05$) hard-coded in the generated analysis code, detected via static regex analysis of Phase~2B Python files (code-level).
\end{itemize}

\paragraph{Auditability.}
\begin{itemize}
    \item \textbf{Evidence grounding rate} (Verif.): fraction of \emph{all} runs that report the four core statistical outputs required for independent verification: statistical test type, per-group sample sizes, effect size, and $p$-value. This metric captures both execution reliability and output completeness in a single number.
\end{itemize}

\paragraph{Additional diagnostics.}
\begin{itemize}
    \item \textbf{L0 feasibility accuracy}: fraction of untestable (L0) hypotheses correctly labeled \textsc{Invalid}, with subset accuracy excluding early Phase~1 stops.
    \item \textbf{Execution reliability}: early feasibility stops (Phase~1 \textsc{Untestable}) reported separately from true parse/runtime failures.
    \item \textbf{Phase~2B alignment}: fraction of Phase~3 verdicts consistent with Phase~2B statistical outputs.
    \item \textbf{Coding trial count}: mean and maximum code execution attempts per hypothesis.
    \item \textbf{Sample-size consistency} ($n$ mismatch): checks whether reported total $n$ equals the sum of per-group sample sizes.
\end{itemize}

\subsection{Implementation Details}
\label{sec:implementation-supp-detail}

\paragraph{Models.}
We evaluate two model deployments. \textbf{Open-weight deployment (local, via Ollama~\cite{ollama2023}):} \texttt{gpt-oss:20b} for PI and imaging roles, \texttt{qwen3:8b} for statistician discussion and critic, and \texttt{qwen3-coder:30b} for coding. \textbf{Frontier deployment (remote, via OpenRouter~\cite{openrouter2024}):} \texttt{openai/gpt-5.2} for PI/imaging/coding and \texttt{openai/gpt-5-mini} for statistician discussion and critic. All other workflow settings are held fixed.

\paragraph{Segmentation.}
The SAT backend~\cite{zhao2025large} provides neural segmentation across domains: cardiac structures (LV, RV, myocardium) for ACDC, and brain tumor subregions (whole tumor, necrotic core, edema, enhancing tumor) for UCSF-PDGM following BraTS label conventions.

\paragraph{Statistical Libraries.}
Code execution phases have access to numpy, pandas, scipy, statsmodels, and lifelines~\cite{Davidson-Pilon2019} for survival analysis.

\paragraph{Agent Orchestration.}
The workflow is implemented using LangGraph~\cite{langgraph2024}, a graph-based state machine framework that enables conditional routing and state persistence.

\subsection{Power Analysis Details}
\label{sec:power-supp}

Power is computed in two independent contexts to avoid the post-hoc power fallacy~\cite{goodman1994use, hoenig2001abuse}:

\paragraph{Agent-side (Phase 1).}
During planning, the Statistician computes \textit{a priori} power from queried sample sizes and a fixed planning SESOI. This informs study feasibility and test selection but is not used for evidence labeling.

\paragraph{Evaluator-side (evidence labeling).}
After Phase~2B completes, the evaluation pipeline independently recomputes power from realized sample sizes:
\begin{equation}
    \pi = P\left(|T| > t_{\alpha/2} \mid \delta = \delta_0, n_1, n_2\right)
\end{equation}
For evaluator-side evidence labeling, SESOI is hypothesis-specific. Each hypothesis stores a \texttt{sesoi\_profile} in the benchmark JSON, and the evaluator resolves this to a profile-specific target by test family: for group differences, Cohen's $d \in \{0.2, 0.5, 0.8\}$ for \texttt{strict}/\texttt{standard}/\texttt{loose}; for correlations, $r \in \{0.2, 0.3, 0.4\}$; for regression, standardized effect sizes $\in \{0.2, 0.3, 0.4\}$; and for survival analyses, hazard-ratio departures from the null corresponding to \{1.2, 1.5, 2.0\}. Evaluator-computed power is used solely for evidence labeling and is never revealed to Phase~3 agents.

\paragraph{Power calculation formulas.}
For \textbf{two-sample group differences}, power is computed using the noncentral $t$-distribution:
\begin{equation}
    \pi = P\left(|T_{n_1+n_2-2, \lambda}| > t_{\alpha/2, n_1+n_2-2}\right), \quad \lambda = d_0 \sqrt{\frac{n_1 n_2}{n_1 + n_2}}
\end{equation}
where $d_0$ is the SESOI (Cohen's $d$) and $\lambda$ is the noncentrality parameter.
For \textbf{correlations}, Fisher-$z$ transform power:
\begin{equation}
    z_r = \text{arctanh}(r_0), \quad \text{se} = \frac{1}{\sqrt{n-3}}, \quad \pi = \Phi\left(\frac{z_r}{\text{se}} - z_{\alpha/2}\right) + \Phi\left(-\frac{z_r}{\text{se}} - z_{\alpha/2}\right)
\end{equation}
For \textbf{survival analyses} (log-rank and Cox PH), power is approximated via the normal method with the number of events:
\begin{equation}
    \pi \approx P\left(|Z| > z_{\alpha/2} \mid \lambda = \log(\text{HR}_0) \sqrt{d \cdot p_1 p_2}\right)
\end{equation}
where $d$ is the total number of events and $p_1, p_2$ are group proportions.
For \textbf{OLS regression}, $F$-test power via the noncentral $F$-distribution:
\begin{equation}
    \pi = P\left(F > F_{\alpha, p_{\text{test}}, n-k-1} \mid \lambda = n \cdot f^2\right)
\end{equation}
where $f^2 = 0.15$ (medium effect per Cohen) and $p_{\text{test}}$ is the number of tested predictors.

\subsection{Compute Infrastructure}
\label{sec:compute_infra-supp}

Local open-weight experiments used single NVIDIA L40 or A40 GPUs.

\section{Complete Hypothesis Bank}
\label{sec:hypotheses}

This section provides the complete 64-hypothesis tiered benchmark with per-hypothesis specifications, ground-truth labels, and clinical rationale. Table~\ref{tab:datasets} summarizes the per-dataset tier distribution.

\begin{table}[t]
\centering
\caption{Evaluation datasets and tiered hypothesis coverage.}
\label{tab:datasets}
\resizebox{\textwidth}{!}{
\begin{tabular}{llccccccccl}
\toprule
 & & & & \multicolumn{6}{c}{\textbf{Hypotheses per tier}} & \\
\cmidrule{5-10}
\textbf{Dataset} & \textbf{Modality} & \textbf{Subj.} & \textbf{Total} & \textbf{L0} & \textbf{L1} & \textbf{L2} & \textbf{L3} & \textbf{L4} & \textbf{L5} & \textbf{Analysis types} \\
\midrule
ACDC~\cite{bernard2018deep} & Cardiac cine MRI & 150 & 32 & 3 & 3 & 18 & 3 & 5 & 0 & GD, Corr \\
UCSF-PDGM~\cite{calabrese2022university} & Brain MRI (multi-seq.) & 501 & 32 & 2 & 6 & 5 & 3 & 11 & 5 & GD, Corr, Surv, Reg \\
\midrule
\textbf{Combined} & & \textbf{651} & \textbf{64} & \textbf{5} & \textbf{9} & \textbf{23} & \textbf{6} & \textbf{16} & \textbf{5} & \\
\bottomrule
\end{tabular}
}
\\
\footnotesize{GD = Group Difference; Corr = Correlation; Surv = Survival analysis; Reg = Regression with covariates. ACDC has no survival endpoint (no L5). UCSF-PDGM covers all six tiers including Cox PH and OLS regression.}
\vspace{-0.4cm}
\end{table}

\subsection{Hypothesis Bank Design Principles}

Each released hypothesis record contains both a \texttt{sesoi\_profile} and a dataset-derived ground-truth block with \texttt{power\_at\_sesoi}. Throughout the per-hypothesis descriptions below, quoted power values refer to this evaluator-side, hypothesis-specific SESOI unless explicitly marked as Phase~1 planning power.

The bank is designed to probe five distinct capabilities:
\begin{enumerate}
    \item \textbf{Positive controls}: well-established clinical findings that should be \textsc{Supported} (37 hypotheses)
    \item \textbf{Negative controls}: reversed-direction hypotheses that should be \textsc{Refuted} (12 hypotheses)
    \item \textbf{No-effect / nonsense controls}: hypotheses with no biological basis, testing false-positive propensity (4 hypotheses)
    \item \textbf{Underpowered controls}: plausible hypotheses in small subgroups where power $< 0.80$ (6 hypotheses)
    \item \textbf{Untestable controls}: hypotheses requiring unavailable data, testing feasibility detection (5 hypotheses)
\end{enumerate}

\subsection{ACDC Cardiac MRI Hypotheses (32)}

\subsubsection{L0: Untestable (3)}

\paragraph{cardiac\_27: HCM have larger LA volume than NOR.}
\textsc{Invalid} (untestable). Left atrium not in ACDC label set (only LV, RV, myocardium).

\paragraph{cardiac\_28: DCM have lower GLS than NOR.}
\textsc{Invalid} (untestable). Global longitudinal strain not derivable from static segmentation.

\paragraph{cardiac\_32: DCM have higher systolic BP than NOR.}
\textsc{Invalid} (untestable). Blood pressure not in ACDC metadata.

\subsubsection{L1: Metadata-Only Correlations (3)}

\paragraph{cardiac\_24: Height positively correlates with weight.}
Spearman correlation, all 150 subjects. Positive control (well-known anthropometric relationship). SESOI profile: \texttt{standard} ($r_0=0.3$). Power: 0.96.

\paragraph{cardiac\_25: ED frame index correlates with weight.}
Spearman correlation, all 150 subjects. No-effect control (no physiological link between cardiac timing and weight). SESOI profile: \texttt{standard} ($r_0=0.3$). Power: 0.96.

\paragraph{cardiac\_26: Cine frame count correlates with weight.}
Spearman correlation, all 150 subjects. No-effect control (frame count is protocol-dependent). SESOI profile: \texttt{standard} ($r_0=0.3$). Power: 0.96.

\subsubsection{L2: Single Imaging Metric (18)}

All L2 hypotheses use Mann-Whitney U tests with 30 vs 30 subjects. Each carries SESOI profile \texttt{loose} ($d_0{=}0.8$), yielding power 0.86.

\begin{table}[h]
\centering
\caption{ACDC L2 hypotheses: single imaging metric group comparisons.}
\label{tab:acdc_l2}
\scriptsize
\begin{tabular}{lp{6.5cm}lll}
\toprule
\textbf{ID} & \textbf{Hypothesis} & \textbf{Groups} & \textbf{Control} & \textbf{Label} \\
\midrule
cardiac\_01 & DCM have significantly lower LVEF than NOR & DCM vs NOR & Positive & Supported \\
cardiac\_02 & DCM have significantly higher LV myocardial mass than NOR & DCM vs NOR & Positive & Supported \\
cardiac\_03 & DCM have significantly higher LVEF than NOR & DCM vs NOR & Negative & Refuted \\
cardiac\_04 & NOR have significantly larger RVEDV than DCM & NOR vs DCM & Negative & Refuted \\
cardiac\_07 & DCM have significantly larger LVEDV than NOR & DCM vs NOR & Positive & Supported \\
cardiac\_08 & DCM have significantly larger LVESV than NOR & DCM vs NOR & Positive & Supported \\
cardiac\_09 & HCM have significantly higher LV myocardial mass than NOR & HCM vs NOR & Positive & Supported \\
cardiac\_10 & HCM have significantly higher LVEF than DCM & HCM vs DCM & Positive & Supported \\
cardiac\_11 & MINF have significantly lower LVEF than NOR & MINF vs NOR & Positive & Supported \\
cardiac\_12 & RV patients have significantly lower RVEF than NOR & RV vs NOR & Positive & Supported \\
cardiac\_13 & HCM have significantly lower LV myocardial mass than NOR & HCM vs NOR & Negative & Refuted \\
cardiac\_14 & DCM have significantly smaller LVEDV than NOR & DCM vs NOR & Negative & Refuted \\
cardiac\_15 & DCM have significantly smaller LVESV than NOR & DCM vs NOR & Negative & Refuted \\
cardiac\_16 & RV patients have significantly higher RVEF than NOR & RV vs NOR & Negative & Refuted \\
cardiac\_19 & HCM have significantly smaller LVEDV than NOR & HCM vs NOR & Positive & Refuted \\
cardiac\_20 & HCM have significantly smaller LVESV than NOR & HCM vs NOR & Positive & Supported \\
cardiac\_30 & RV patients have significantly larger RVEDV than NOR & RV vs NOR & Positive & Supported \\
cardiac\_31 & RV patients have significantly smaller RVEDV than NOR & RV vs NOR & Negative & Refuted \\
\bottomrule
\end{tabular}
\end{table}

\subsubsection{L3: Engineered Features (3)}

\paragraph{cardiac\_21: DCM have higher LVESV/LVEDV ratio than NOR.}
Mann-Whitney U, 30 vs 30. Positive control. Inverse of EF; reduced systolic function elevates ratio. SESOI profile: \texttt{loose} ($d_0=0.8$). Power: 0.86.

\paragraph{cardiac\_22: HCM have higher mass-to-volume ratio than DCM.}
Mann-Whitney U, 30 vs 30. Positive control. HCM: thick walls + small cavity; DCM: dilated cavity. SESOI profile: \texttt{loose} ($d_0=0.8$). Power: 0.86.

\paragraph{cardiac\_23: DCM have lower mass-to-volume ratio than NOR.}
Mann-Whitney U, 30 vs 30. Positive control. DCM dilation increases LVEDV more than mass. SESOI profile: \texttt{loose} ($d_0=0.8$). Power: 0.86.

\subsubsection{L4: Mixed Metadata + Imaging (5)}

\paragraph{cardiac\_05: Weight correlates with LVEF in DCM+NOR.}
Spearman, $n=60$. Underpowered control (weak/confounded relationship). SESOI profile: \texttt{standard} ($r_0=0.3$). Power: 0.65.

\paragraph{cardiac\_06: Height correlates with LVEDV.}
Spearman, $n=150$. Positive control (body-size scaling). SESOI profile: \texttt{standard} ($r_0=0.3$). Power: 0.96.

\paragraph{cardiac\_17: In NOR: weight correlates with LVEF.}
Spearman, $n=30$. Underpowered control. SESOI profile: \texttt{standard} ($r_0=0.3$). Power: 0.36.

\paragraph{cardiac\_18: In DCM: height correlates with LVEF.}
Spearman, $n=30$. Underpowered control. SESOI profile: \texttt{standard} ($r_0=0.3$). Power: 0.36.

\paragraph{cardiac\_29: DCM have higher LVEDV/BSA than NOR.}
Mann-Whitney U, 30 vs 30. Positive control. BSA = $\sqrt{h_{\text{cm}} \times w_{\text{kg}} / 3600}$. SESOI profile: \texttt{loose} ($d_0=0.8$). Power: 0.86.

\subsection{UCSF-PDGM Glioma MRI Hypotheses (32)}

\subsubsection{L0: Untestable (2)}

\paragraph{glioma\_30: Mean tumor ADC associated with survival.}
\textsc{Invalid}. ADC/DWI modality not available through the Imaging Analysis API.

\paragraph{glioma\_32: IDH-mutant gliomas more often frontal.}
\textsc{Invalid}. No tumor location or atlas features available.

\subsubsection{L1: Metadata-Only (6)}

\paragraph{glioma\_29: In GBM, men are older than women.}
Mann-Whitney U, M (239) vs F (157) in Grade IV ($n=396$). Nonsense control (no biological basis for sex$\to$age shift in GBM). SESOI profile: \texttt{standard} ($d_0=0.5$). Power: $>$0.99.

\paragraph{glioma\_11: MGMT methylated GBM survive longer.}
Log-rank + Cox HR, methylated (273) vs unmethylated (105), Grade IV ($n=378$, 217 events). Positive control. SESOI profile: \texttt{loose} (HR-scale target 2.0). Power: $>$0.99.

\paragraph{glioma\_13: GTR patients survive longer than STR.}
Log-rank + Cox HR, GTR (227) vs STR (127), Grade IV. Positive control. SESOI profile: \texttt{loose} (HR-scale target 2.0). Power: $>$0.99.

\paragraph{glioma\_14: IDH-mutant Grade IV survive longer.}
Log-rank + Cox HR, mutant (28) vs wildtype (367), Grade IV. Underpowered control (few IDH-mutant in Grade IV). SESOI profile: \texttt{standard} (HR-scale target 1.5). Power: 0.72.

\paragraph{glioma\_16: GTR patients have shorter survival.}
Log-rank (reuses glioma\_13). Negative control (reversed direction). SESOI profile: \texttt{loose} (HR-scale target 2.0). Power: $>$0.99.

\paragraph{glioma\_28: In GBM, men survive longer than women.}
Log-rank + Cox HR, M (238) vs F (157), Grade IV. Nonsense control. SESOI profile: \texttt{standard} (HR-scale target 1.5). Power: $>$0.99.

\subsubsection{L2: Imaging-Only (5)}

All L2 hypotheses use Mann-Whitney U tests, Grade IV ($n{=}396$) vs Grade II ($n{=}56$). Each carries SESOI profile \texttt{loose} ($d_0{=}0.8$), with power $>$0.99.

\paragraph{glioma\_01: Grade IV have larger whole tumor volume than Grade II.}
Positive control. Well-established: higher-grade gliomas grow more aggressively.

\paragraph{glioma\_05: Grade II have larger enhancing tumor volume than Grade IV.}
Negative control (reversed direction). Enhancement is a hallmark of high-grade tumors.

\paragraph{glioma\_07: Grade IV have larger enhancing tumor volume than Grade II.}
Positive control. Contrast enhancement reflects blood-brain barrier disruption in GBM.

\paragraph{glioma\_08: Grade IV have larger necrotic core volume than Grade II.}
Positive control. Necrosis is a defining feature of Grade~IV (GBM).

\paragraph{glioma\_09: Grade IV have larger edema volume than Grade II.}
Positive control. Higher-grade tumors cause more peritumoral edema.

\subsubsection{L3: Engineered Features (3)}

\paragraph{glioma\_19/20/21: Grade IV have higher enhancing / necrotic / edema fraction.}
Mann-Whitney U, Grade IV (396) vs Grade II (56). Positive controls. Each uses SESOI profile \texttt{standard} ($d_0=0.5$). Power: 0.94. Ground truth: only the enhancing-fraction hypothesis (glioma\_19) is \textsc{Supported}; the necrotic (glioma\_20) and edema (glioma\_21) fraction hypotheses are \textsc{Refuted}.

\subsubsection{L4: Mixed (11)}

\paragraph{glioma\_02: IDH-wildtype tumors have larger enhancing tumor volumes than IDH-mutant.}
Mann-Whitney U, all grades ($n{=}495$), grouped by IDH status. Positive control. SESOI profile: \texttt{loose} ($d_0=0.8$). Power: $>$0.99.

\paragraph{glioma\_04: IDH-mutant tumors have more necrotic core than IDH-wildtype.}
Mann-Whitney U, all grades ($n{=}495$). Negative control (reversed direction). SESOI profile: \texttt{loose} ($d_0=0.8$). Power: $>$0.99.

\paragraph{glioma\_06: Age correlates with whole tumor volume in GBM.}
Spearman, Grade IV ($n{=}396$). Negative control (no expected relationship). SESOI profile: \texttt{standard} ($r_0{=}0.3$). Power: $>$0.99.

\paragraph{glioma\_10: IDH-wildtype tumors have larger necrotic core than IDH-mutant.}
Mann-Whitney U, all grades ($n{=}495$). Positive control. SESOI profile: \texttt{loose} ($d_0{=}0.8$). Power: $>$0.99.

\paragraph{glioma\_17: In Grade~II, IDH-wildtype have more enhancement than IDH-mutant.}
Mann-Whitney U, Grade II only ($n{=}56$; IDH-wt $n{=}10$, IDH-mut $n{=}46$). Underpowered control (very small IDH-wt subgroup). SESOI profile: \texttt{loose} ($d_0{=}0.8$). Power: 0.61.

\paragraph{glioma\_22: IDH-wildtype tumors have higher enhancing fraction than IDH-mutant.}
Mann-Whitney U, all grades ($n{=}495$). Positive control. SESOI profile: \texttt{standard} ($d_0{=}0.5$). Power: 0.99.

\paragraph{glioma\_31: In lower-grade gliomas, 1p/19q codeleted tumors have lower enhancing fraction.}
Mann-Whitney U, Grade II/III ($n{=}99$; codeleted $n{=}13$, intact $n{=}86$). Underpowered control. SESOI profile: \texttt{standard} ($d_0{=}0.5$). Power: 0.38.

\paragraph{glioma\_03: Enhancing tumor volume negatively correlates with survival in GBM.}
Cox PH / log-rank, Grade IV ($n{=}395$, 228 events). Imaging predictor: ET\_ml. Positive control. SESOI profile: \texttt{standard} (HR-scale target 1.5). Power: $>$0.99.

\paragraph{glioma\_15: Enhancing tumor volume positively correlates with survival in GBM.}
Cox PH / log-rank, Grade IV ($n{=}395$). Negative control (reversed direction of glioma\_03). SESOI profile: \texttt{standard} (HR-scale target 1.5). Power: $>$0.99.

\paragraph{glioma\_24: Higher necrotic fraction correlates with worse survival in GBM.}
Cox PH, Grade IV ($n{=}395$). Positive control (expected but borderline). SESOI profile: \texttt{standard} (HR-scale target 1.5). Power: $>$0.99.

\paragraph{glioma\_12: CET volume $>$10\,mL associated with shorter survival in GBM.}
Log-rank, Grade IV ($n{=}395$). Derived binary grouping: ET $>$10\,mL vs $\leq$10\,mL. Positive control. SESOI profile: \texttt{standard} (HR-scale target 1.5). Power: $>$0.99.

\subsubsection{L5: Multivariate / Advanced (5)}

\paragraph{glioma\_25: MGMT protective after adjusting for age + EoR.}
Cox PH multivariate, Grade IV ($n=378$, 217 events). Primary: mgmt\_binary (methylated=1). Covariates: age, eor\_binary (GTR=1). Positive control. SESOI profile: \texttt{standard} (HR-scale target 1.5). Power: $>$0.99.

\paragraph{glioma\_26: Older age worsens survival adjusting MGMT + EoR.}
Cox PH multivariate, Grade IV. Primary: age (continuous). Covariates: mgmt\_binary, eor\_binary. Positive control. SESOI profile: \texttt{standard} (HR-scale target 1.5). Power: $>$0.99.

\paragraph{glioma\_27: IDH-wt increases enhancing fraction after grade adjustment.}
OLS regression, all grades ($n=495$). Dependent: enhancing\_fraction. Predictors: idh\_wildtype + grade dummies (ref=Grade II). Positive control. SESOI profile: \texttt{standard} ($f^2$ target 0.3). Power: $>$0.99.

\paragraph{glioma\_18: ET volume predicts survival adjusting for EoR.}
Cox PH multivariate, Grade IV. Primary: ET\_ml (continuous). Covariate: eor\_binary. Positive control. SESOI profile: \texttt{standard} (HR-scale target 1.5). Power: $>$0.99.

\paragraph{glioma\_23: IDH protective in full multivariate model.}
Cox PH, all grades ($n=494$, 248 events). Primary: idh\_binary (mutant=1). Covariates: age, grade\_III, grade\_IV, mgmt\_binary, eor\_binary, ET\_ml. 7-covariate model. Positive control. SESOI profile: \texttt{standard} (HR-scale target 1.5). Power: $>$0.99.

\section{Ground-Truth Computation Pipeline}
\label{sec:ground_truth}

\subsection{Methodology}

Ground-truth evidence labels are computed independently of the agent pipeline, using the same imaging data and statistical conventions. For each hypothesis, we:

\begin{enumerate}
    \item Load ground-truth segmentation masks (expert-annotated for ACDC; BraTS-convention masks for UCSF-PDGM)
    \item Compute metrics using the same analysis-interface functions available to agents (\texttt{calculate\_volume}, \texttt{calculate\_mass}, \texttt{calculate\_ejection\_fraction})
    \item Execute the canonical statistical test (Mann-Whitney U for group differences, Spearman for correlations, log-rank/Cox PH for survival, OLS for regression)
    \item Record complete statistics: sample sizes, effect size with 95\% CI, p-value, group medians/means
    \item Resolve the hypothesis-specific \texttt{sesoi\_profile} and compute power at that SESOI
    \item Assign the evidence label mechanically using the rules in Table~\ref{tab:evidence}
\end{enumerate}

Power is computed per test family using the formulas in \cref{sec:power-supp}. The implementation uses \texttt{statsmodels} power functions for group differences and regression, and analytic normal approximations for correlations and survival analyses.

\section{Agent System Prompts}
\label{sec:prompts}

All agents are defined as structured objects with four fields: \texttt{title}, \texttt{expertise}, \texttt{goal}, and \texttt{role}. The system prompt for each agent is rendered as: ``You are a \{title\}. Your expertise is in \{expertise\}. Your goal is to \{goal\}. Your role is to \{role\}.'' Below we reproduce the complete role definitions; expertise and goal fields are shown inline.

\subsection{Principal Investigator}
\textit{Expertise:} running a science research lab. \textit{Goal:} perform research that maximizes scientific impact.

\begin{promptbox}[Principal Investigator Role Prompt]
\begin{lstlisting}[basicstyle=\ttfamily\scriptsize,breaklines=true,numbers=none,xleftmargin=0pt,breakindent=0pt,breakautoindent=false]
Lead a team of experts to solve important scientific problems. Listen to team member input, synthesize their recommendations into clear, actionable decisions. Make decisive choices when there are trade-offs. Follow any output format requirements specified.
\end{lstlisting}
\end{promptbox}

\subsection{Medical Imaging Specialist (Discussion)}
\textit{Expertise:} medical image analysis, image segmentation, and SAT foundation model usage. \textit{Goal:} provide expertise on imaging analysis capabilities and segmentation approaches.

\begin{promptbox}[Medical Imaging Specialist (Discussion) Role Prompt]
\begin{lstlisting}[basicstyle=\ttfamily\scriptsize,breaklines=true,numbers=none,xleftmargin=0pt,breakindent=0pt,breakautoindent=false]
Provide imaging expertise for research planning. Recommend which anatomical structures to segment, which observations to capture, and assess segmentation feasibility and expected quality.

Available tools:
- list_dataset_patients(dataset, group=None, metadata_filters=None):
  Discover cohort sizes. Returns: {patients: [{patient_id, group}], total_count}
- list_available_structures(category): Verify segmentable structures.
  Categories: cardiac, abdominal, vascular, spine, brain, urogenital, etc.
- check_sat_status(): Verify SAT model availability

Use tools to gather information needed for recommendations.
\end{lstlisting}
\end{promptbox}

\subsection{Medical Imaging Specialist (Interpretation)}
\textit{Expertise:} medical image analysis, segmentation quality assessment, and technical limitations. \textit{Goal:} interpret segmentation results and assess their impact on study conclusions.

\begin{promptbox}[Medical Imaging Specialist (Interpretation) Role Prompt]
\begin{lstlisting}[basicstyle=\ttfamily\scriptsize,breaklines=true,numbers=none,xleftmargin=0pt,breakindent=0pt,breakautoindent=false]
Interpret imaging results from a technical perspective. NOTE: Segmentation quality metrics (Dice scores) are NOT available in the current framework since there is no ground truth - assume segmentations are adequate for analysis. Discuss potential technical limitations (partial volumes, motion artifacts, frame selection) that could affect interpretation, but do NOT request unavailable metrics or make their absence a reason for inconclusive verdict. Focus on whether technical factors would plausibly invalidate the statistical findings.
\end{lstlisting}
\end{promptbox}

\subsection{ML Statistician (Discussion)}
\textit{Expertise:} statistical experimental design, power analysis, hypothesis testing methodology. \textit{Goal:} design rigorous statistical experiments and provide methodological guidance.

\begin{promptbox}[ML Statistician (Discussion) Role Prompt]
\begin{lstlisting}[basicstyle=\ttfamily\scriptsize,breaklines=true,numbers=none,xleftmargin=0pt,breakindent=0pt,breakautoindent=false]
Provide statistical guidance for research planning. Recommend appropriate statistical tests (t-test, Mann-Whitney, ANOVA, etc.), assess sample size adequacy, and discuss study feasibility.

Focus on TEXT-BASED discussion. The meeting output is a PLAN, not code.

**Power analysis (REQUIRED in Phase 1):**
Use the sample sizes reported by the Imaging Specialist to compute a priori power. Run this ONCE and report your conclusion:

  from statsmodels.stats.power import TTestIndPower
  analysis = TTestIndPower()
  power = analysis.power(effect_size=0.5, nobs1=<n1>, alpha=0.05, ratio=<n2/n1>)

For correlation hypotheses, use NormalIndPower with effect_size=0.3.
State clearly whether the study is adequately powered (threshold: 0.80).
\end{lstlisting}
\end{promptbox}

\subsection{Coding ML Statistician}
\textit{Expertise:} statistical analysis, ML, Python programming with scipy, numpy, sklearn, matplotlib. \textit{Goal:} test hypotheses through code-driven statistical analysis.

\begin{promptbox}[ML Statistician (Coding) Role Prompt]
\begin{lstlisting}[basicstyle=\ttfamily\scriptsize,breaklines=true,numbers=none,xleftmargin=0pt,breakindent=0pt,breakautoindent=false]
Execute statistical analysis using Python code. Write code in steps if needed - each code block runs immediately and you see the output before continuing.

**Pre-loaded Imaging Analysis API:**
- sat.list_patients(db_path) -> list of patient_ids
- sat.get_patient_metadata(patient_id) -> dict with 'group', etc.
- sat.load_structure_mask(db_path, patient_id, structure,
    source_image_contains=observation) -> list of {mask, spacing}
- sat.calculate_volume(mask, spacing) -> float (mL)

**Data requirements:**
- Use REAL data from the Imaging Analysis API only (no mock/random data)
- Do NOT use raw filesystem crawling as a substitute for the Imaging Analysis API
- Prefer the planned hypothesis test; switch only if assumptions violated

**Statistical standards:**
- Sign convention: positive effect = group1 has HIGHER values
- Cohen's d = (mean1 - mean2) / pooled_std
- Mann-Whitney U effect size: rank-biserial r
- CI: bootstrap for nonparametric, parametric CI for t-test
- Survival: logrank_test or CoxPHFitter, effect size = hazard ratio

**Required outputs:**
- data/statistical_results.json (exact schema in agenda)
- plots/*.png (at least one visualization)
\end{lstlisting}
\end{promptbox}

\subsection{Phase-Aware Critic}
\textit{Expertise:} providing phase-appropriate critical feedback for multi-phase research workflows. \textit{Goal:} ensure work meets the specific goals of the current phase without demanding out-of-scope analysis.

\begin{promptbox}[Phase-Aware Critic Role Prompt]
\begin{lstlisting}[basicstyle=\ttfamily\scriptsize,breaklines=true,numbers=none,xleftmargin=0pt,breakindent=0pt,breakautoindent=false]
Provide phase-specific critique only. Identify the current phase from workflow instruction and evaluate only that phase.

General critic behavior:
- Prioritize blocking errors first, then high-value warnings.
- Cite concrete evidence for each issue and give one actionable fix.
- Do not request extra analyses outside the plan.

PHASE 1 (planning):
- Verify plan feasibility reasoning is coherent.
- Verify core contract fields: groups, structures, observations, metrics.
- Do not demand code execution in Phase 1.

PHASE 2A (segmentation request):
- Verify code writes segmentation_request.json.
- Verify structures/observations align with Phase 1 plan.
- Do not demand statistical testing in Phase 2A.

PHASE 2B (statistical analysis):
Blocking checks:
1) No fabricated/synthetic/mock data
2) No off-contract primary data loading
3) No manual sample capping/subsampling
4) data/statistical_results.json must exist with required keys
5) No silent deviation from planned groups/test/adjust_for
6) If plan requires adjustment, verify implementation

Test policy: Prefer the planned test. If assumptions invalidate it, allow one valid alternative when justified.

PHASE 3 (interpretation):
- Verify interpretation uses Phase 2B results (no re-analysis).
- Verify final verdict is clear and supported.
- Verify final verdict JSON block is present.
\end{lstlisting}
\end{promptbox}

\section{Phase Agenda Templates}
\label{sec:phase_agendas}

Each phase begins with a structured \emph{agenda} that is dynamically generated from the dataset configuration, hypothesis text, and outputs of preceding phases. Below we reproduce the core template for each phase, using angle-bracket placeholders (\texttt{<...>}) for values injected at runtime. These agendas are the actual prompts given to agents: no additional instructions are provided beyond the agent system prompts (Section~\ref{sec:prompts}) and the agenda below.

\subsection{Phase 1: Planning Agenda}

The planning agenda provides the hypothesis, dataset context, and role-specific instructions for the three discussion agents.

\begin{promptbox}[Analysis Planning Phase Prompt]
\begin{lstlisting}[basicstyle=\ttfamily\scriptsize,breaklines=true,numbers=none]
**Research Question:** <hypothesis>

**Dataset Context:**
<dataset_info_text>

**Task:** produce one executable analysis plan.

**Core rules (all agents):**
- Keep the target quantity exact; no proxy substitution unless hypothesis explicitly allows it.
- If prerequisites are missing in principle, mark UNTESTABLE with subtype + missing requirements.
- Use only dataset observations/timepoints in observations; never place metadata there.

**Imaging specialist:**
- Choose exact structure names via list_available_structures().
- Query cohort/sample sizes with list_dataset_patients() for planned groups.
- If hypothesis is metadata-only (no image-derived measurements), set structures: [].

**Statistician:**
- Choose one primary test that matches the hypothesis.
- Compute a priori power using queried sample sizes (d=0.5 group tests, r=0.3 correlations); report adequacy at 0.80.
- Survival: use analysis_type: "survival"; use log-rank for unadjusted two-group survival and Cox PH for adjusted/continuous-predictor survival.
- Mixed-cohort correlation/regression: consider confounding control via adjust_for or stratify_by.
- For correlation/regression, declare exact tested variables in target_variables (one outcome + predictors).

**PI synthesis:**
- Use only metadata fields listed in dataset context.
- Metadata-value groups belong in groups; use grouping_field and restrict_to when needed.
- Derived groups require group_spec: {type: "derived", rule: "..."}.

**Output:** a valid Phase 1 JSON plan block with feasibility + analysis contract fields.
\end{lstlisting}
\end{promptbox}

The PI then produces a structured JSON output containing: feasibility status, groups, structures, observations, metrics, statistical test, analysis type, grouping field, predictors, covariates, and target variables. The full output schema is defined in \texttt{build\_phase1\_summary\_instructions()}.

\subsection{Phase 2A: Segmentation Request Agenda}

Phase 2A translates the Phase 1 plan into a concrete segmentation request by querying the Imaging Analysis API for patient identifiers.

\begin{promptbox}[Segmentation Phase Prompt]
\begin{lstlisting}[basicstyle=\ttfamily\scriptsize,breaklines=true,numbers=none]
Build segmentation_request.json from the Phase 1 plan.

**Plan contract:**
- Dataset: <dataset_name>
- Groups: <groups>
- Structures: <structures>
- Observations: <observations>
- Cohort mode: <cohort_mode>

**Rules:**
- Use exact structure names from plan, unchanged.
- Build identifiers in format <dataset>:<patient_id>:<observation>.
- Include only plan observations.
- If groups are ["ALL"], iterate real dataset groups; never pass "ALL" to API.
- For metadata-group hypotheses, map plan labels to canonical metadata values.

**API (pre-loaded):**
  result = list_dataset_patients("<dataset>", group="<group>")
  meta = get_patient_metadata("<dataset>", "patient001")
  for obs_name in <observations>:
      identifier = meta["identifiers"].get(obs_name)

**Required output (segmentation_request.json):**
  {
    "identifiers": ["<dataset>:patient001:<obs>", ...],
    "structures": <structures>,
    "results_database": "<results_db>",
    "modality": "<modality>",
    "model_variant": "nano",
    "chunk_size": 64
  }

Write complete code now and save as segmentation_request.json.
\end{lstlisting}
\end{promptbox}

\subsection{Phase 2B: Statistical Analysis Agenda}

The statistical analysis agenda is the most detailed, specifying the exact analysis contract, Imaging Analysis API usage, hard constraints, and required output schema.

\begin{promptbox}[Statistical Analysis Phase Prompt]
\begin{lstlisting}[basicstyle=\ttfamily\scriptsize,breaklines=true,numbers=none]
Execute Phase 2B statistical analysis using Phase 2A outputs.

**Plan contract (must match exactly):**
- results_db: <phase2a_results_db>
- groups: <groups>
- structures: <structures>
- observations: <observations>
- metrics: <metrics>
- statistical_test: <test>
- analysis_type: <group_difference|correlation|regression|survival>
- cohort_mode: <groups|all>
- predictors: <predictors>
- adjust_for: <covariates> -- REQUIRED|RECOMMENDED|OPTIONAL
- stratify_by: <variables> -- REQUIRED|RECOMMENDED|OPTIONAL
- target_variables: {outcome: "<var>", predictors: ["<var>", ...]}
- available metadata fields: <metadata_fields>

**Hard constraints:**
- Use the Imaging Analysis API as primary data source; no synthetic/mock placeholders.
- Use full eligible cohort after planned restrictions (no slicing/sample caps).
- Prefer the planned primary analysis. If assumptions invalidate it, switch only to a statistically valid alternative and document the reason.
- Implement planned adjustment/stratification in code and in statistical_results.json.
- If any planned group has n=0, raise an error and fix loading/filter logic.
- Before finishing, ensure data/statistical_results.json exists.

**Analysis track (strict):**
<track-specific guidance based on analysis_type and adjustment level>

**Task:** Write Python code to:
1. Load segmentation results via sat.list_patients(results_db_path)
2. Filter patients by group (<group_filter_instruction>)
3. Load masks for structures at observations
4. Calculate metrics from segmentation masks
5. Perform <statistical_test> according to analysis_type=<type>
6. Create visualizations
7. Save results to data/statistical_results.json and plots/*.png

**Imaging Analysis API (pre-loaded):**
- sat.list_patients(results_db_path) -> all patient IDs
- sat.get_patient_metadata(patient_id) -> metadata
- sat.get_observation_identifiers(patient_id) -> per-patient obs map
- sat.load_structure_mask(results_db_path, patient_id, structure, source_image_contains=...)
- sat.calculate_volume(mask, spacing)

**Required output (data/statistical_results.json):**
  {
    "analysis_type": "<type>",
    "test_performed": "<test>",
    "p_value": <float>,
    "effect_size": <float>,
    "effect_size_type": "<cohens_d|rank_biserial|hazard_ratio|...>",
    "n_total": <int>,
    "sample_sizes": {"<group1>": <int>, "<group2>": <int>},
    "variables_tested": {outcome: "<var>", predictors: ["<var>"]}
  }

Write complete analysis code now.
\end{lstlisting}
\end{promptbox}

\subsection{Phase 3: Interpretation Agenda}

The interpretation agenda provides the results snapshot from prior phases and guides the discussion toward a final verdict.

\begin{promptbox}[Interpretation Phase Prompt]
\begin{lstlisting}[basicstyle=\ttfamily\scriptsize,breaklines=true,numbers=none]
**Hypothesis:** <hypothesis>

**Results from Previous Phases:**
<results_snapshot>

**Task:** Interpret the results and reach a verdict.

**Discussion Focus:**
This is an INTERPRETATION phase - focus on TEXT-BASED discussion.
- Review what was found in Phase 2A (segmentation) and Phase 2B (statistics)
- Discuss implications, limitations, and confidence in the findings
- Reach a final verdict: YES (supported), NO (rejected), or INCONCLUSIVE

**Verdict Guidelines:**
- YES: p < 0.05 AND effect in direction claimed by hypothesis. Verify direction from group_statistics means (NOT effect_size sign).
- NO: Not statistically significant, OR effect direction is opposite. Non-significant result in adequately powered study (power >= 0.80) is evidence AGAINST the hypothesis.
- INCONCLUSIVE: ONLY when BOTH: (1) Phase 1 found study underpowered (power < 0.80) AND (2) result is not significant (p >= 0.05).
- INVALID: fundamental methodological failures (NaN results, p=1.0 exactly, completely wrong analysis type) -- not for assumption violations.

**Required output:**
  {
    "verdict": "<YES|NO|INCONCLUSIVE>",
    "evidence_label": "<SUPPORTED|REFUTED|UNDERPOWERED|INVALID>",
    "p_value": <number from Phase 2B>,
    "effect_size": <number from Phase 2B>,
    "test_used": "<test name>",
    "sample_sizes": {"group1": N, "group2": N},
    "confidence": "<high|medium|low>",
    "reasoning": "<one sentence conclusion>"
  }
\end{lstlisting}
\end{promptbox}

\vspace{4pt}

\section{Ethics and Broader Impact}
\label{sec:ethics}

\begin{enumerate}
    \item \textbf{Research acceleration, not clinical deployment}: \framework{} is designed to accelerate hypothesis screening in research settings. It produces evidence labels, not medical advice, and human expert oversight remains essential before any clinical or policy decision.
    \item \textbf{Generalization}: Our evaluation spans two MRI modalities (cardiac cine and brain MRI) with a single segmentation backend. Performance on other imaging modalities, pathologies, or segmentation models requires future validation.
    \item \textbf{Statistical assumptions}: The evidence-label framework relies on conventional SESOI thresholds. Domain-specific calibration of these thresholds may be necessary for applications where clinically meaningful effect sizes differ substantially from Cohen's conventions.
    \item \textbf{Code correctness}: While code-driven execution eliminates natural-language statistical hallucination (fabricated $p$-values), agents can generate syntactically correct but methodologically flawed code (\eg, the mediator-adjustment error in Section~\ref{sec:failures}). The automated validity checks (Section~\ref{sec:validity-supp}) catch many but not all such errors.
    \item \textbf{Automation bias}: Users may over-trust automated conclusions, particularly when the system produces detailed statistical outputs and visualizations. The auditability framework (Section~\ref{sec:artifact-trail}) mitigates this by making every reasoning step and code artifact inspectable, but appropriate training on system limitations remains important.
\end{enumerate}

%% file: main.bib
@article{swanson2025virtual,
  title={The Virtual Lab of AI agents designs new SARS-CoV-2 nanobodies},
  author={Swanson, Kyle and Wu, Wesley and Bulaong, Nash L and Pak, John E and Zou, James},
  journal={Nature},
  volume={646},
  number={8085},
  pages={716--723},
  year={2025},
  publisher={Nature Publishing Group UK London}
}

@article{zhang2026virtual,
  title={The Virtual Biotech: A Multi-Agent AI Framework for Therapeutic Discovery and Development},
  author={Zhang, Harrison G and Eckmann, Peter and Miao, Jiacheng and Mahon, Andrew B and Zou, James},
  journal={bioRxiv},
  pages={2026--02},
  year={2026},
  publisher={Cold Spring Harbor Laboratory}
}

@inproceedings{hong2023metagpt,
  title={MetaGPT: Meta programming for a multi-agent collaborative framework},
  author={Hong, Sirui and Zhuge, Mingchen and Chen, Jonathan and Zheng, Xiawu and Cheng, Yuheng and Wang, Jinlin and Zhang, Ceyao and Wang, Zili and Yau, Steven Ka Shing and Lin, Zijuan and others},
  booktitle={The twelfth international conference on learning representations},
  year={2023}
}

@article{hoopes2024voxelprompt,
  title={VoxelPrompt: A Vision Agent for End-to-End Medical Image Analysis},
  author={Hoopes, Andrew and Dey, Neel and Butoi, Victor Ion and Guttag, John V and Dalca, Adrian V},
  journal={arXiv preprint arXiv:2410.08397},
  year={2024}
}

@article{zhao2024biomedparse,
  title={Biomedparse: a biomedical foundation model for image parsing of everything everywhere all at once},
  author={Zhao, Theodore and Gu, Yu and Yang, Jianwei and Usuyama, Naoto and Lee, Ho Hin and Naumann, Tristan and Gao, Jianfeng and Crabtree, Angela and Abel, Jacob and Moung-Wen, Christine and others},
  journal={arXiv preprint arXiv:2405.12971},
  year={2024}
}

@article{chen2025radfabric,
  title={RadFabric: Agentic AI System with Reasoning Capability for Radiology},
  author={Chen, Wenting and Dong, Yi and Ding, Zhaojun and Shi, Yucheng and Zhou, Yifan and Zeng, Fang and Luo, Yijun and Lin, Tianyu and Su, Yihang and Wu, Yichen and others},
  journal={arXiv preprint arXiv:2506.14142},
  year={2025}
}

@inproceedings{tang2024medagents,
  title={Medagents: Large language models as collaborators for zero-shot medical reasoning},
  author={Tang, Xiangru and Zou, Anni and Zhang, Zhuosheng and Li, Ziming and Zhao, Yilun and Zhang, Xingyao and Cohan, Arman and Gerstein, Mark},
  booktitle={Findings of the Association for Computational Linguistics: ACL 2024},
  pages={599--621},
  year={2024}
}

@article{jin2025agentmd,
  title={AgentMD: Empowering language agents for risk prediction with large-scale clinical tool learning},
  author={Jin, Qiao and Wang, Zhizheng and Yang, Yifan and Zhu, Qingqing and Wright, Donald and Huang, Thomas and Khandekar, Nikhil and Wan, Nicholas and Ai, Xuguang and Wilbur, W John and others},
  journal={Nature Communications},
  volume={16},
  number={1},
  pages={9377},
  year={2025},
  publisher={Nature Publishing Group UK London}
}

@article{kim2024mdagents,
  title={Mdagents: An adaptive collaboration of llms for medical decision-making},
  author={Kim, Yubin and Park, Chanwoo and Jeong, Hyewon and Chan, Yik S and Xu, Xuhai and McDuff, Daniel and Lee, Hyeonhoon and Ghassemi, Marzyeh and Breazeal, Cynthia and Park, Hae W},
  journal={Advances in Neural Information Processing Systems},
  volume={37},
  pages={79410--79452},
  year={2024}
}

@article{li2024agent,
  title={Agent hospital: A simulacrum of hospital with evolvable medical agents},
  author={Li, Junkai and Lai, Yunghwei and Li, Weitao and Ren, Jingyi and Zhang, Meng and Kang, Xinhui and Wang, Siyu and Li, Peng and Zhang, Ya-Qin and Ma, Weizhi and others},
  journal={arXiv preprint arXiv:2405.02957},
  year={2024}
}

@inproceedings{su2025many,
  title={Many heads are better than one: Improved scientific idea generation by a llm-based multi-agent system},
  author={Su, Haoyang and Chen, Renqi and Tang, Shixiang and Yin, Zhenfei and Zheng, Xinzhe and Li, Jinzhe and Qi, Biqing and Wu, Qi and Li, Hui and Ouyang, Wanli and others},
  booktitle={Proceedings of the 63rd Annual Meeting of the Association for Computational Linguistics (Volume 1: Long Papers)},
  pages={28201--28240},
  year={2025}
}

@article{ghafarollahi2025sciagents,
  title={SciAgents: automating scientific discovery through bioinspired multi-agent intelligent graph reasoning},
  author={Ghafarollahi, Alireza and Buehler, Markus J},
  journal={Advanced Materials},
  volume={37},
  number={22},
  pages={2413523},
  year={2025},
  publisher={Wiley Online Library}
}

@inproceedings{zhang2025multi,
  title={Multi-agent reasoning for cardiovascular imaging phenotype analysis},
  author={Zhang, Weitong and Qiao, Mengyun and Zang, Chengqi and Niederer, Steven and Matthews, Paul M and Bai, Wenjia and Kainz, Bernhard},
  booktitle={International Conference on Medical Image Computing and Computer-Assisted Intervention},
  pages={429--439},
  year={2025},
  organization={Springer}
}

@article{yang2025qwen3,
  title={Qwen3 technical report},
  author={Yang, An and Li, Anfeng and Yang, Baosong and Zhang, Beichen and Hui, Binyuan and Zheng, Bo and Yu, Bowen and Gao, Chang and Huang, Chengen and Lv, Chenxu and others},
  journal={arXiv preprint arXiv:2505.09388},
  year={2025}
}

@article{cao2026qwen3,
  title={Qwen3-Coder-Next Technical Report},
  author={Cao, Ruisheng and Chen, Mouxiang and Chen, Jiawei and Cui, Zeyu and Feng, Yunlong and Hui, Binyuan and Jing, Yuheng and Li, Kaixin and Li, Mingze and Lin, Junyang and others},
  journal={arXiv preprint arXiv:2603.00729},
  year={2026}
}

@misc{openai2025gpt52,
  author = {{OpenAI}},
  title  = {Introducing GPT-5.2},
  year   = {2025},
  month  = {12},
  day    = {11},
  url    = {https://openai.com/index/introducing-gpt-5-2/},
  note   = {Accessed: 2026-03-03}
}

@misc{openai2025gptoss,
  author = {{OpenAI}},
  title  = {Introducing gpt-oss},
  year   = {2025},
  month  = {8},
  day    = {5},
  url    = {https://openai.com/index/introducing-gpt-oss/},
  note   = {Accessed: 2026-03-03}
}

@misc{openai2025gpt5mini,
  author = {{OpenAI}},
  title  = {GPT-5 Mini},
  year   = {2025},
  month  = {1},
  day    = {23},
  url    = {https://openai.com/index/gpt-5-mini/},
  note   = {Accessed: 2026-03-03}
}

@article{schick2023toolformer,
  title={Toolformer: Language models can teach themselves to use tools},
  author={Schick, Timo and Dwivedi-Yu, Jane and Dess{\`\i}, Roberto and Raileanu, Roberta and Lomeli, Maria and Hambro, Eric and Zettlemoyer, Luke and Cancedda, Nicola and Scialom, Thomas},
  journal={Advances in neural information processing systems},
  volume={36},
  pages={68539--68551},
  year={2023}
}

@inproceedings{suris2023vipergpt,
  title={Vipergpt: Visual inference via python execution for reasoning},
  author={Sur{\'\i}s, D{\'\i}dac and Menon, Sachit and Vondrick, Carl},
  booktitle={Proceedings of the IEEE/CVF international conference on computer vision},
  pages={11888--11898},
  year={2023}
}

@article{qin2023toolllm,
  title={Toolllm: Facilitating large language models to master 16000+ real-world apis},
  author={Qin, Yujia and Liang, Shihao and Ye, Yining and Zhu, Kunlun and Yan, Lan and Lu, Yaxi and Lin, Yankai and Cong, Xin and Tang, Xiangru and Qian, Bill and others},
  journal={arXiv preprint arXiv:2307.16789},
  year={2023}
}

@inproceedings{wang2024executable,
  title={Executable code actions elicit better llm agents},
  author={Wang, Xingyao and Chen, Yangyi and Yuan, Lifan and Zhang, Yizhe and Li, Yunzhu and Peng, Hao and Ji, Heng},
  booktitle={Forty-first International Conference on Machine Learning},
  year={2024}
}

@article{patil2024gorilla,
  title={Gorilla: Large language model connected with massive apis},
  author={Patil, Shishir G and Zhang, Tianjun and Wang, Xin and Gonzalez, Joseph E},
  journal={Advances in Neural Information Processing Systems},
  volume={37},
  pages={126544--126565},
  year={2024}
}

@inproceedings{mall2025disciple,
  title={Disciple: Learning interpretable programs for scientific visual discovery},
  author={Mall, Utkarsh and Phoo, Cheng Perng and Chiquier, Mia and Hariharan, Bharath and Bala, Kavita and Vondrick, Carl},
  booktitle={Proceedings of the Computer Vision and Pattern Recognition Conference},
  pages={29258--29267},
  year={2025}
}

@article{nori2025sequential,
  title={Sequential diagnosis with language models},
  author={Nori, Harsha and Daswani, Mayank and Kelly, Christopher and Lundberg, Scott and Ribeiro, Marco Tulio and Wilson, Marc and Liu, Xiaoxuan and Sounderajah, Viknesh and Carlson, Jonathan and Lungren, Matthew P and others},
  journal={arXiv preprint arXiv:2506.22405},
  year={2025}
}

@article{li2023llava,
  title={Llava-med: Training a large language-and-vision assistant for biomedicine in one day},
  author={Li, Chunyuan and Wong, Cliff and Zhang, Sheng and Usuyama, Naoto and Liu, Haotian and Yang, Jianwei and Naumann, Tristan and Poon, Hoifung and Gao, Jianfeng},
  journal={Advances in Neural Information Processing Systems},
  volume={36},
  pages={28541--28564},
  year={2023}
}

@inproceedings{bannur2023learning,
  title={Learning to exploit temporal structure for biomedical vision-language processing},
  author={Bannur, Shruthi and Hyland, Stephanie and Liu, Qianchu and Perez-Garcia, Fernando and Ilse, Maximilian and Castro, Daniel C and Boecking, Benedikt and Sharma, Harshita and Bouzid, Kenza and Thieme, Anja and others},
  booktitle={Proceedings of the IEEE/CVF conference on computer vision and pattern recognition},
  pages={15016--15027},
  year={2023}
}

@article{wang2025medagent,
  title={Medagent-pro: Towards evidence-based multi-modal medical diagnosis via reasoning agentic workflow},
  author={Wang, Ziyue and Wu, Junde and Cai, Linghan and Low, Chang Han and Yang, Xihong and Li, Qiaxuan and Jin, Yueming},
  journal={arXiv preprint arXiv:2503.18968},
  year={2025}
}

@article{korbak2025chain,
  title={Chain of thought monitorability: A new and fragile opportunity for ai safety},
  author={Korbak, Tomek and Balesni, Mikita and Barnes, Elizabeth and Bengio, Yoshua and Benton, Joe and Bloom, Joseph and Chen, Mark and Cooney, Alan and Dafoe, Allan and Dragan, Anca and others},
  journal={arXiv preprint arXiv:2507.11473},
  year={2025}
}

@article{chen2022program,
  title={Program of thoughts prompting: Disentangling computation from reasoning for numerical reasoning tasks},
  author={Chen, Wenhu and Ma, Xueguang and Wang, Xinyi and Cohen, William W},
  journal={arXiv preprint arXiv:2211.12588},
  year={2022}
}

@inproceedings{gao2023pal,
  title={Pal: Program-aided language models},
  author={Gao, Luyu and Madaan, Aman and Zhou, Shuyan and Alon, Uri and Liu, Pengfei and Yang, Yiming and Callan, Jamie and Neubig, Graham},
  booktitle={International conference on machine learning},
  pages={10764--10799},
  year={2023},
  organization={PMLR}
}

@article{qiao2023taskweaver,
  title={Taskweaver: A code-first agent framework},
  author={Qiao, Bo and Li, Liqun and Zhang, Xu and He, Shilin and Kang, Yu and Zhang, Chaoyun and Yang, Fangkai and Dong, Hang and Zhang, Jue and Wang, Lu and others},
  journal={arXiv preprint arXiv:2311.17541},
  year={2023}
}

@inproceedings{chen2024teaching,
    title={Teaching Large Language Models to Self-Debug},
    author={Xinyun Chen and Maxwell Lin and Nathanael Sch{\"a}rli and Denny Zhou},
    booktitle={The Twelfth International Conference on Learning Representations},
    year={2024}
}

@article{turpin2023language,
  title={Language models don't always say what they think: Unfaithful explanations in chain-of-thought prompting},
  author={Turpin, Miles and Michael, Julian and Perez, Ethan and Bowman, Samuel},
  journal={Advances in Neural Information Processing Systems},
  volume={36},
  pages={74952--74965},
  year={2023}
}

@article{song2026statllm,
  title={Statllm: A dataset for evaluating the performance of large language models in statistical analysis},
  author={Song, Xinyi and Lee, Lina and Xie, Kexin and Liu, Xueying and Deng, Xinwei and Hong, Yili},
  journal={Scientific Data},
  year={2026},
  publisher={Nature Publishing Group UK London}
}

@article{lu2025stateval,
  title={Stateval: A comprehensive benchmark for large language models in statistics},
  author={Lu, Yuchen and Yang, Run and Zhang, Yichen and Yu, Shuguang and Dai, Runpeng and Wang, Ziwei and Xiang, Jiayi and Gao, Siran and Ruan, Xinyao and Huang, Yirui and others},
  journal={arXiv preprint arXiv:2510.09517},
  year={2025}
}

@inproceedings{yao2022react,
  title={React: Synergizing reasoning and acting in language models},
  author={Yao, Shunyu and Zhao, Jeffrey and Yu, Dian and Du, Nan and Shafran, Izhak and Narasimhan, Karthik R and Cao, Yuan},
  booktitle={The eleventh international conference on learning representations},
  year={2022}
}

@article{bernard2018deep,
  title={Deep learning techniques for automatic MRI cardiac multi-structures segmentation and diagnosis: is the problem solved?},
  author={Bernard, Olivier and Lalande, Alain and Zotti, Clement and Cervenansky, Frederick and Yang, Xin and Heng, Pheng-Ann and Cetin, Irem and Lekadir, Karim and Camara, Oscar and Ballester, Miguel Angel Gonzalez and others},
  journal={IEEE transactions on medical imaging},
  volume={37},
  number={11},
  pages={2514--2525},
  year={2018},
  publisher={ieee}
}

@article{calabrese2022university,
  title={The University of California San Francisco preoperative diffuse glioma MRI dataset},
  author={Calabrese, Evan and Villanueva-Meyer, Javier E and Rudie, Jeffrey D and Rauschecker, Andreas M and Baid, Ujjwal and Bakas, Spyridon and Cha, Soonmee and Mongan, John T and Hess, Christopher P},
  journal={Radiology: Artificial Intelligence},
  volume={4},
  number={6},
  pages={e220058},
  year={2022},
  publisher={Radiological Society of North America}
}

@inproceedings{kirillov2023segment,
  title={Segment anything},
  author={Kirillov, Alexander and Mintun, Eric and Ravi, Nikhila and Mao, Hanzi and Rolland, Chloe and Gustafson, Laura and Xiao, Tete and Whitehead, Spencer and Berg, Alexander C and Lo, Wan-Yen and others},
  booktitle={Proceedings of the IEEE/CVF international conference on computer vision},
  pages={4015--4026},
  year={2023}
}

@article{ravi2024sam,
  title={Sam 2: Segment anything in images and videos},
  author={Ravi, Nikhila and Gabeur, Valentin and Hu, Yuan-Ting and Hu, Ronghang and Ryali, Chaitanya and Ma, Tengyu and Khedr, Haitham and R{\"a}dle, Roman and Rolland, Chloe and Gustafson, Laura and others},
  journal={arXiv preprint arXiv:2408.00714},
  year={2024}
}

@article{carion2025sam,
  title={Sam 3: Segment anything with concepts},
  author={Carion, Nicolas and Gustafson, Laura and Hu, Yuan-Ting and Debnath, Shoubhik and Hu, Ronghang and Suris, Didac and Ryali, Chaitanya and Alwala, Kalyan Vasudev and Khedr, Haitham and Huang, Andrew and others},
  journal={arXiv preprint arXiv:2511.16719},
  year={2025}
}

@article{ma2024segment,
  title={Segment anything in medical images},
  author={Ma, Jun and He, Yuting and Li, Feifei and Han, Lin and You, Chenyu and Wang, Bo},
  journal={Nature communications},
  volume={15},
  number={1},
  pages={654},
  year={2024},
  publisher={Nature Publishing Group UK London}
}

@article{ma2025medsam2,
  title={Medsam2: Segment anything in 3d medical images and videos},
  author={Ma, Jun and Yang, Zongxin and Kim, Sumin and Chen, Bihui and Baharoon, Mohammed and Fallahpour, Adibvafa and Asakereh, Reza and Lyu, Hongwei and Wang, Bo},
  journal={arXiv preprint arXiv:2504.03600},
  year={2025}
}

@article{zhu2024medical,
  title={Medical sam 2: Segment medical images as video via segment anything model 2},
  author={Zhu, Jiayuan and Hamdi, Abdullah and Qi, Yunli and Jin, Yueming and Wu, Junde},
  journal={arXiv preprint arXiv:2408.00874},
  year={2024}
}

@article{zhao2025large,
  title={Large-vocabulary segmentation for medical images with text prompts},
  author={Zhao, Ziheng and Zhang, Yao and Wu, Chaoyi and Zhang, Xiaoman and Zhou, Xiao and Zhang, Ya and Wang, Yanfeng and Xie, Weidi},
  journal={NPJ Digital Medicine},
  volume={8},
  number={1},
  pages={566},
  year={2025},
  publisher={Nature Publishing Group UK London}
}

@article{rokuss2025voxtell,
  title={Voxtell: Free-text promptable universal 3d medical image segmentation},
  author={Rokuss, Maximilian and Langenberg, Moritz and Kirchhoff, Yannick and Isensee, Fabian and Hamm, Benjamin and Ulrich, Constantin and Regnery, Sebastian and Bauer, Lukas and Katsigiannopulos, Efthimios and Norajitra, Tobias and others},
  journal={arXiv preprint arXiv:2511.11450},
  year={2025}
}

@article{isensee2021nnu,
  title={nnU-Net: a self-configuring method for deep learning-based biomedical image segmentation},
  author={Isensee, Fabian and Jaeger, Paul F and Kohl, Simon AA and Petersen, Jens and Maier-Hein, Klaus H},
  journal={Nature methods},
  volume={18},
  number={2},
  pages={203--211},
  year={2021},
  publisher={Nature Publishing Group US New York}
}

@article{lu2024ai,
  title={The ai scientist: Towards fully automated open-ended scientific discovery},
  author={Lu, Chris and Lu, Cong and Lange, Robert Tjarko and Foerster, Jakob and Clune, Jeff and Ha, David},
  journal={arXiv preprint arXiv:2408.06292},
  year={2024}
}

@article{gottweis2025towards,
  title={Towards an AI co-scientist},
  author={Gottweis, Juraj and Weng, Wei-Hung and Daryin, Alexander and Tu, Tao and Palepu, Anil and Sirkovic, Petar and Myaskovsky, Artiom and Weissenberger, Felix and Rong, Keran and Tanno, Ryutaro and others},
  journal={arXiv preprint arXiv:2502.18864},
  year={2025}
}

@article{ghareeb2025robin,
  title={Robin: A multi-agent system for automating scientific discovery},
  author={Ghareeb, Ali Essam and Chang, Benjamin and Mitchener, Ludovico and Yiu, Angela and Szostkiewicz, Caralyn J and Laurent, Jon M and Razzak, Muhammed T and White, Andrew D and Hinks, Michaela M and Rodriques, Samuel G},
  journal={arXiv preprint arXiv:2505.13400},
  year={2025}
}

@article{schmidgall2025agent,
  title={Agent laboratory: Using llm agents as research assistants},
  author={Schmidgall, Samuel and Su, Yusheng and Wang, Ze and Sun, Ximeng and Wu, Jialian and Yu, Xiaodong and Liu, Jiang and Moor, Michael and Liu, Zicheng and Barsoum, Emad},
  journal={Findings of the Association for Computational Linguistics: EMNLP 2025},
  pages={5977--6043},
  year={2025},
  publisher={Association for Computational Linguistics}
}

@inproceedings{wu2024autogen,
  title={Autogen: Enabling next-gen LLM applications via multi-agent conversations},
  author={Wu, Qingyun and Bansal, Gagan and Zhang, Jieyu and Wu, Yiran and Li, Beibin and Zhu, Erkang and Jiang, Li and Zhang, Xiaoyun and Zhang, Shaokun and Liu, Jiale and others},
  booktitle={First conference on language modeling},
  year={2024}
}

@article{yang2025moose,
  title={Moose-chem2: Exploring llm limits in fine-grained scientific hypothesis discovery via hierarchical search},
  author={Yang, Zonglin and Liu, Wanhao and Gao, Ben and Liu, Yujie and Li, Wei and Xie, Tong and Bing, Lidong and Ouyang, Wanli and Cambria, Erik and Zhou, Dongzhan},
  journal={arXiv preprint arXiv:2505.19209},
  year={2025}
}

@inproceedings{wang2024scimon,
  title={Scimon: Scientific inspiration machines optimized for novelty},
  author={Wang, Qingyun and Downey, Doug and Ji, Heng and Hope, Tom},
  booktitle={Proceedings of the 62nd Annual Meeting of the Association for Computational Linguistics (Volume 1: Long Papers)},
  pages={279--299},
  year={2024}
}

@article{boiko2023autonomous,
  title={Autonomous chemical research with large language models},
  author={Boiko, Daniil A and MacKnight, Robert and Kline, Ben and Gomes, Gabe},
  journal={Nature},
  volume={624},
  number={7992},
  pages={570--578},
  year={2023},
  publisher={Nature Publishing Group UK London}
}

@article{m2024augmenting,
  title={Augmenting large language models with chemistry tools},
  author={M. Bran, Andres and Cox, Sam and Schilter, Oliver and Baldassari, Carlo and White, Andrew D and Schwaller, Philippe},
  journal={Nature machine intelligence},
  volume={6},
  number={5},
  pages={525--535},
  year={2024},
  publisher={Nature Publishing Group UK London}
}

@inproceedings{hong2025data,
  title={Data interpreter: An llm agent for data science},
  author={Hong, Sirui and Lin, Yizhang and Liu, Bang and Liu, Bangbang and Wu, Binhao and Zhang, Ceyao and Li, Danyang and Chen, Jiaqi and Zhang, Jiayi and Wang, Jinlin and others},
  booktitle={Findings of the Association for Computational Linguistics: ACL 2025},
  pages={19796--19821},
  year={2025}
}

@article{yang2024swe,
  title={Swe-agent: Agent-computer interfaces enable automated software engineering},
  author={Yang, John and Jimenez, Carlos E and Wettig, Alexander and Lieret, Kilian and Yao, Shunyu and Narasimhan, Karthik and Press, Ofir},
  journal={Advances in Neural Information Processing Systems},
  volume={37},
  pages={50528--50652},
  year={2024}
}

@article{agrawal2026can,
  title={Can AI Conduct Autonomous Scientific Research? Case Studies on Two Real-World Tasks},
  author={Agrawal, Shreyansh and Anadkat, Harsh B and Athimoolam, Kiran K and Bhardwaj, Harsh and Chowdhury, Trishul and Gao, Shengtao and Kamat, Purva K and Makwana, Vishwadeepsinh and Shariff, Mohammed H and Badkul, Amitesh and others},
  journal={bioRxiv},
  pages={2026--01},
  year={2026},
  publisher={Cold Spring Harbor Laboratory}
}

@article{wei2025ai,
  title={From ai for science to agentic science: A survey on autonomous scientific discovery},
  author={Wei, Jiaqi and Yang, Yuejin and Zhang, Xiang and Chen, Yuhan and Zhuang, Xiang and Gao, Zhangyang and Zhou, Dongzhan and Wang, Guangshuai and Gao, Zhiqiang and Cao, Juntai and others},
  journal={arXiv preprint arXiv:2508.14111},
  year={2025}
}

@article{gao2024empowering,
  title={Empowering biomedical discovery with AI agents},
  author={Gao, Shanghua and Fang, Ada and Huang, Yepeng and Giunchiglia, Valentina and Noori, Ayush and Schwarz, Jonathan Richard and Ektefaie, Yasha and Kondic, Jovana and Zitnik, Marinka},
  journal={Cell},
  volume={187},
  number={22},
  pages={6125--6151},
  year={2024},
  publisher={Elsevier}
}

@article{huang2025automated,
  title={Automated hypothesis validation with agentic sequential falsifications},
  author={Huang, Kexin and Jin, Ying and Li, Ryan and Li, Michael Y and Cand{\`e}s, Emmanuel and Leskovec, Jure},
  journal={arXiv preprint arXiv:2502.09858},
  year={2025}
}

@article{baek2024agent,
    author={Baek, Jinheon and Jauhar, Sujay Kumar and Cucerzan, Silviu and Hwang, Sung Ju},
    title={ResearchAgent: Iterative Research Idea Generation over Scientific Literature with Large Language Models},
    journal={arXiv preprint arXiv:2404.07738},
    year={2024}
}

@article{anthropic2024claude,
    author={{Anthropic}},
    title={The Claude 3 Model Family: Opus, Sonnet, Haiku},
    journal={Anthropic},
    year={2024}
}

@article{comanici2025gemini,
  title={Gemini 2.5: Pushing the frontier with advanced reasoning, multimodality, long context, and next generation agentic capabilities},
  author={Comanici, Gheorghe and Bieber, Eric and Schaekermann, Mike and Pasupat, Ice and Sachdeva, Noveen and Dhillon, Inderjit and Blistein, Marcel and Ram, Ori and Zhang, Dan and Rosen, Evan and others},
  journal={arXiv preprint arXiv:2507.06261},
  year={2025}
}

@article{Davidson-Pilon2019,
  doi = {10.21105/joss.01317},
  url = {https://doi.org/10.21105/joss.01317},
  year = {2019},
  publisher = {The Open Journal},
  volume = {4},
  number = {40},
  pages = {1317},
  author = {Cameron Davidson-Pilon},
  title = {lifelines: survival analysis in Python},
  journal = {Journal of Open Source Software}
}

@misc{ollama2023,
  author       = {{Ollama Team}},
  title        = {Ollama: Get up and running with large language models},
  howpublished = {\url{https://ollama.com/}},
  year         = {2023},
  note         = {Accessed: 2026-03-03}
}

@misc{langgraph2024,
  author       = {{LangChain AI}},
  title        = {LangGraph: Building stateful, multi-actor applications with LLMs},
  howpublished = {\url{https://github.com/langchain-ai/langgraph}},
  year         = {2024},
  note         = {Accessed: 2026-03-03}
}

@misc{openrouter2024,
  author       = {{OpenRouter Team}},
  title        = {OpenRouter: A unified interface for LLMs},
  howpublished = {\url{https://openrouter.ai/}},
  year         = {2024},
  note         = {Accessed: 2026-03-03}
}

@article{goodman1994use,
  title={The use of predicted confidence intervals when planning experiments and the misuse of power when interpreting results},
  author={Goodman, Steven N and Berlin, Jesse A},
  journal={Annals of internal medicine},
  volume={121},
  number={3},
  pages={200--206},
  year={1994},
  publisher={American College of Physicians}
}

@article{hoenig2001abuse,
  title={The abuse of power: the pervasive fallacy of power calculations for data analysis},
  author={Hoenig, John M and Heisey, Dennis M},
  journal={The American Statistician},
  volume={55},
  number={1},
  pages={19--24},
  year={2001},
  publisher={Taylor \& Francis}
}

@article{westreich2013table,
  title={The table 2 fallacy: presenting and interpreting confounder and modifier coefficients},
  author={Westreich, Daniel and Greenland, Sander},
  journal={American journal of epidemiology},
  volume={177},
  number={4},
  pages={292--298},
  year={2013},
  publisher={Oxford University Press}
}

@book{Pearl2009causality,
	author = {Judea Pearl},
	editor = {},
	publisher = {Cambridge University Press},
	title = {Causality: Models, Reasoning, and Inference},
	year = {2009}
}

@article{cohen1994earth,
  title={The earth is round (p<. 05).},
  author={Cohen, Jacob},
  journal={American psychologist},
  volume={49},
  number={12},
  pages={997},
  year={1994},
  publisher={American Psychological Association}
}

@article{ioannidis2005most,
  title={Why most published research findings are false},
  author={Ioannidis, John PA},
  journal={PLoS medicine},
  volume={2},
  number={8},
  pages={e124},
  year={2005},
  publisher={Public Library of Science}
}

@article{benjamin2018redefine,
  title={Redefine statistical significance},
  author={Benjamin, Daniel J and Berger, James O and Johannesson, Magnus and Nosek, Brian A and Wagenmakers, E-J and Berk, Richard and Bollen, Kenneth A and Brembs, Bj{\"o}rn and Brown, Lawrence and Camerer, Colin and others},
  journal={Nature human behaviour},
  volume={2},
  number={1},
  pages={6--10},
  year={2018},
  publisher={Nature Publishing Group UK London}
}

@misc{wasserstein2019moving,
  title={Moving to a world beyond “p< 0.05”},
  author={Wasserstein, Ronald L and Schirm, Allen L and Lazar, Nicole A},
  journal={The American Statistician},
  volume={73},
  number={sup1},
  pages={1--19},
  year={2019},
  publisher={Taylor \& Francis}
}

@article{lakens2017equivalence,
  title={Equivalence tests: A practical primer for t tests, correlations, and meta-analyses},
  author={Lakens, Dani{\"e}l},
  journal={Social psychological and personality science},
  volume={8},
  number={4},
  pages={355--362},
  year={2017},
  publisher={Sage Publications Sage CA: Los Angeles, CA}
}

@article{lakens2018equivalence,
  title={Equivalence testing for psychological research: A tutorial},
  author={Lakens, Dani{\"e}l and Scheel, Anne M and Isager, Peder M},
  journal={Advances in methods and practices in psychological science},
  volume={1},
  number={2},
  pages={259--269},
  year={2018},
  publisher={Sage Publications Sage CA: Los Angeles, CA}
}

@article{zwanenburg2020image,
  title={The image biomarker standardization initiative: standardized quantitative radiomics for high-throughput image-based phenotyping},
  author={Zwanenburg, Alex and Valli{\`e}res, Martin and Abdalah, Mahmoud A and Aerts, Hugo JWL and Andrearczyk, Vincent and Apte, Aditya and Ashrafinia, Saeed and Bakas, Spyridon and Beukinga, Roelof J and Boellaard, Ronald and others},
  journal={Radiology},
  volume={295},
  number={2},
  pages={328--338},
  year={2020},
  publisher={Radiological Society of North America}
}

@article{van2017computational,
  title={Computational radiomics system to decode the radiographic phenotype},
  author={Van Griethuysen, Joost JM and Fedorov, Andriy and Parmar, Chintan and Hosny, Ahmed and Aucoin, Nicole and Narayan, Vivek and Beets-Tan, Regina GH and Fillion-Robin, Jean-Christophe and Pieper, Steve and Aerts, Hugo JWL},
  journal={Cancer research},
  volume={77},
  number={21},
  pages={e104--e107},
  year={2017},
  publisher={American Association for Cancer Research}
}

@article{button2013power,
  title={Power failure: why small sample size undermines the reliability of neuroscience},
  author={Button, Katherine S and Ioannidis, John PA and Mokrysz, Claire and Nosek, Brian A and Flint, Jonathan and Robinson, Emma SJ and Munaf{\`o}, Marcus R},
  journal={Nature reviews neuroscience},
  volume={14},
  number={5},
  pages={365--376},
  year={2013},
  publisher={Nature Publishing Group UK London}
}

@article{menze2014multimodal,
  title={The Multimodal Brain Tumor Image Segmentation Benchmark ({BRATS})},
  author={Menze, Bjoern H and Jakab, Andras and Bauer, Stefan and Kalpathy-Cramer, Jayashree and Farahani, Keyvan and Kirby, Justin and Burren, Yuliya and Porz, Nicole and Slotboom, Johannes and Wiest, Roland and others},
  journal={IEEE Transactions on Medical Imaging},
  volume={34},
  number={10},
  pages={1993--2024},
  year={2015},
  publisher={IEEE},
  doi={10.1109/TMI.2014.2377694}
}

@article{bakas2018identifying,
  title={Identifying the Best Machine Learning Algorithms for Brain Tumor Segmentation, Progression Assessment, and Overall Survival Prediction in the {BRATS} Challenge},
  author={Bakas, Spyridon and Reyes, Mauricio and Jakab, Andras and Bauer, Stefan and Rempfler, Markus and Crimi, Alessandro and Shinohara, Russell Takeshi and Berger, Christoph and Ha, Sung Min and Rozycki, Martin and others},
  journal={arXiv preprint arXiv:1811.02629},
  year={2018}
}
